\documentclass[12pt,preprint]{aastex}

\newcommand{\etal}{et~al.\ }

\newcommand{\feka}{\hbox{Fe\,K$\alpha$}}

\newcommand{\simgt}{\lower 2pt \hbox{$\, \buildrel {\scriptstyle >}\over {\scriptstyle\sim}\,$}}
\newcommand{\simlt}{\lower 2pt \hbox{$\, \buildrel {\scriptstyle <}\over {\scriptstyle\sim}\,$}}

\newcommand{\rxj}{RX~J0911.4+0551}

\newcommand{\cxo}{{\emph{Chandra X-ray Observatory}}}

\newcommand{\chandra}{{\emph{Chandra}}}

\newcommand{\hst}{{\emph{HST}}}

\shorttitle{\emph{CHANDRA} OBSERVATIONS of GRAVITATIONAL LENSES}
\shortauthors{CHEN ET AL.}

\tighten

\begin{document}

\def\sarc{$^{\prime\prime}\!\!.$}
\def\arcsec{$^{\prime\prime}$}

\title{X-ray Monitoring of Gravitational Lenses With \emph{Chandra}}

\author{Bin Chen\altaffilmark{1}, Xinyu Dai\altaffilmark{1},  Christopher S. Kochanek\altaffilmark{2}, George Chartas\altaffilmark{3},  \\ Jeffrey A. Blackburne\altaffilmark{2}, Christopher W. Morgan\altaffilmark{4}  }

\altaffiltext{1}{Homer L. Dodge Department of Physics and Astronomy, The University of Oklahoma,
Norman, OK, 73019, USA, bchen@ou.edu}

\altaffiltext{2}{Department of Astronomy, The Ohio State University, Columbus, OH 43210, USA}

\altaffiltext{3}{Department of Physics and Astronomy, College of Charleston, SC 29424, USA}

\altaffiltext{4}{Department of Physics, United States Naval Academy, 572C Holloway Road, Annapolis, MD 21402, USA}

\begin{abstract}
We present \emph{Chandra}  monitoring data for six gravitationally lensed quasars: QJ~0158$-$4325, HE~0435$-$1223,  HE~1104$-$1805, SDSS~0924+0219, SDSS~1004+4112, and Q~2237+0305. 
We detect X-ray microlensing variability in all six lenses with high confidence. 
We detect energy dependent microlensing in HE~0435$-$1223, SDSS~1004+4112, SDSS~0924+0219 and  Q~2237+0305. 
We present a detailed spectral analysis for each lens, and find that simple power-law models plus Gaussian emission lines give good fits to the spectra. 
We detect intrinsic spectral variability in two epochs of Q~2237+0305. 
We detect differential absorption between  images in four lenses. 
We also detect the \feka\ emission line in all six lenses, and the Ni XXVII~K$\alpha$ line in two images of Q~2237+0305. 
The rest frame equivalent widths of the \feka\ lines are measured to be 0.4--1.2 keV, significantly higher than those measured in typical active galactic nuclei of similar X-ray luminosities. 
This suggests that the \feka\  emission region is more compact or centrally concentrated than the continuum emission region.
\end{abstract}


\section{Introduction}

The concept of quasar microlensing, the effect of gravitational lensing by stellar mass objects in a lens galaxy on background quasars, was proposed  almost immediately (e.g., Chang \& Refsdal 1979; Gott 1981) after the discovery of the first gravitational lens (Walsh et al.\ 1979).  
However, it was only ten years later that the effect was observationally detected in Q~0957+561 (Vanderriest et al.\ 1989) and Q~2237+0305 (Irwin et al.\ 1989).  
Recently, quasar microlensing has become an important tool for probing the structure of quasar accretion disks and distinguishing models for their non-thermal emission (see the reviews  by Wambsganss 2006; Kochanek et al.\ 2007). 
The ensemble of microlensing stars produces a magnification pattern in the source plane, and as the source moves across this pattern due to relative motions between the source, the lens, the stars in the lens, and the observer, the source magnification varies with time. 
More importantly for studying accretion disk models, the (time varying) microlensing magnification depends on the size and spatial structure of the emission source in the central engine of quasar. 
Larger emission regions smooth the magnification patterns more heavily and will have less microlensing variability than smaller emission regions.
Additionally, the magnification diverges on caustics, which makes it possible to resolve emission regions arbitrarily smaller than the Einstein radius of the stars.
Unlike almost any other astrophysical probe, microlensing becomes a more powerful tool as the source becomes smaller. 
As a result,  quasar microlensing is a unique tool for probing the innermost structure of accretion disks. 

There are two major challenges in utilizing this approach: first, to measure the microlensing light curves and second, to interpret the curves.
There are several effects contributing to the light curves of quasar images and their variability. 
First, the average fluxes of the images are affected by the macro lens model, milli-lensing by substructures in the lens (e.g., Kochanek \& Dalal 2004; Cohn \& Kochanek 2004; Keeton \& Moustakas 2009), extinction by dust (e.g., Falco et al.\ 1999; El{\'{\i}}asd{\'o}ttir et al.\ 2006), and X-ray absorption by gas in the lens (e.g., Dai et al.\ 2006; Dai \& Kochanek 2009). 
Second, they vary with time due to  a combination of microlensing and the intrinsic variability of quasars.
Of these effects, the intrinsic variability is usually relatively easy to remove by measuring the time delays between the images and then shifting the light curves.
Many of other effects are also energy/wavelength dependent, but the best way to distinguish microlensing variability from other effects is through long term monitoring programs at the relevant energies. This approach also has the advantages of ameliorating systematic uncertainties due to the mean stellar mass in the lens galaxy.  
Limited by observational resources, early microlensing studies were confined to either single epoch, multiple wavelength (e.g., Falco et al.\ 1996; Mediavilla et al.\ 1998; Agol et al.\ 2000), or multiple epoch, single wavelength studies (e.g., Wo$\acute{\rm{z}}$niak \etal 2000).
Only recently have there been intensive studies spanning both the time and wavelength (e.g., Poindexter \etal 2008; Chen \etal 2011; Blackburne \etal 2011; Mosquera et al. 2011; Mu$\rm \tilde{n}$oz et al. 2011).

Theoretically, the importance of the clustering of caustics on interpretation of  quasar microlensing light curves was realized quite early (e.g., Wambsganss et al.\ 1990).  
Because of the high optical depth of quasar microlensing in lens systems, quasar microlensing light curves are generally affected by many stars, thus requiring theoretical models constructed for these complicated conditions.
The only real solution has been to simply model the data using variants of the Bayesian Monte Carlo method (Kochanek 2004). 
This technique and its variants have led to unique constraints on the emission sizes of accretion disks in different wavelengths (Morgan et al.\ 2008; Eigenbrod et al. 2008; Chartas et al.\ 2009; Dai et al.\ 2010; Mediavilla et al.\ 2011, Poindexter \& Kochanek 2010b; Blackburne et al. 2011), and on the stellar/dark matter distribution of the foreground lensing galaxies (e.g., Mediavilla et al.\ 2009; Poindexter \& Kochanek 2010a; Bate et al.\ 2011; Pooley et al. 2012).

We are particularly interested in using microlensing to study the non-thermal, X-ray emission regions of quasar accretion disks.  
The X-ray continuum emission of quasars is observed to follow a power-law, and it is believed to be generated through the unsaturated inverse Compton scattering of UV disk photons on hot electrons in a corona above the disk (see  the review by Reynolds \& Nowak 2003). 
The origin of the corona, its geometrical size and detailed structure are still topics under debate.  
In addition to the X-ray continuum, the disk-corona model also predicts the existence of a reflection component including emission lines (e.g., \feka) formed through irradiation of the colder accretion disk by the continuum (e.g., George \& Fabian 1991).  
However, the location of this reflection component, whether it comes from the disk, the disk-wind, or the ``torus", has been debated for more than two decades.
In principle, with enough observations, the quasar microlensing technique can resolve all these issues.

Microlensing of X-rays was first tentatively suggested in \rxj\ (Morgan et al.\ 2001).  
Because of the lack of multi-epoch observations and the large intrinsic X-ray variability in quasars, later studies focused on  microlensing of the \feka\ line (Chartas et al.\ 2002, 2004; Dai et al.\ 2003; Ota et al.\ 2006), where dramatic changes in  the \feka\ line equivalent widths  were most easily interpreted. 
Studies of the X-ray continuum resumed with the report of the dramatic flux ratio differences between the X-ray and optical bands in RXJ1131$-$1231 (Blackburne et al.\ 2006) and the single epoch, optical-to-X-ray flux ratio comparison for a sample of lenses by Pooley et al.\ (2006).
Short \chandra\ monitoring programs of a few lenses start to emerge at this stage (e.g., Chartas et al.\ 2009), and combining them with the Bayesian Monte Carlo microlensing analysis technique (Kochanek 2004), we have generally found that the X-ray continuum is emitted on scales of roughly $10\,r_g$ (Morgan et al.\ 2008; Chartas et al.\ 2009; Dai et al.\ 2010; Blackburne et al. 2011), 7--10 times smaller than the observed optical emission region (Morgan et al.\ 2010).
However,  the current lower limits on the sizes of the X-ray emission regions  are not well constrained, because strong lower limits require more densely sampled light curves than any presently available. 
Nonetheless, these microlensing results have ruled out disk-corona models where the corona covers extensive areas of the optical accretion disk.
The large equivalent widths of the \feka\ line measured in a few lensed quasars suggest that a significant fraction of the \feka\ line flux originates from a region smaller than the X-ray continuum emission region (Dai et al.\ 2003; Popovi$\rm\acute{c}$ et al. 2003, 2006).
However, these studies were limited by the lack of deep, better sampled X-ray microlensing light curves.

In this paper, we report results from monitoring six gravitationally lensed quasars with \chandra: QJ~0158$-$4325 (Maza et al. 1995; Morgan et al.\ 1999),  SDSS~0924+0219 (Inada et al.\ 2003a), SDSS~1004+4112 (Inada et al.\ 2003b), HE~0435$-$1223 (Wisotzki et al.\ 2002), HE~1104$-$1805 (Wisotzki \etal 1993), and Q~2237+0305 (Huchra et al.\ 1985)\footnote{Basic information for each lens system (e.g., the lens/source redshifts, celestial coordinates, and relative image positions) can be found on the CfA-Arizona Space Telescope Lens Survey (CASTLES) website (http://cfa-www.harvard.edu/castles/).}.
This is the largest quasar microlensing X-ray monitoring program to date.
QJ~0158$-$4325 and HE~1104$-$1805 are doubly imaged  while the other four systems have four bright images. 
Five of the foreground lenses are galaxies, while SDSS~1004+4112 is lensed by a galaxy cluster.  
Since microlensing in the optical bands has been extensively studied in these systems (e.g., Morgan et al.\ 2006; Kochanek et al.\ 2006; Lamer et al. 2006; Fohlmeister et al. 2007, 2008;  Eigenbrod et al. 2008; Faure et al. 2009; Poindexter \& Kochanek 2010a, 2010b;  Ricci et al. 2011; Blackburne et al. 2011), our X-ray microlensing light curves provide complementary data to constrain non-thermal emission processes in accretion disks.
For example, a simple comparison of the microlensing variability between the soft and the hard X-ray bands in Q~2237+0305 has already led us to conclude that the hard X-ray continuum emission region is smaller than that of the soft in this quasar, suggesting a temperature gradient in the X-ray emitting corona (Chen et al.\ 2011). 
In our original proposals, only the observations of Q~2237+0305  and RXJ1131$-$1231 (which will be discussed in Chartas et al. 2012) were explicitly designed to have the count rates needed to obtain accurate measurements in multiple energy bands. 

We present the \chandra\ observations and the data reduction techniques in Section~2, the image analysis in Section~3, and the spectral analysis in Section~4. The microlensing analysis is in Section~5, and the discussion and conclusions in Section~6. 
Throughout this paper, we assume a flat $\Lambda$CDM cosmology with $\Omega_\Lambda=0.73,$ $\Omega_M=0.27,$ and $\rm H_0=70\, km\, s^{-1}\, Mpc^{-1} .$

\section{Observations And Data Reduction}

 All observations presented in this paper were made  with ACIS (Garmire et al. 2003) on the \cxo\ (Weisskopf et al.\ 2002). 
 With an on-axis point spread function (PSF) of 0\sarc 5, \chandra\ is the only X-ray telescope that can resolve lensed images typically separated by 1--2\arcsec.  
 The majority of the observations are from our large \chandra\ Cycle 11 program to monitor seven gravitational lenses. 
 Here, we report on six of the lenses, QJ~0158$-$4325,  SDSS~0924+0219, SDSS~1004+4112,  HE~0435$-$1223, HE~1104$-$1805, and Q~2237+0305. 
 Although the absorption corrected count rates of Q~2237+0305 were published in Chen et al. (2011), we present the detailed spectral analysis in this paper. 
 The analysis of  RXJ1131$-$1231 will be presented in a companion paper (Chartas et al. 2012, in preparation). 
 We also analyzed the archival \chandra\ data for the six lenses.  
 We list the basic information for these observations, such as observation dates and exposure times, in Tables  \ref{tab:0158count}$-$\ref{tab:2237count}.

 We reprocessed all data using the CIAO 4.3 software tools  by first removing the pixel randomization and applying a sub-pixel algorithm for event positions (Tsunemi et al.\ 2001). 
 We then filtered all events using the standard ASCA grades of 0, 2, 3, 4  and 6 with energies between 0.4 and 8.0 keV (observer's frame), and further separated the events into soft and hard  bands for subsequent analyses.   
 We selected the energy boundary between the soft and hard band for each lensing system to balance the soft and hard band count rates  (see e.g., Table \ref{tab:0158count}) in order to produce the best statistics for detecting  differences between  the two energy bands. 
 For QJ~0158$-$4325, SDSS~0924+0219, and SDSS~1004+4112  we  split the soft and hard bands at 3.0~keV (rest frame), for HE~0435$-$1223 and Q~2237+0305 we split the soft and hard bands at 3.5~keV (rest frame), and for HE~1104$-$1805, we split the soft and hard bands at 4.0~keV (rest frame).  
In Chen et al. (2011), we used (observed frame) full, soft, and hard bands at 0.2--8.0 keV,  0.2--2.0 keV, and 2.0--8.0 keV respectively. In the current paper, we define the full band as 0.4--8.0 keV, and divide the soft and hard bands at 1.3 keV (3.5 keV in rest frame). 
This energy division provides better statistical results than our definitions in Chen et al. (2011), but it alters none of our conclusions. 
The pile-up effect is negligible for all our sources ($<$2\% in all images), since they are generally faint. 
 Nevertheless, we used the 1/2 subarray mode in our observations of  QJ~0158$-$4325,  SDSS~0924+0219,  HE~0435$-$1223, HE~1104$-$1805, and Q~2237+0305 in order to minimize any residual pile-up effects. 
 In our program, RXJ1131$-$1231 is the only  case where the pile-up effect needs to be extensively modeled (Chartas et al.\ 2009).
\section{Imaging  Analysis}\label{image}

\begin{figure*}
\begin{center}$
\begin{array}{cc}
\includegraphics[width=0.45\textwidth,height=0.3\textheight]{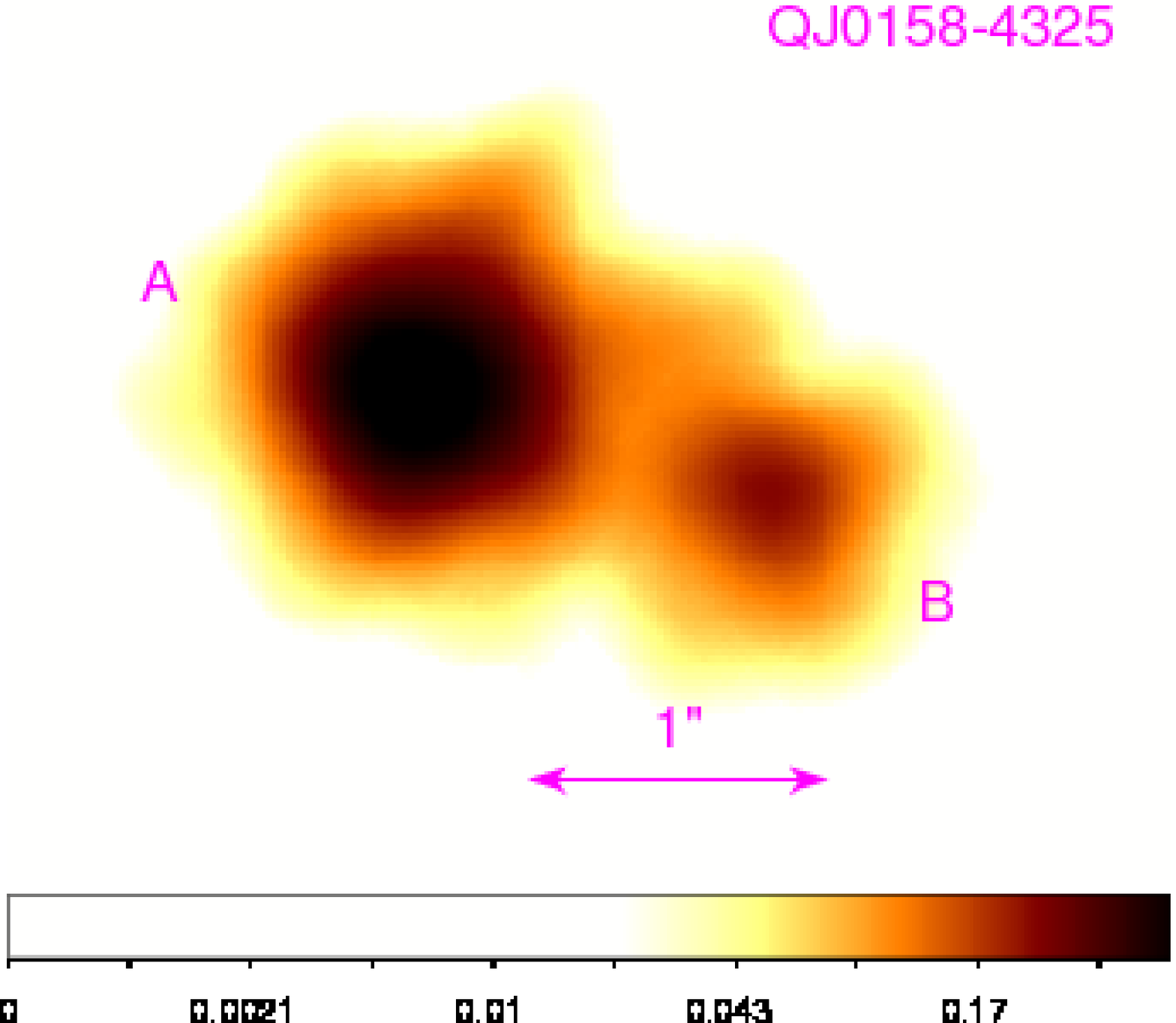}
\hspace{15pt}
\includegraphics[width=0.45\textwidth,height=0.3\textheight]{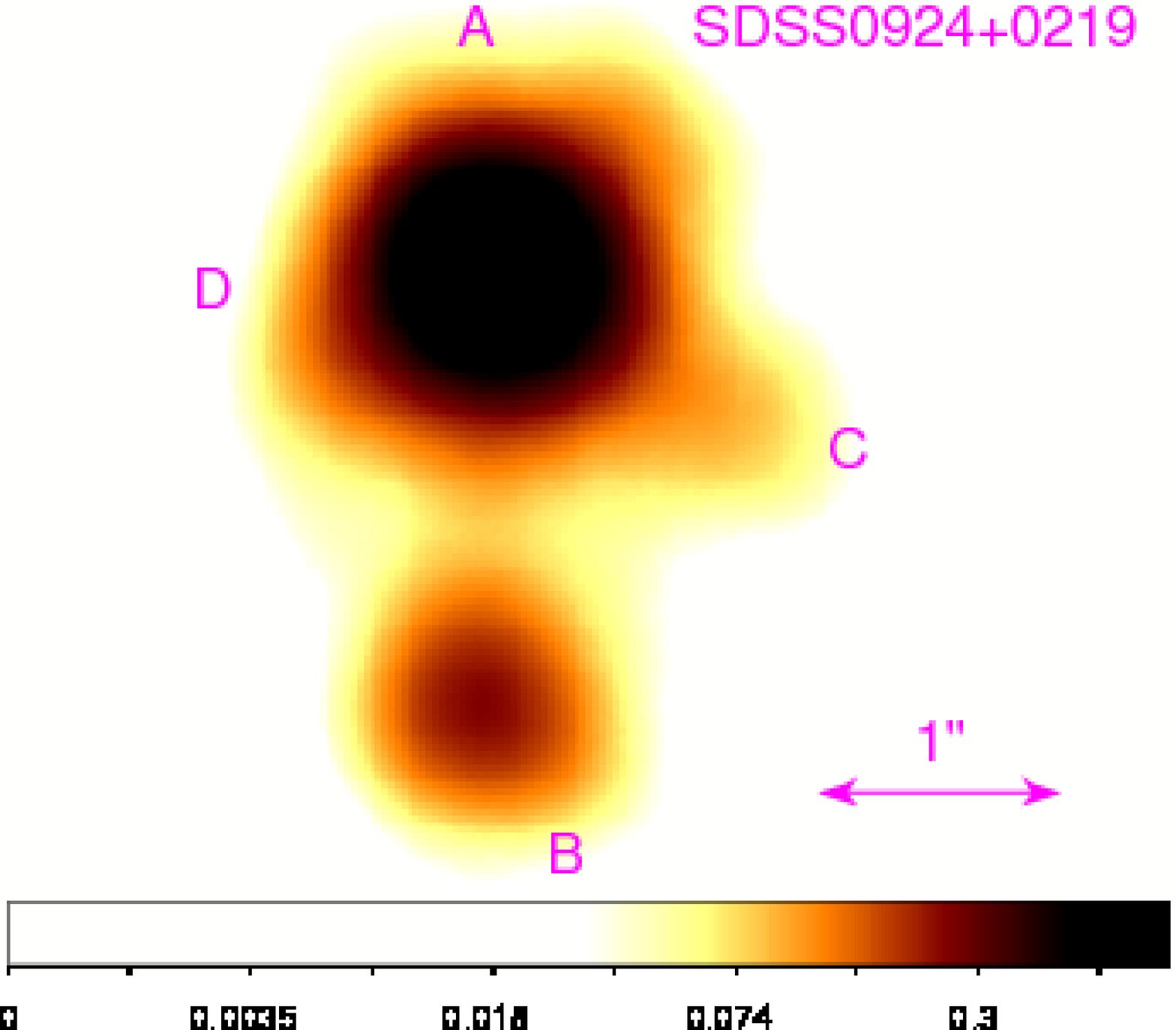}\\
\vspace{10pt}
\includegraphics[width=0.45\textwidth,height=0.3\textheight]{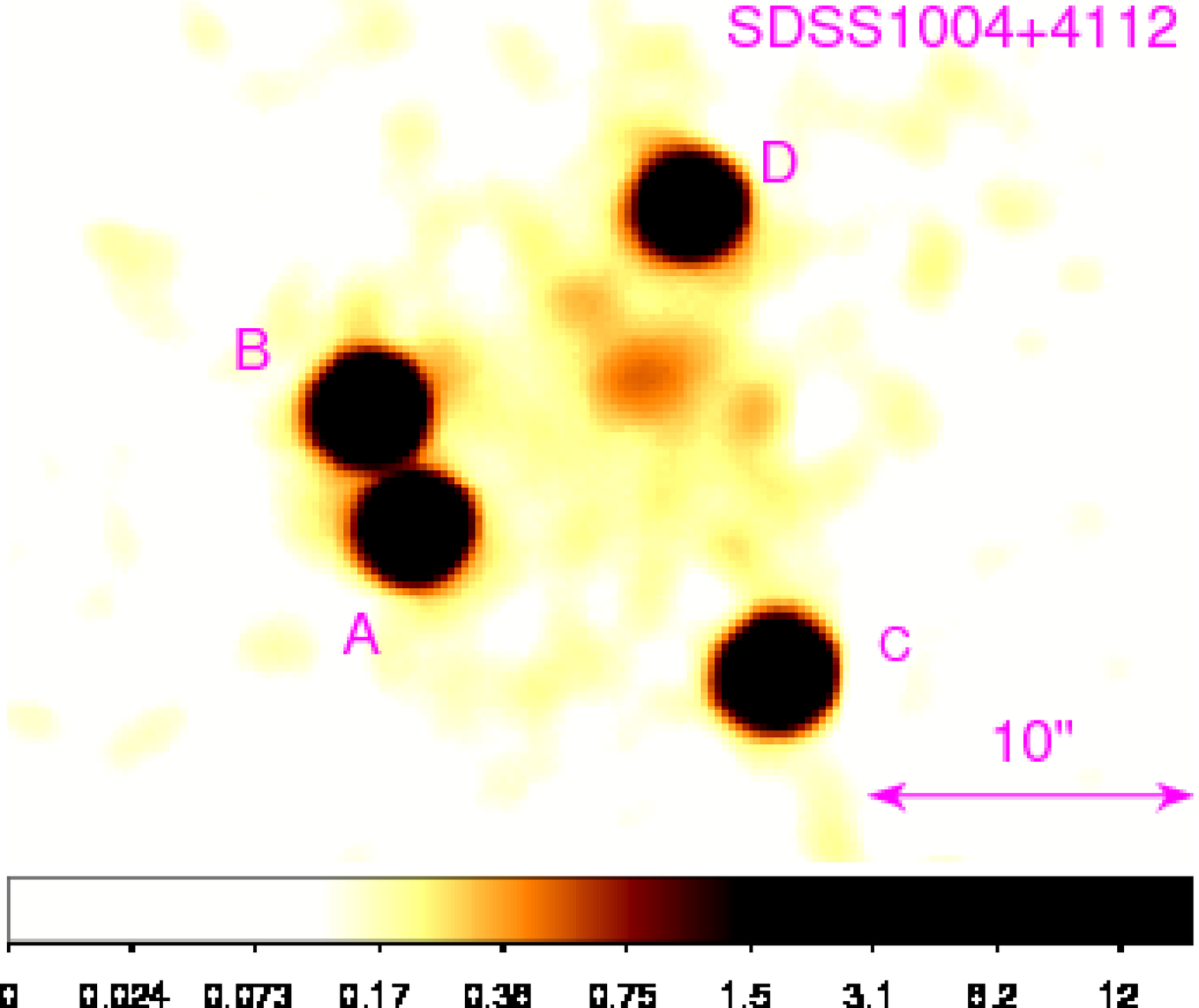}
\hspace{15pt}
\includegraphics[width=0.45\textwidth,height=0.3\textheight]{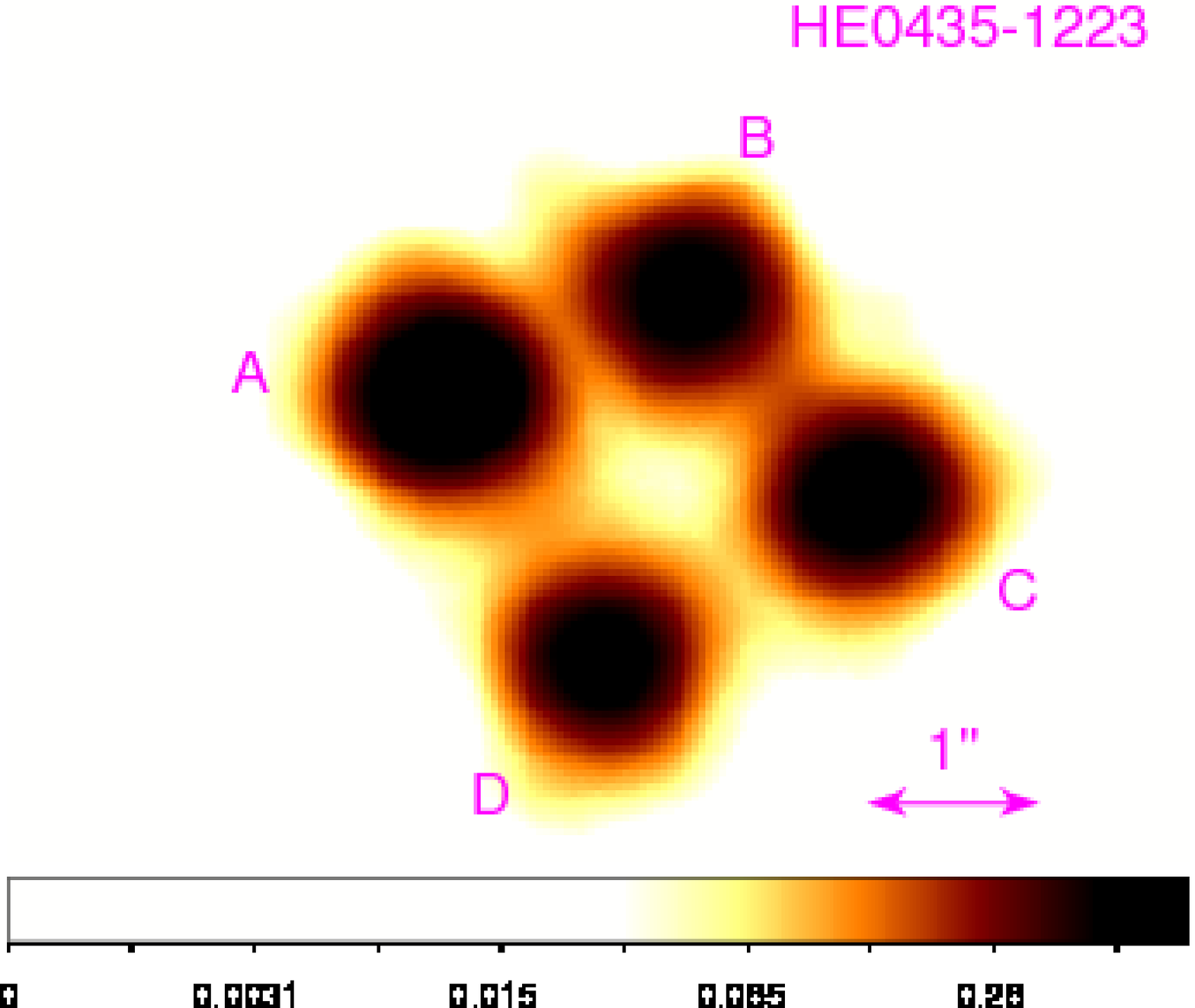}\\
\vspace{10pt}
\includegraphics[width=0.45\textwidth,height=0.3\textheight]{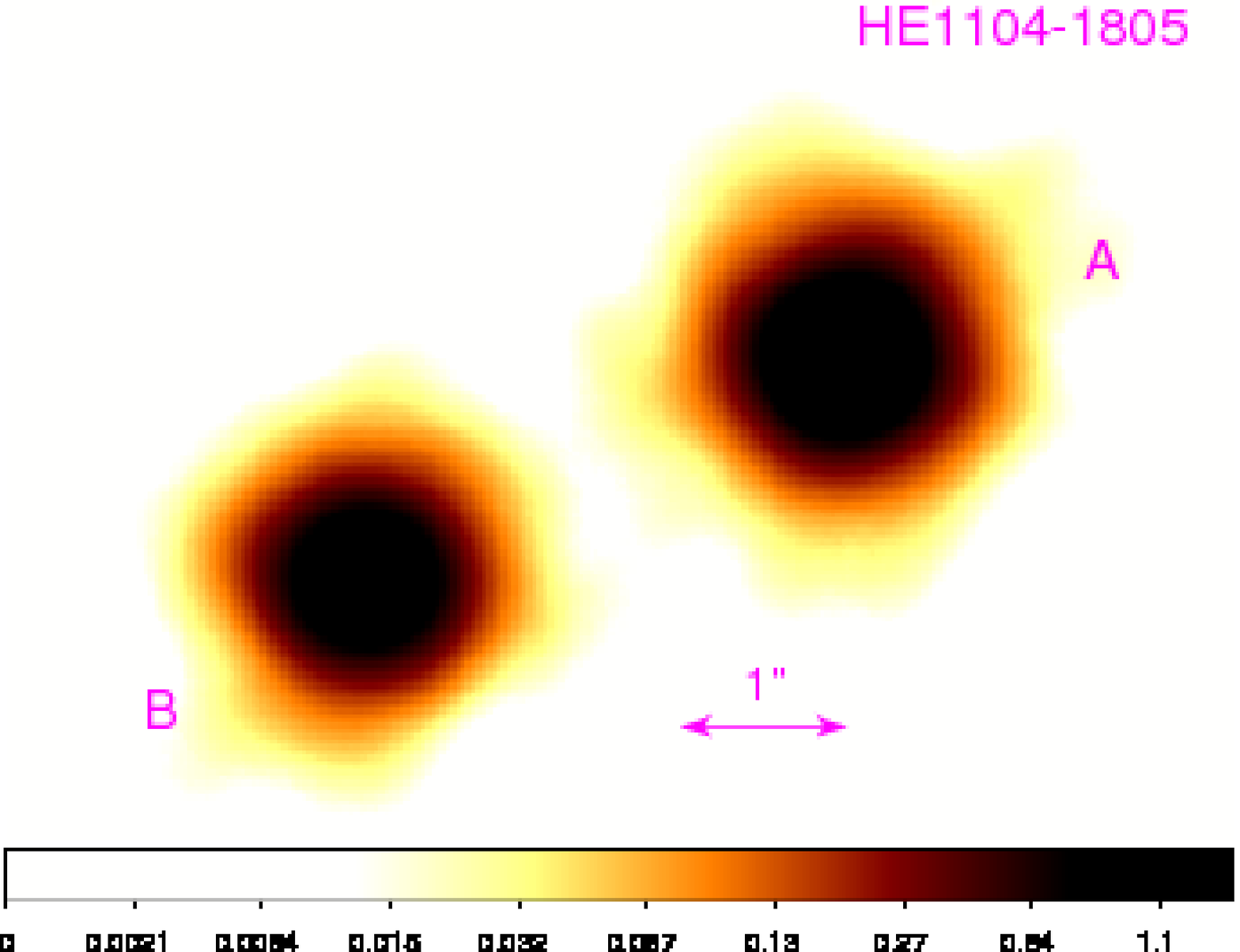}
\hspace{15pt}
\includegraphics[width=0.45\textwidth,height=0.3\textheight]{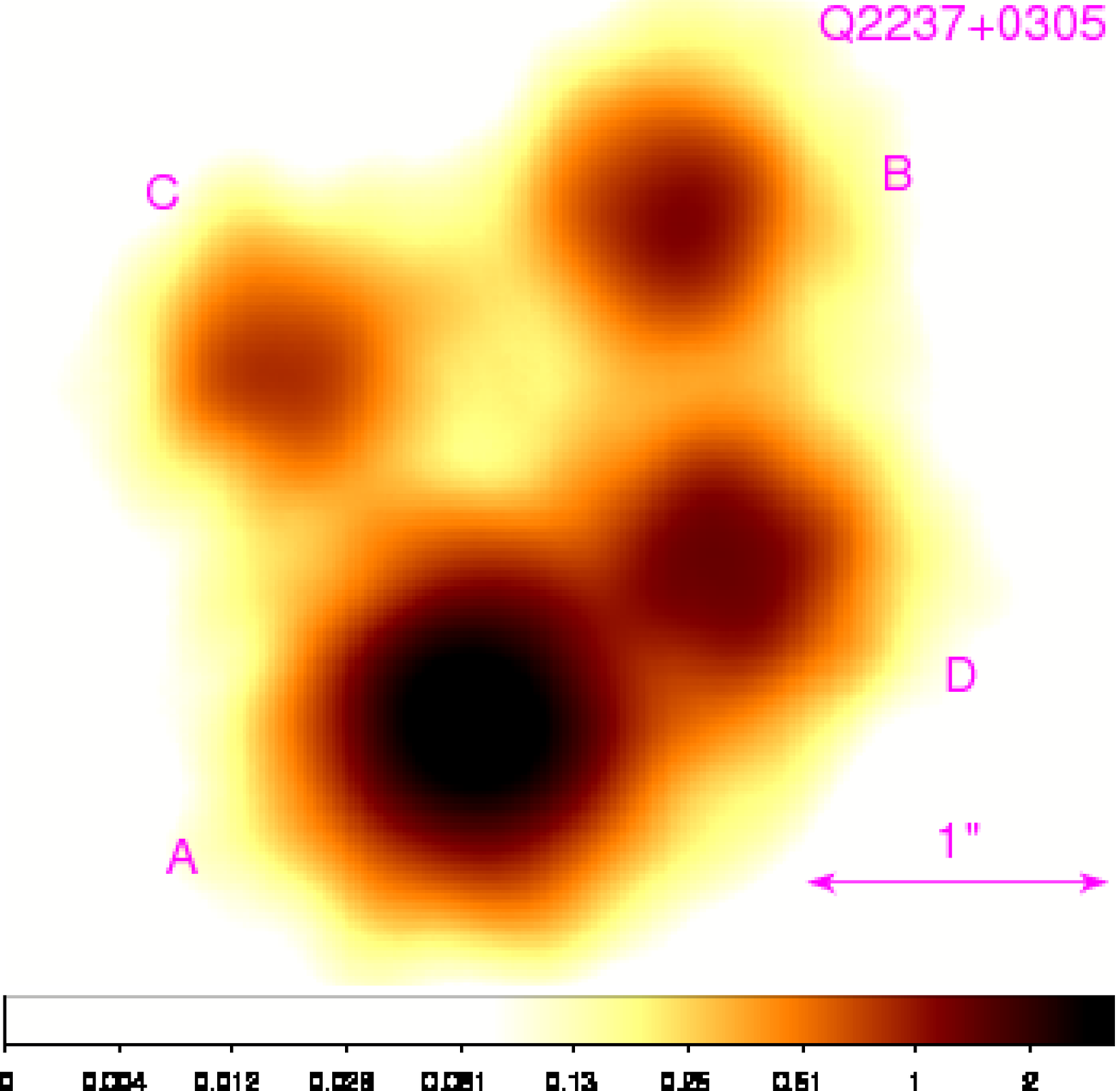}
\end{array}$
\end{center}
\caption{Stacked images of  QJ~0158$-$4325, SDSS~0924+0219, SDSS~1004+4112, HE~0435$-$1223, HE~1104$-$1805, and Q2237+0305. Note the extended emission from the cluster lens SDSS~1004+4112.
\label{fig:stacked_image}}
\end{figure*}

 Beside the cluster lens SDSS1004+4112, whose four bright images are well separated from each other (Inada et al. 2003b), the images in the five galaxy-scale lenses are separated by only 1--2 arcsec and their PSFs are therefore blended. 
 In each of these systems, we first extracted the total counts in each band (full, soft, and hard)  within circular apertures  3--$4''$  in radius that contain most of the source counts (see Figure \ref{fig:stacked_image}). We estimated the background using concentric annular regions with inner and outer radii of $\sim$$10''$ and $\sim$$20''$, respectively.   
 We then used PSF fitting to model the relative count rates between the images, fixing the relative image positions to the \hst\ positions from the CASTLES website. 
 We binned the X-ray images in 0\sarc0984 pixels, and fit these binned images using \verb+Sherpa+ to minimize the Cash C-statistic between the observed and model images.  
 The count rates were then normalized by partitioning the background-subtracted total source counts based on the relative count rates for each image obtained from the PSF fits. 
 Although some images in these five lenses are well-separated and their count rates could also be estimated using aperture photometry, 
 we find that the count rates from the two methods are consistent and report only the results based on the PSF fits.
 
 In SDSS~1004+4112, the four images are well separated (from $\sim$$4''$ to $\sim$$15''$),  so we computed the photon count rates using aperture photometry.  
 Since the foreground lens is a cluster, there is non-negligible X-ray emission from the cluster at the positions of the lensed images (Ota et al. 2006).
 We chose a circular region of  $\sim$1\sarc5 radius around each image as the source region. 
 We estimated the local background  in partial annuli at the same radius from the cluster center as the quasar images in order to estimate, and then subtract the contamination from the lensing cluster. 
 We then corrected for the finite aperture size to determine the total count rates in Table \ref{tab:1004count}.   

In all six lenses, the count rate of each image was corrected for Galactic absorption and absorption in the lens (see Section \ref{sec:spectrum}). The final results are presented  in Tables  \ref{tab:0158count}--\ref{tab:2237count}. 
We note that the total count rates can differ slightly from the sum of  the soft and hard band count rates because the PSF fits were done separately for the full, soft, and hard band images, and because of the absorption corrections.  
Finally, we also constructed a stacked image (combining all epochs) using the best-fitting image positions obtained from the fits to  register the epochs and then smoothing with a suitably sized Gaussian kernel. These images are shown in Figure \ref{fig:stacked_image}.

\section{Spectral Analysis}\label{sec:spectrum}

We extracted three sets of spectra from each lens: the spectra of the individual images for the individual epochs, the stacked spectra of the individual images combining all the epochs, and the stacked spectra of all images.
First, we extracted spectra of the individual images  using the CIAO tool ``psextract" in circles of radii about 0\sarc5--0\sarc8  (for SDSS~1004+4112 we used 1\sarc5) centered on the  positions from the PSF fits to each observation.
Because of the superb angular resolution of \chandra, the normal X-ray background contributes little to the source spectrum, and the ``background" is dominated by contaminating emission from adjacent images. 
To estimate the contamination of image A by image B, we measured the spectrum in a circular aperture of the same size at the position X that A would have if reflected through image B (i.e., X$-$B$-$A with $\rm \overline{XB}=\overline{BA}$).
For the cluster lens SDSS~1004+4112, we  extracted the source spectra using 1\sarc5 radius apertures and the background spectra from concentric partial rings at the same radius from the cluster center to model the cluster contamination.
Images C and D of SDSS~0924+0219 are faint and poorly resolved (see Figure \ref{fig:stacked_image}),  so we extracted a single spectrum from a polygonal region containing images A, C and D. 
After extracting the spectra of the individual images for each epoch, we also obtained combined spectra for each image by stacking the spectra from the different epochs and the corresponding rmf and arf files weighted by their exposure times to increase the S/N ratio of the spectra. 
Finally, we obtained a spectrum combining  all the images and all the epochs for each lens. 

We analyzed the spectra using XSPEC V12.   
We first modeled the spectra of the individual images extracted from the individual epochs using a simple power-law model modified by Galactic absorption (Dickey \& Lockman 1990) and  neutral absorption at the lens over the 0.4--8.0 keV band. 
In the fitting process, we assumed the same fixed Galactic absorption for all the images of a lens, but allowed the absorption from the lens galaxy to vary independently. 
We further assumed that the power-law index $\Gamma$ of the source spectrum was  the same for all images.
In general, this may not be true either because of  the intrinsic spectral variability modulated by the time delays between the images, or because of energy-dependent microlensing, but it is a good average approximation.  
The spectral fitting results are presented in Tables \ref{tab:0158fit}--\ref{tab:2237fit}, and Figure \ref{fig:Gamma} shows the evolution of the power-law index $\Gamma$ in each lens. 
Besides the power-law index $\Gamma$ and the column density $\rm N_H$ of the neutral absorption in the lens, we also give the estimated absorption-corrected energy flux for the soft and hard bands.

To better constrain the lens absorption and power-law indices of the quasar, we fit the stacked spectra combining the individual epochs for each image. These spectra are shown in Figures \ref{fig:0158_stack_spec}--\ref{fig:2237_stack_spec}.
When fitting these stacked spectra, we also added Gaussian emission lines to the model.  
We detected differential absorption between some of the images in all the lenses except for SDSS~0924+0219 and HE~1104$-$1805. 
We detected \feka\ emission lines in all lenses and the Ni XXVII~K$\alpha$  line in Q~2237+0305.
We list the spectral fitting results for the stacked spectra in Tables \ref{tab:0158fit}--\ref{tab:2237fit} for the photon indices and lens absorption and in Table \ref{tab:EW}  for the emission lines. 
After obtaining the best fits to the stacked spectra, we calculated the ratios of the count rates between the absorbed and unabsorbed models in the full, soft, and hard bands, respectively. 
These mean ratios were then applied to the absorbed count rates of the individual epochs and images to estimate the absorption corrected count rates. 
For SDSS~0924+0219 we use the same absorption correction for images A, C and D. 
We also modeled the stacked spectra allowing different power-law indices between the images.
Finally, we fit the stacked spectrum combining all images and all epochs for each lens, and  these results are also presented in Tables \ref{tab:0158fit}--\ref{tab:2237fit}.

\subsection{Photon Indices}
 
\begin{figure}
	\epsscale{1.0}
	\plotone{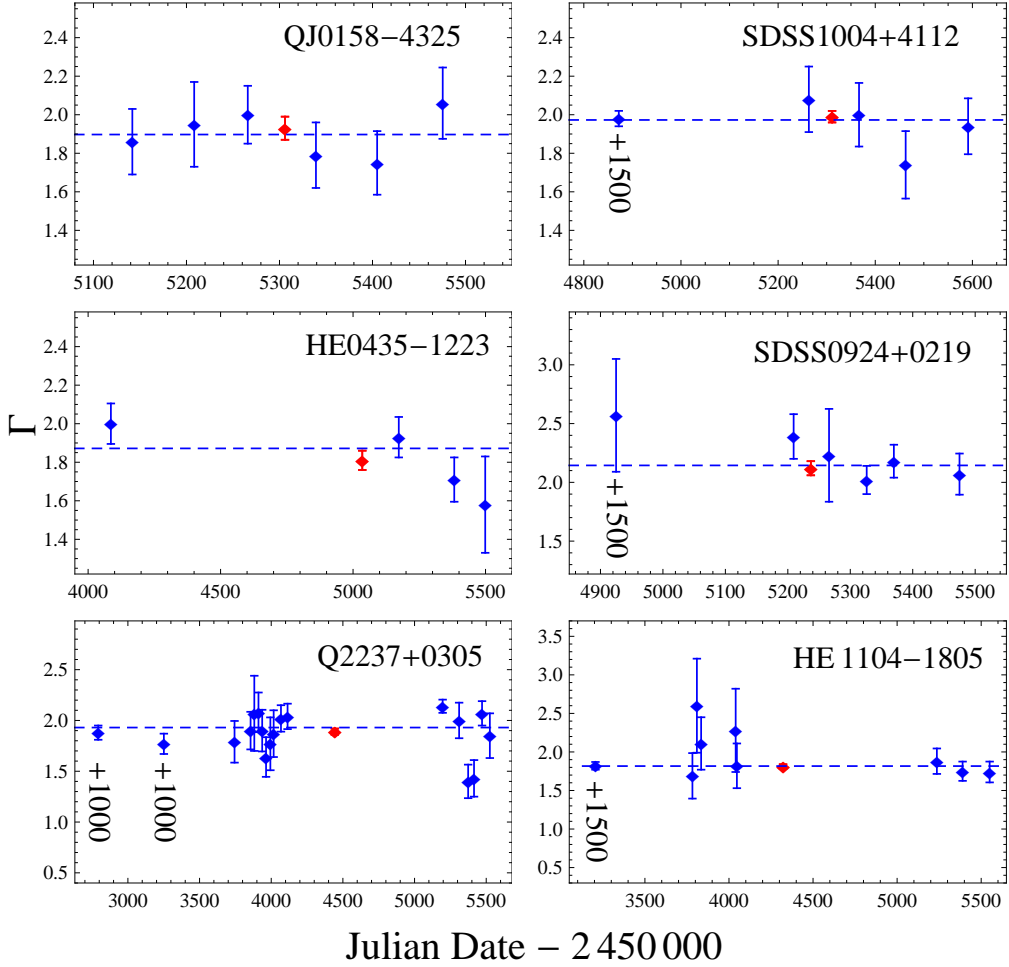}
	\caption{Evolution of the spectral power-law index $\Gamma$ with time for QJ~0158$-$4325, SDSS~0924+0219,  HE~0435$-$1223, SDSS~1004+4112,  Q~2237+0305 and HE~1104--1805.  For clarity, the first points for  SDSS~0924+0219,   SDSS~1004+4112 and HE~1104--1805 are shifted by +1500 days, and the first two events of Q~2237+0305 are shifted by +1000 days. In each panel, the dashed horizontal line is the best fit $\Gamma$ to the measurements from the individual epochs, and the red point is the power-law index $\Gamma$ obtained from fitting the stacked spectra. 
         \label{fig:Gamma}}
\end{figure}

\begin{figure}
	\epsscale{0.6}
	\plotone{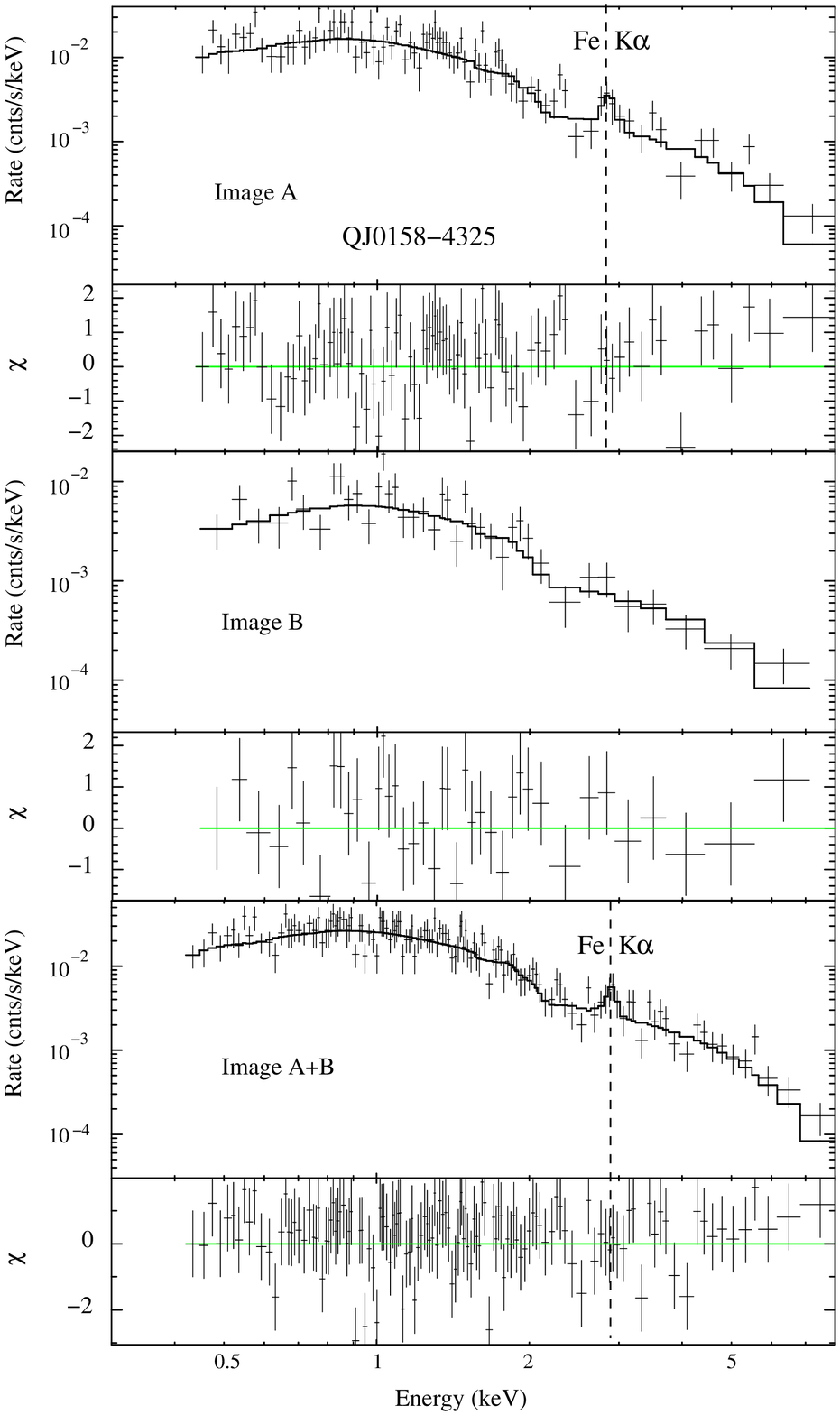}
	\caption{Stacked spectra of QJ~0158$-$4325 (observed frame) fit with a simple power-law plus Gaussian line model modified by Galactic and lens galaxy absorption. The top, middle, and bottom panels show results for image A, image B, and their combined spectrum, in each case showing the data and the model fit in one sub-panel, and the statistical residuals in a second sub-panel. 	        \label{fig:0158_stack_spec}}
\end{figure}

\begin{figure}
	\epsscale{0.6}
	\plotone{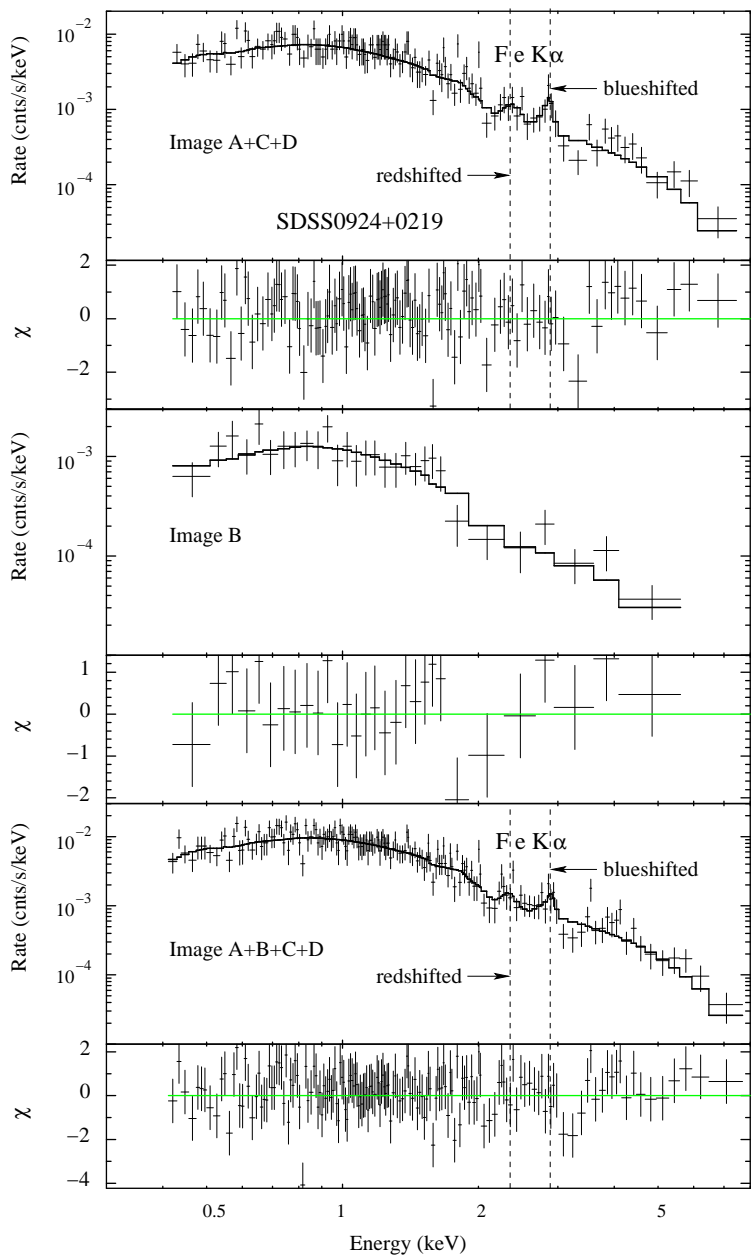}
	\caption{Stacked spectra of SDSS~0924+0219 (observed frame) fit with a simple power-law plus Gaussian line model modified by Galactic and lens galaxy absorption. The top, middle, and bottom panels show results for image A+C+D, image B, and their combined spectrum, in each case showing the data and the model fit in one sub-panel, and the statistical residuals in a second sub-panel. 	We detect both a blueshifted and a redshifted wing of the iron line.    
	        \label{fig:0924_stack_spec}}
\end{figure}

\begin{figure}
	\epsscale{0.6}
	\plotone{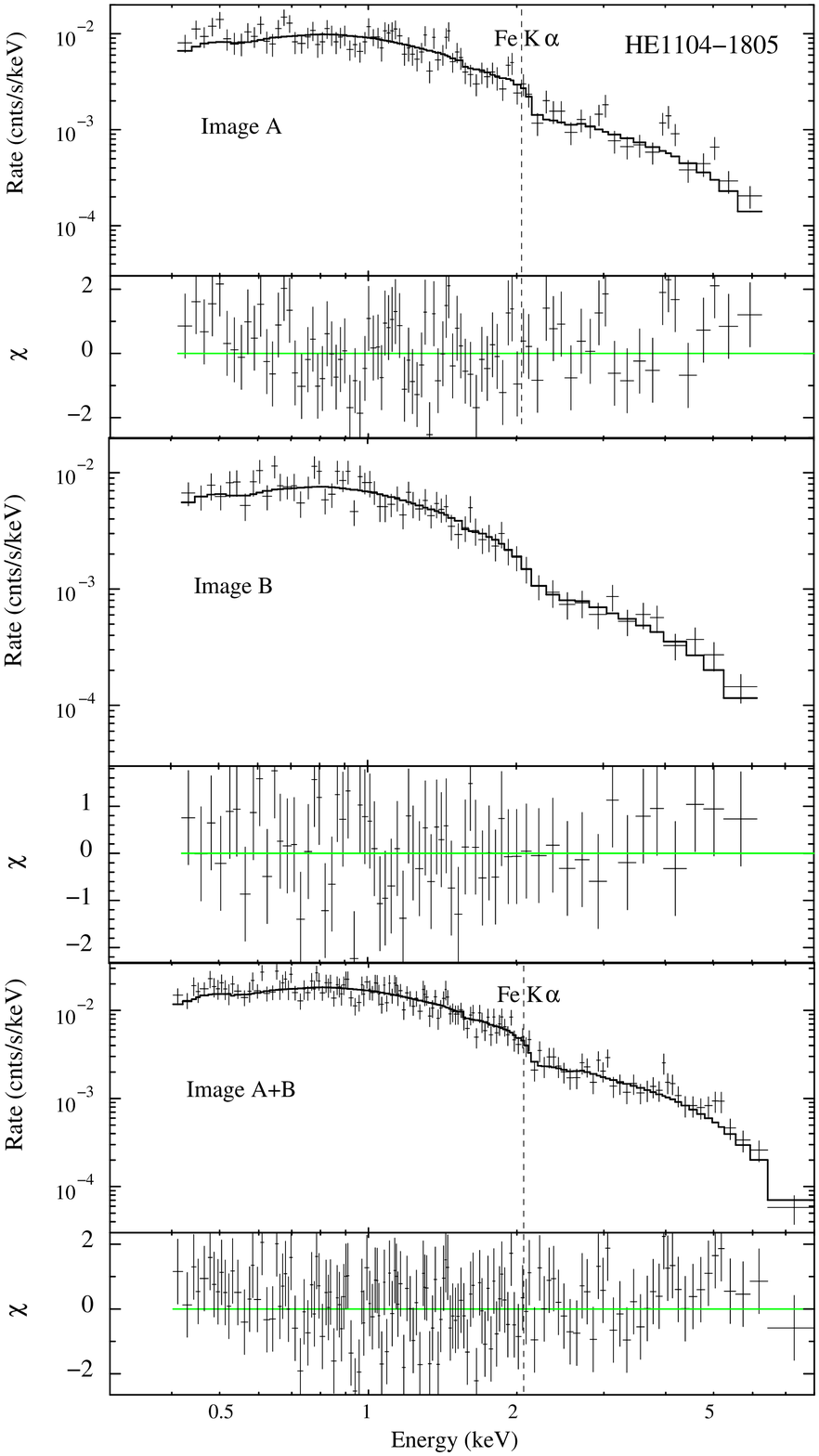}
	\caption{Stacked spectra of HE~1104$-$1805 (observed frame) fit with a simple power-law plus Gaussian line model modified by Galactic and lens galaxy absorption. The top, middle, and bottom panels show results for image A, image B, and their combined spectrum, in each case showing the data and the model fit in one sub-panel, and the statistical residuals in a second sub-panel. 	
		        \label{fig:1104_stack_spec}}
\end{figure}

\begin{figure}
	\epsscale{1.0}
	\plotone{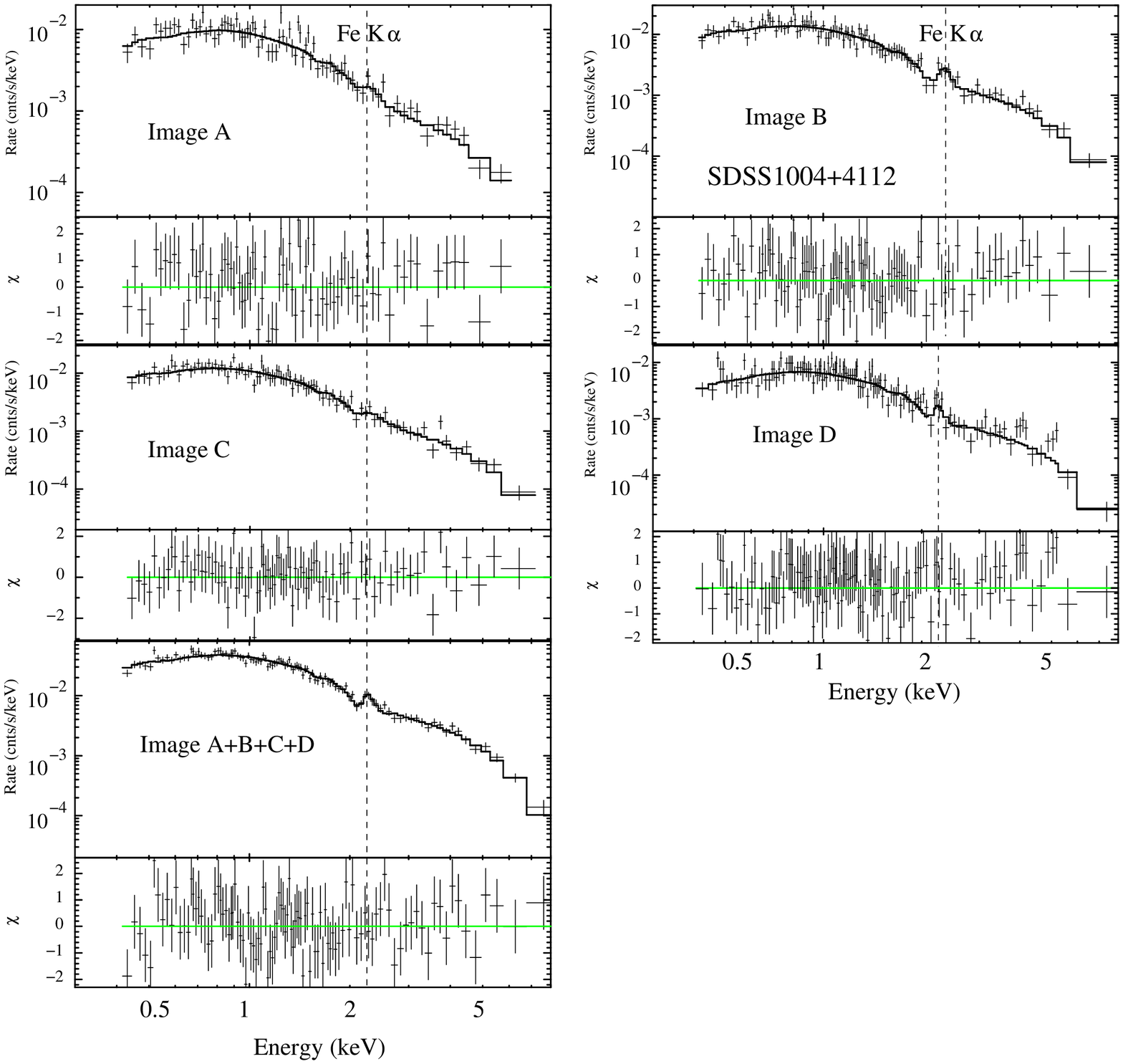}
	\caption{Stacked spectra of SDSS~1004+4112 (observed frame) fit with a simple power-law plus Gaussian line model modified by Galactic and lens galaxy absorption. The individual A, B, C, and D spectra as well as the stacked A+B+C+D spectrum are shown, in each case showing the data and the model fit in one sub-panel, and the statistical residuals in a second sub-panel.
		        \label{fig:1004_stack_spec}}
\end{figure}

\begin{figure}
	\epsscale{1.0}
	\plotone{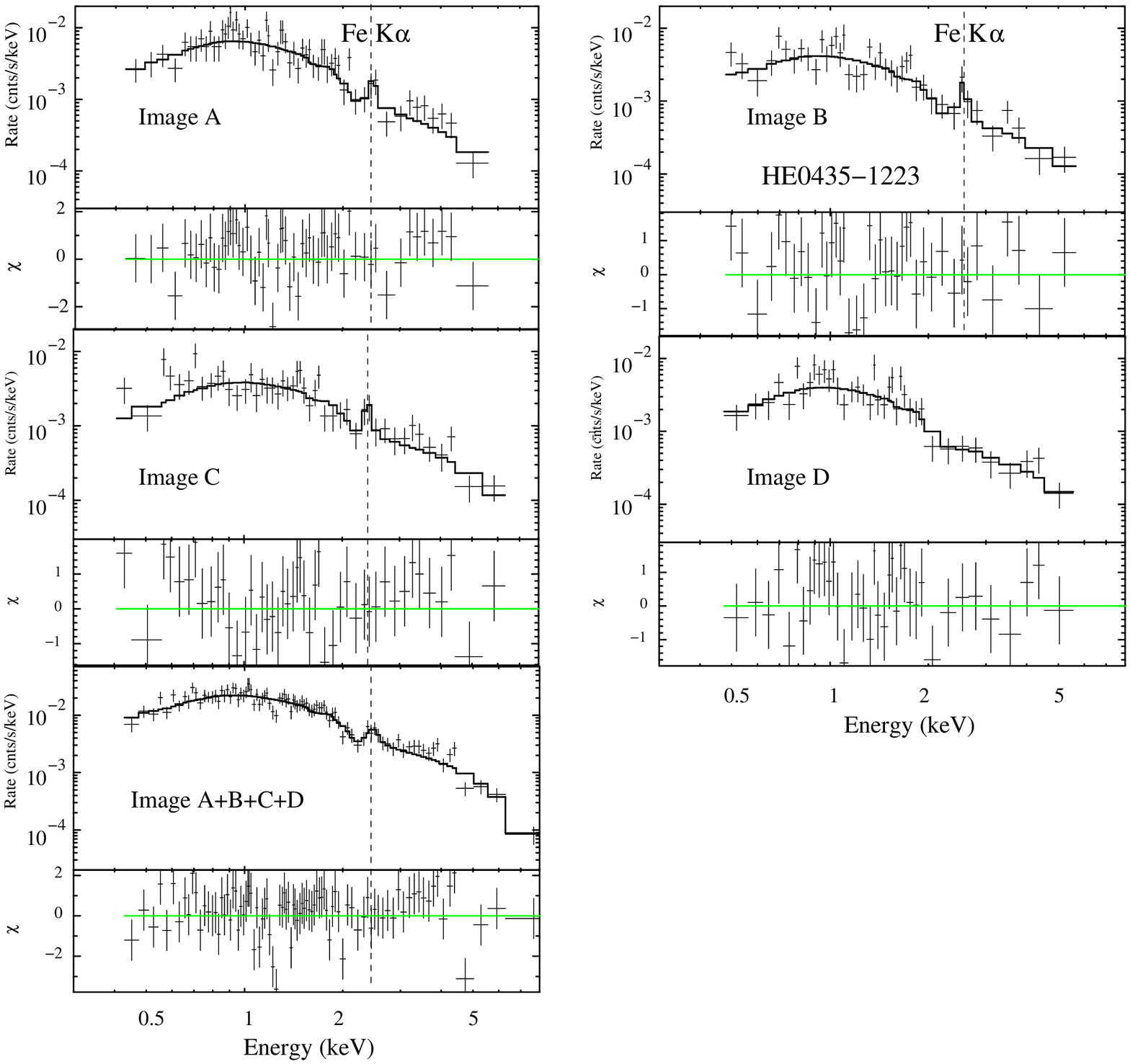}
	\caption{Stacked spectra of HE~0435$-$1223 (observed frame) fit with a power-law plus Gaussian line model modified by Galactic and lens galaxy absorption. The individual A, B, C, and D spectra as well as the stacked A+B+C+D spectrum are shown, in each case showing the data and the model fit in one sub-panel, and the statistical residuals in a second sub-panel.
    \label{fig:0435_stack_spec}}
\end{figure}

\begin{figure}
	\epsscale{1.0}
	\plotone{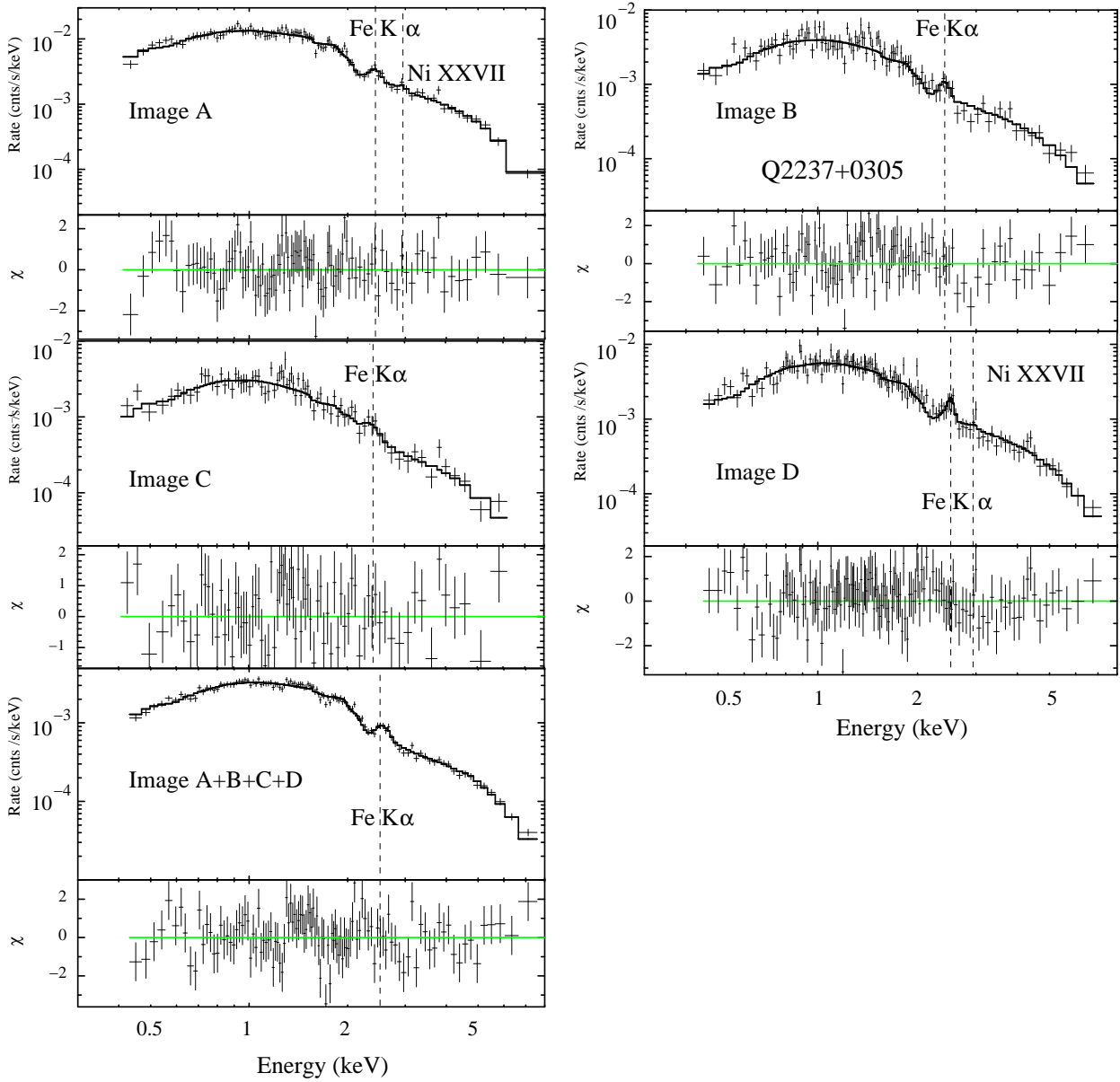}
	\caption{Stacked spectra of Q~2237+0305 (observed frame) fit with a simple power-law plus Gaussian line model modified by Galactic and lens galaxy absorption.The individual A, B, C, and D spectra as well as the stacked A+B+C+D spectrum are shown, in each case showing the data and the model fit in one sub-panel, and the statistical residuals in a second sub-panel.  Iron line is detected in images A, B, C  and D. The nickel line is only detected in images A and D.
	        \label{fig:2237_stack_spec}}
\end{figure}

We show the time evolution of the power-law index $\Gamma$ of each lens in Figure~\ref{fig:Gamma}. 
To test for spectral variability, we can compare the fits to the individual epochs  to their mean and its uncertainty (dashed lines in Figure \ref{fig:Gamma}).
The results for  QJ~0158$-$4325, HE~0435$-$1223, HE~1104$-$1805, SDSS~0924+0219, and SDSS~1004+4112 are all consistent with a constant spectral index. 
Only Q~2237+0305 shows clear evidence for spectral variability, with a constant $\Gamma$ model having  $\chi^2=38.4$ (dof=17), so we detect spectral variability in Q~2237+0305 at $99.8\%$ confidence. 
This spectral variability is mainly caused by epochs 15 and 16, where the power-law index drops suddenly from $\sim$$2.0$ to $\sim$$1.4$.
These two epochs contribute $\sim$$18.0$ out of the total $\chi^2=38.4$, and the spectral variability would not be statistically significant without these two points.

Spectral variability can be either intrinsic variability or caused by chromatic microlensing between the soft and hard X-ray bands. 
For example, the spectrum of a microlensed quasar image more strongly magnified in the hard X-ray band will appear harder than the other images. 
Distinguishing these two possibilities is particularly important for Q~2237+0305, since in this lens we observed more microlensing variability amplitude in the hard X-ray band than in the soft.
As can be seen from Figures 2 and 3 of Chen et al. (2011), around date 5400 (epochs 15 and 16), there is no significant microlensing variability in images A, B, and D based on the relatively flat B/A, D/A, and B/D flux ratio light curves.  
Although there appears to be some microlensing activity in image C around this time, image C is the faintest of the  four images at this time and contributes less than $10\%$ of the total flux (Table \ref{tab:2237fit}, fit No.\ 15, 16), so it cannot  be responsible for the sudden decrease in the power-law index.  
Moreover, if we fit the spectra of the four images independently at these two epochs, we find similar photon indices for all images. 
We conclude that the spectral variability at these two epochs is intrinsic, instead of due to chromatic microlensing.            
We also compare the average power-law indices of  the individual epochs (dashed lines in Figure~\ref{fig:Gamma}) with those obtained from fitting the stacked spectra combining all the epochs (red points in Figure~\ref{fig:Gamma}), and find that they are consistent.

\subsection{Absorption in the Lens}

\begin{figure}
	\epsscale{0.5}
	\plotone{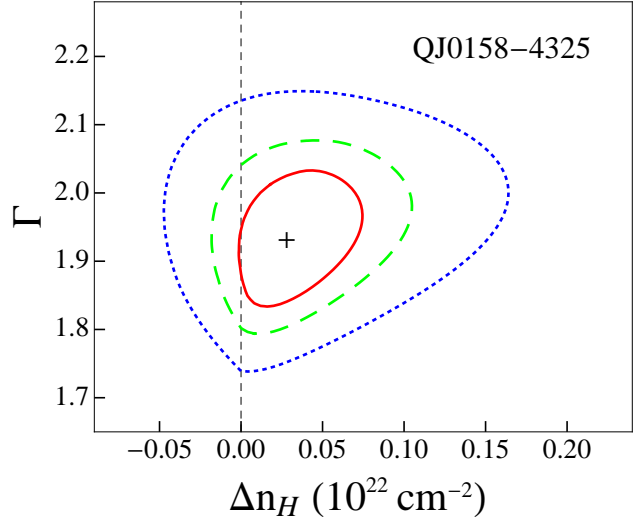}
	\caption{Differential absorption $\rm n_H^{B}-n_H^{A}$ in the lens as a function of the power-law index  $\Gamma$ for QJ~0158$-$4325. The stacked spectra of image A and B are simultaneously fit with the same power-law index but different amounts of neutral absorption $n_H.$ The red solid, green dashed, and blue dotted curves show the two-parameter $68\%,$ $90\%$ and $99\%$ confidence contours respectively.
        \label{fig:0158contour}} 
\end{figure}

\begin{figure}
	\epsscale{0.5}
	\plotone{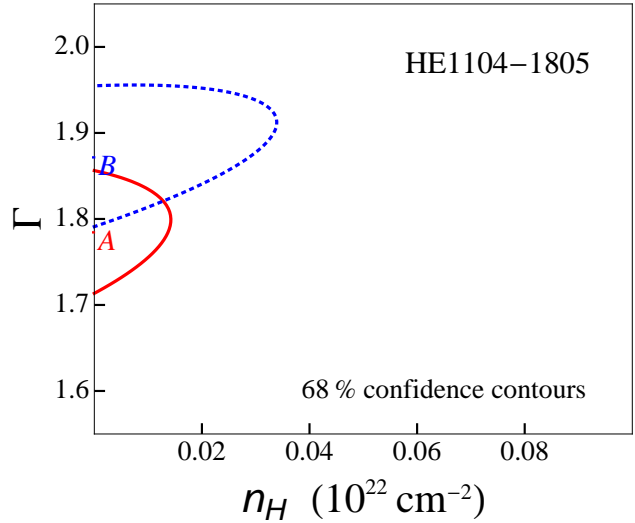}
	\caption{Lens absorption $\rm n_H$ as a function of the  power-law index  $\Gamma$ for HE~1104--1805. The curves show the 68\% confidence level contour for two parameters.
          \label{fig:1104contour}} 
\end{figure}

\begin{figure*}
\begin{center}$
\begin{array}{cc}
\includegraphics[width=0.5\textwidth,height=0.3\textheight]{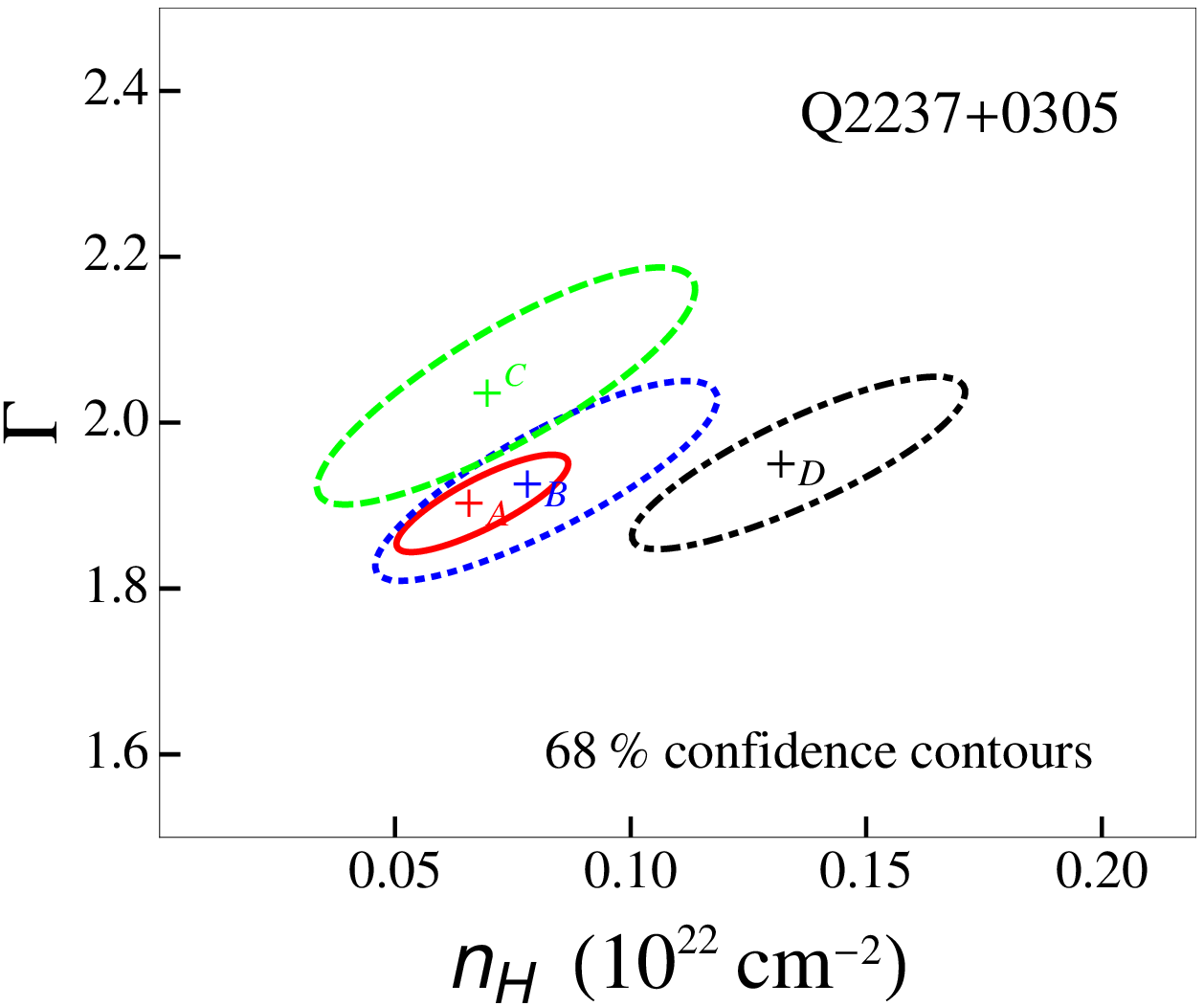}
\hspace{5pt}
\includegraphics[width=0.5\textwidth,height=0.3\textheight]{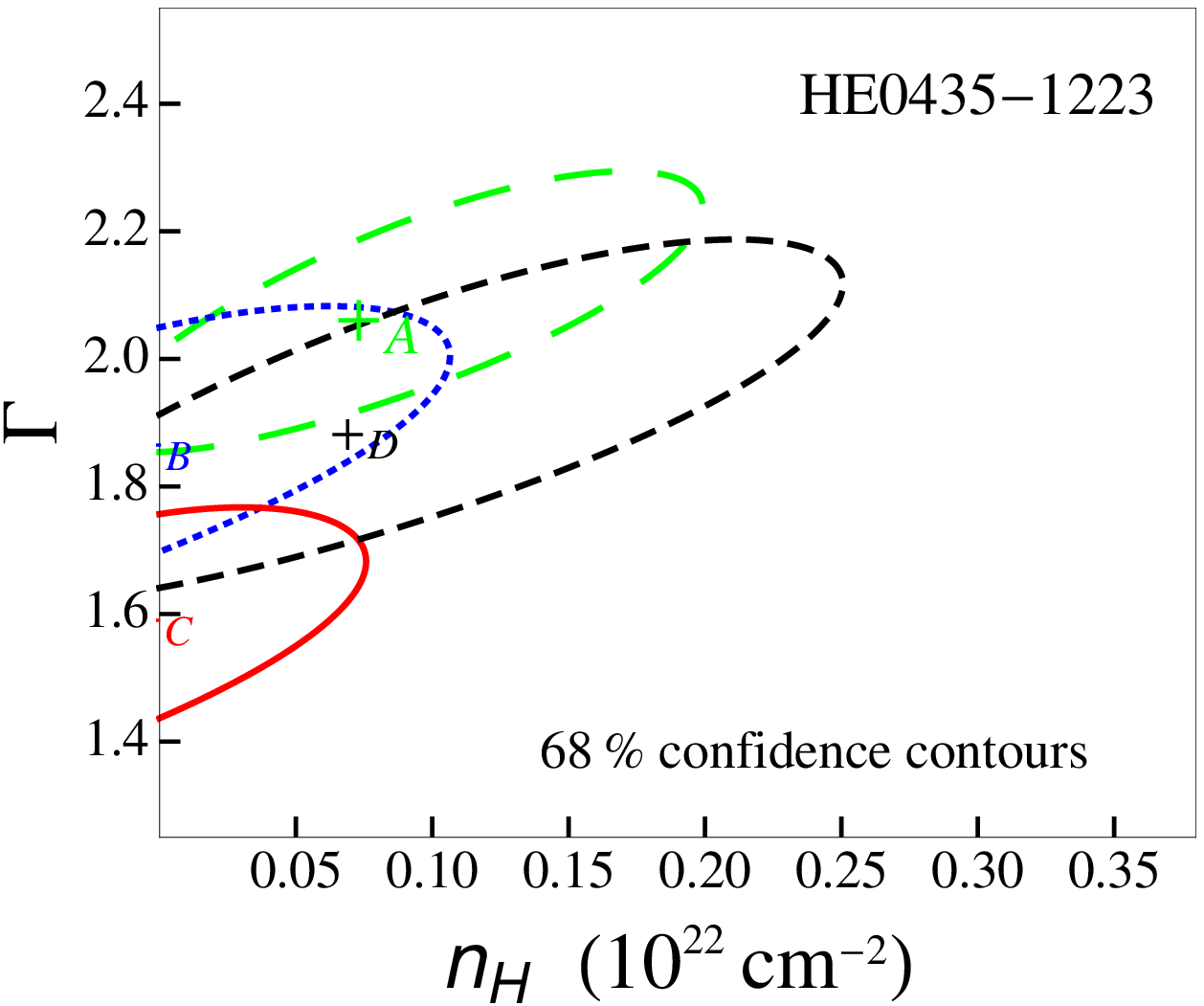}\\
\vspace{20pt}
\includegraphics[width=0.5\textwidth,height=0.3\textheight]{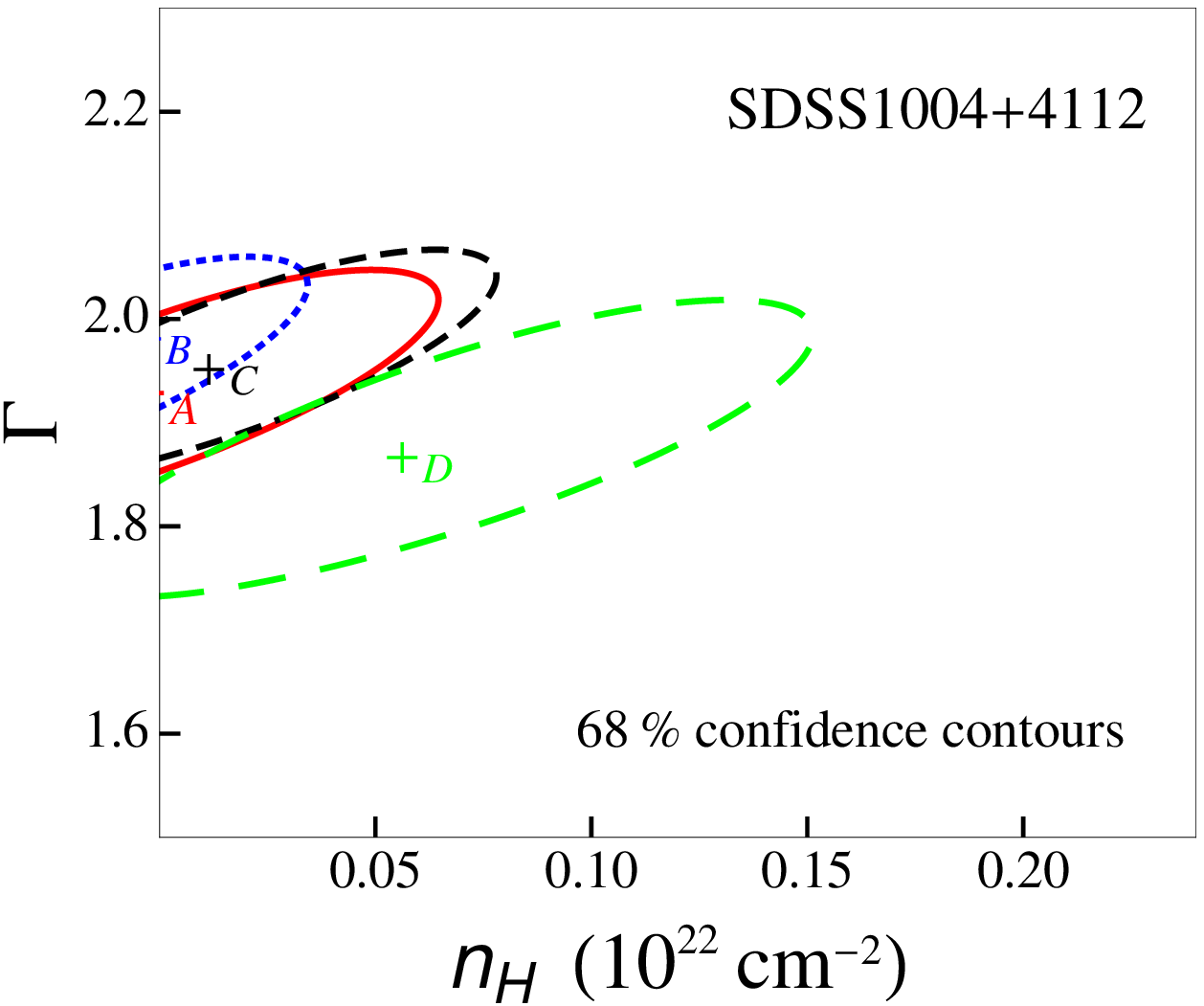}
\hspace{5pt}
\includegraphics[width=0.5\textwidth,height=0.3\textheight]{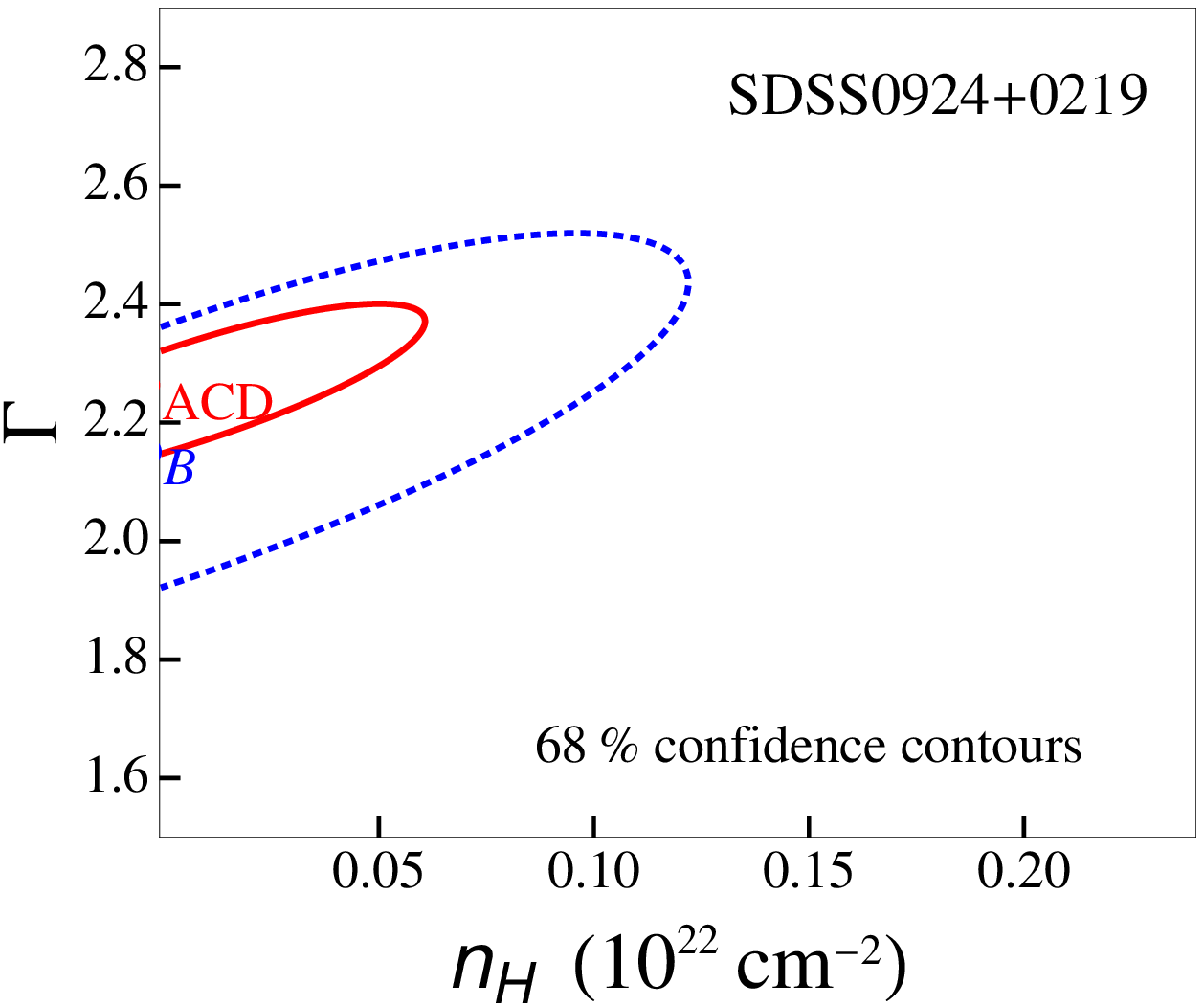}
\end{array}$
\end{center}
\caption{ Absorption $\rm n_H$ by the lens as a function of the power-law index  $\Gamma$ for Q~2237+0305 (upper left), HE~0435--1223 (upper right),  SDSS~1004+4112 (lower left),  and SDSS~0924+0219 (lower right). The stacked spectrum of each image was separately fit with a power-law model modified by Galactic and lens galaxy absorption. In SDSS~0924+0219, the spectra of image A, C and D were combined into one spectrum (labelled by ACD) for these fits. Contours are for two parameters.}
\label{fig:contour}
\end{figure*}

We report our estimates for the amount of absorption in the lens from analyzing the combined spectra of each image in Tables \ref{tab:0158fit}--\ref{tab:2237fit} (the $5^{\rm th}$ column). For absorption we focus on the stacked spectra since the changes in the intrinsic spectra are small or non-existent and the absorption does not change between epochs. 
The stacked spectra of  QJ~0158$-$4325,  SDSS~0924+0219, HE~1104$-$1805, SDSS~1004+4112,  HE~0435$-$1223, and Q~2237+0305 are shown in Figures \ref{fig:0158_stack_spec}--\ref{fig:2237_stack_spec} along with their best-fit models (a power-law modified by Galactic and lens absorption plus Gaussian emission lines).  
Figures ~\ref{fig:0158contour}--\ref{fig:contour} show the two-parameter confidence contours for the estimated column densities $\rm N_H$ and the spectral index $\Gamma$ of the source. 
For the two-image lens QJ~0158--4325, we detect differential absorption between image A and B (Figure~\ref{fig:0158contour}). 
We detect no lens absorption in HE~1104$-$1805 (Figure \ref{fig:1104contour}).
For SDSS~0924+0219, images C and D  are not well separated from image A, so we extracted only one spectrum combining images A, C and D, labeled by ACD in Figure \ref{fig:contour} (bottom right panel). 
We detected no neutral absorption in SDSS~0924+0219. 
We detected differential absorption between images in SDSS~1004+4112, HE~0435--1223, and Q~2237+0305. 
Dai et al.\ (2003) detected neutral absorption in images A and C of Q~2237+0305, but not in images B and D (see also Agol et al. 2009). The latest Chandra data (stacking over 20 epochs with a total exposure time over 290 ks) shows that there is neutral absorption in all 4 images, where image D suffers the most significant absorption from the foreground lens.  

\subsection{Metal Emission Lines}

\begin{figure}
	\epsscale{1.0}
	\plotone{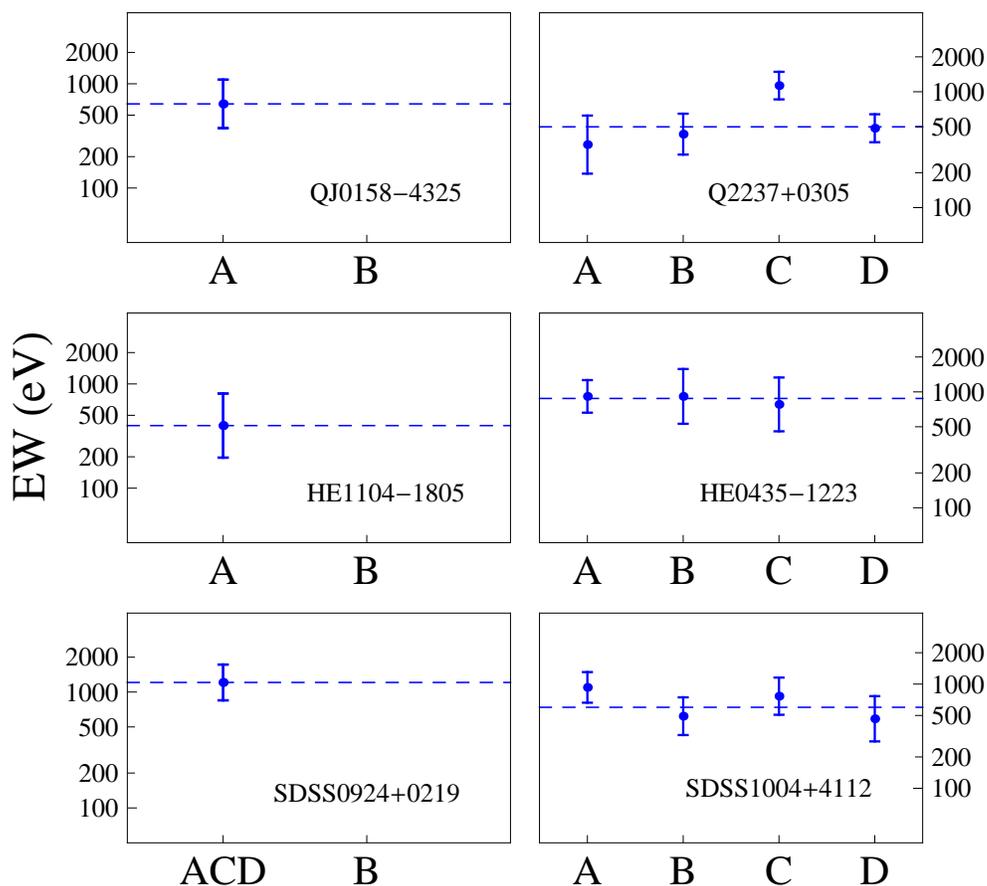}
	\caption{Rest frame equivalent  widths of the \feka\  lines  of QJ~0158$-$4325, Q~2237+0305, HE~1104$-$1805, HE~0435$-$1223, SDSS~0924+0219,  and SDSS~1004+4112. Dashed lines are the best fits for the equivalent widths of the images as constants. For SDSS~0924+0219, we extracted one spectrum for the combination of  images A, C, and D, and the iron line for this spectra is labeled by $\rm ACD$ (A is the brightest image). The iron line was detected in all four images of SDSS~1004+4112, and Q~2237+0305. We did not detect the iron line in image B of  QJ~0158$-$4325,  SDSS~0924+0219, or HE~1104$-$1805,  or image D of  HE~0435$-$1223.
	 \label{fig:EW}}
\end{figure}

\begin{figure}
	\epsscale{1.0}
	\plotone{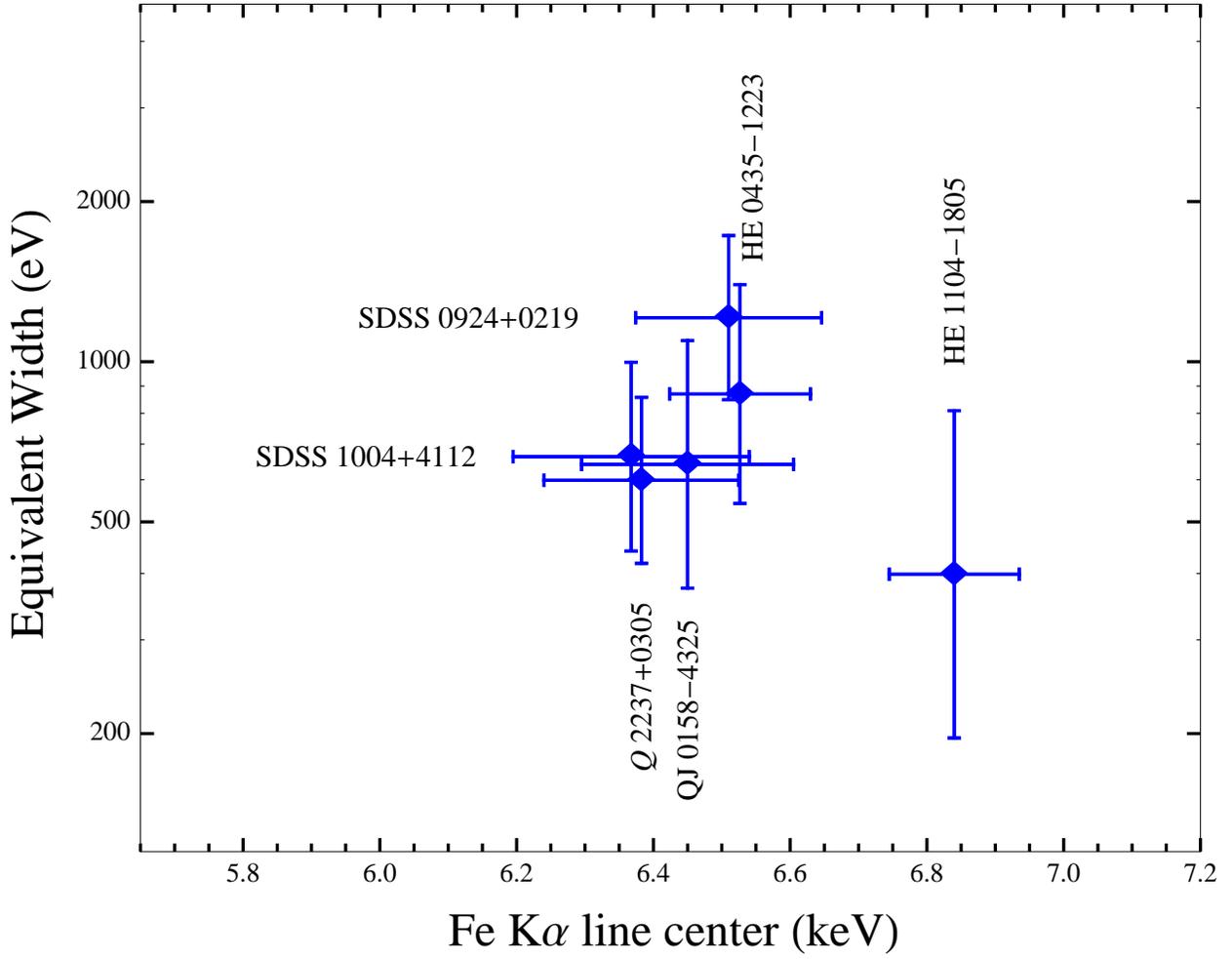}
	\caption{Average rest frame \feka\  line centers and equivalent widths for the six lenses. 
	\label{fig:EW_Eline}}
\end{figure}

We detect \feka\ emission lines in all six lenses. 
Table~\ref{tab:EW} reports the rest frame mean energy, $\rm E_{line},$ Gaussian line width, $\sigma_{\rm line},$ the equivalent width, EW, and significance for each lens. 
Figure~\ref{fig:EW} compares the EWs of the individual images, and Figure~\ref{fig:EW_Eline} shows the average (over all images) of the line centers and equivalent widths. 
We do not detect the \feka\ line for image B of  SDSS~0924+0219, QJ~0158$-$4325, or HE~1104$-$1805. In HE~0435--1223, we detect the \feka\ line in  images A, B, and C, but not  D.  In Q~2237+0305, we detect the \feka\ line in all four images, confirming the earlier detection of the line in the brightest image A (Dai et al.\ 2003).
In addition, we detect  at low significance the Ni XXVII K$\alpha$ line (at $E\sim 7.8$ keV) in images A and D of Q~2237+0305 (Figure~\ref{fig:2237_stack_spec} and  Table~\ref{tab:EW}) for the first time in a lensed quasar, and we list them in Table~\ref{tab:EW} for completeness. 
While adding one \feka\ line component is sufficient for most spectra, in SDSS 0924+0219, we find that two \feka\ line components can significantly improve the fit for both the spectrum ACD and the total spectrum.  We detect one blueshifted (5.9 keV) and one redshifted (7.1 keV) \feka\ line components in both of these spectra with an average line energy of 6.5 keV and total EW=1.2 keV.
Since we cannot completely resolve the A, C, and D images of this lens, it is possible that the two components are from two images.
It is also possible that the two components are from a single image, possibly from the brightest image A,  caused by a caustic dissecting the inner accretion disk.
In HE~1104$-$1805, we formally detect the \feka\ line at the 99\% significance level in the spectra of image A and the total image.
However, the lines are hard to observe by eye, because they are at the energy where the effective area of the mirror changes significantly, and it is difficult to determine the continuum by eye.

The \feka\ line has a rest-frame energy range of 6.4--6.7~keV depending on the ionization state.
Most of the \feka\ lines we detect have rest-frame energies consistent with the neutral line energy at 6.4~keV.
In both HE~0435--1223 and Q~2237+0305, we see shifts in the \feka\ line energy between the images.
We detect a redshifted line, 2.5$\sigma$ from the rest-frame energy, in image C of Q~2237+0305 with $E=6.08\pm0.13$~keV.
We detect both a redshifted component at $E=5.92$ keV,   2.5$\sigma$ from the rest-frame energy, and a blueshifted component at $E=7.12$ keV,  $>5\sigma$ from the rest frame energy, for image A+C+D of SDSS~0924+0219.  
We also detect blue-shifted or ionized lines in image B of HE~0435--1223 with $E=6.72\pm0.08$~keV and image D of Q~2237+0305 with $E=6.62\pm0.06$~keV.
In HE~1104$-$1805 we detect a blueshifted iron line for image A at $E=6.84$ keV, 3$\sigma$ from the rest-frame energy.
A small fraction of the lines are resolved as broad lines, while most are unresolved, with lower limits consistent with narrow lines.  
The equivalent widths of the lines are significantly larger than those detected in unlensed AGN with similar X-ray luminosities, which we will discuss in Section \ref{sec:iron}.


\section{Microlensing Analysis}\label{sec:microlensing}

\begin{figure}
	\epsscale{0.7}
	\plotone{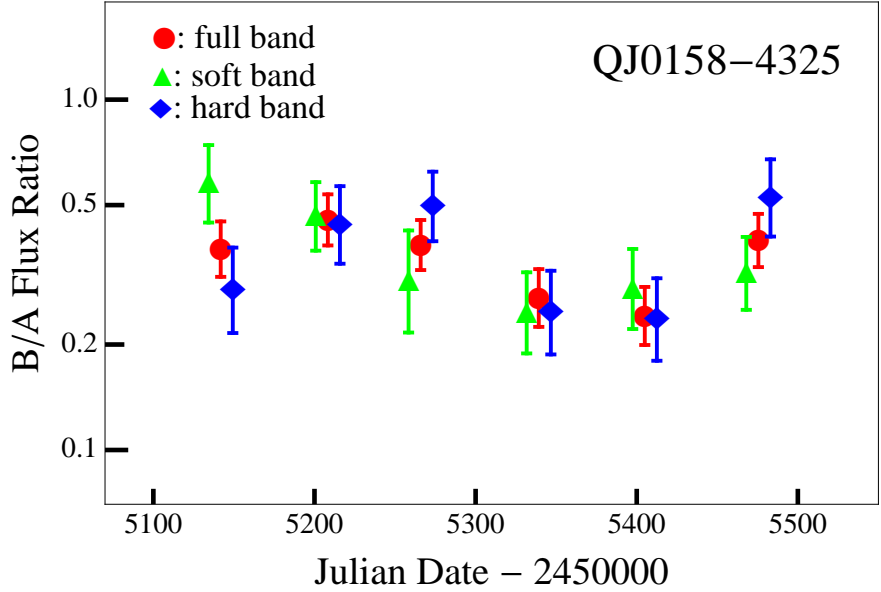}
	\caption{Chromatic X-ray microlensing light curves for QJ~0158$-$4325 in the full (0.4--8.0 keV, red circles), soft (0.4--1.3 keV, green triangles), and hard (1.3--8.0 keV, blue diamonds) X-ray bands. The observation dates  for the soft and hard bands are slightly offset for clarity. Uncertainties are  $1\,\sigma.$ 
	        \label{fig:0158_lightcurve}}
\end{figure}

\begin{figure}
	\epsscale{1.0}
	\plotone{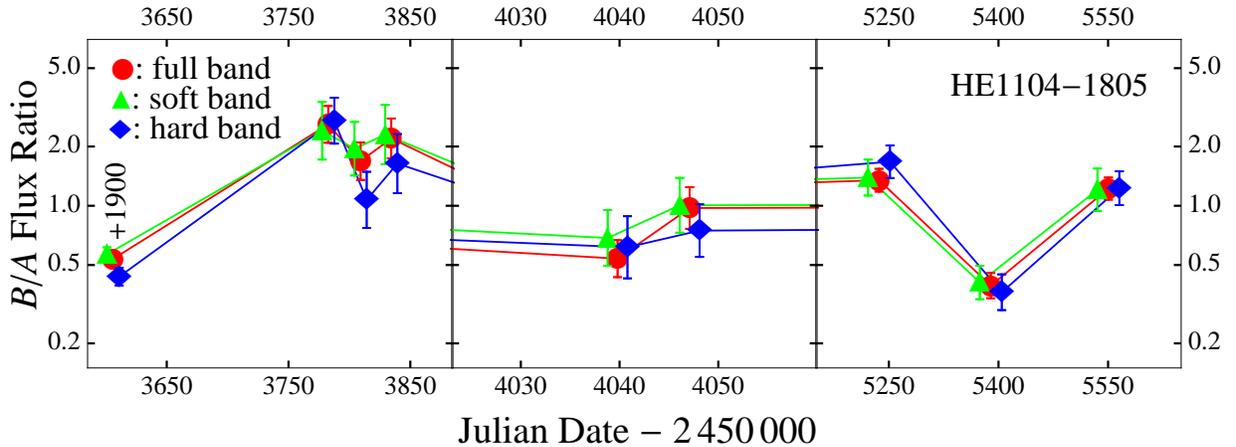}
	\caption{Chromatic X-ray microlensing light curves for HE~1104$-$1805 in the full (0.4--8.0 keV, red circles), soft (0.4--1.2 keV, green triangles), and hard (1.2--8.0 keV, blue diamonds) X-ray bands. Uncertainties are  $1\,\sigma.$ The observation dates for the soft and hard bands are slightly offset for clarity.  The first observation (made in 2000) was shifted by +1,900 days to better resolve other observations.	  \label{fig:1104_lightcurve}}
\end{figure}

\begin{figure}
	\epsscale{1.0}
	\plotone{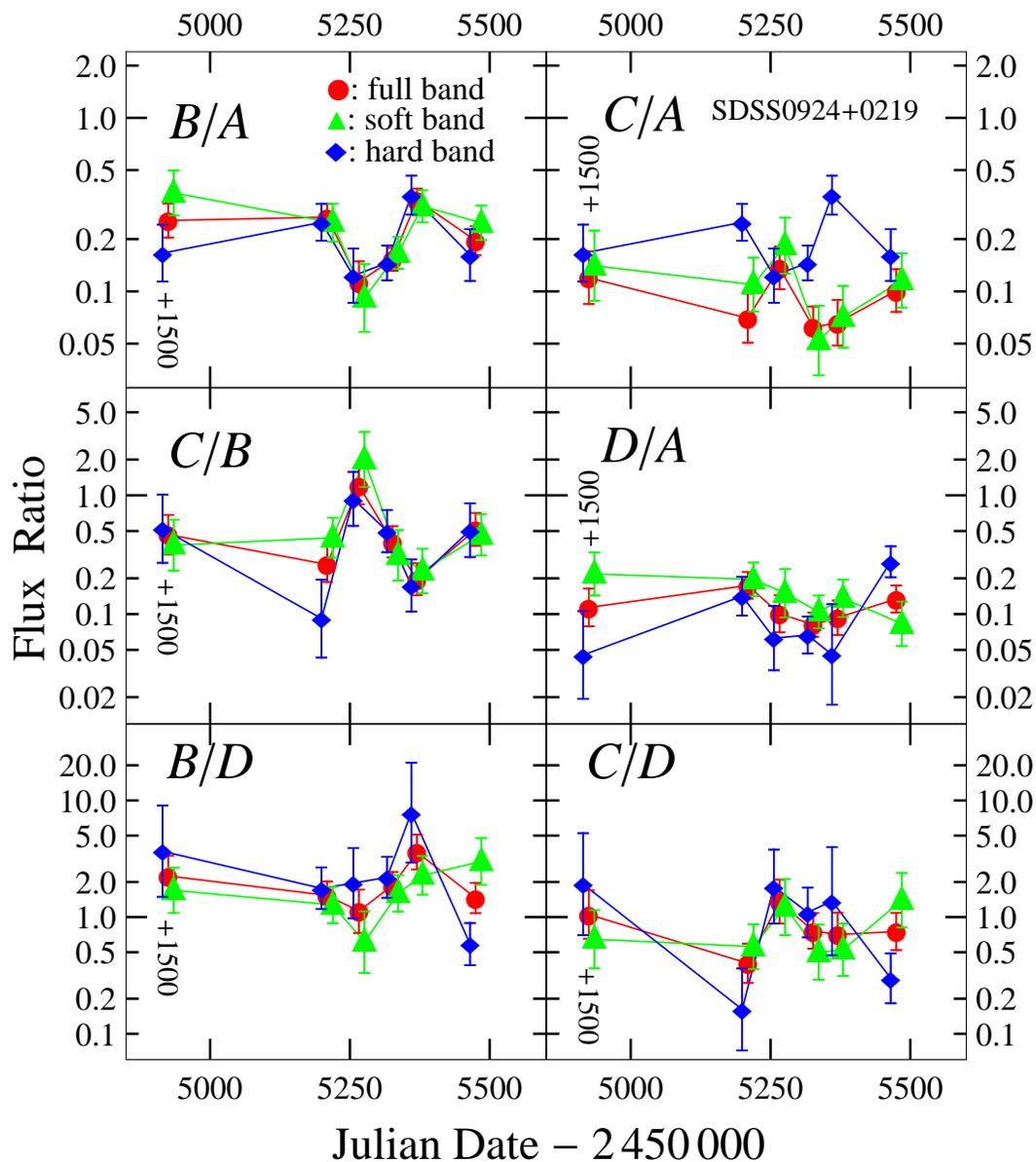}
	\caption{Chromatic X-ray microlensing light curves for SDSS~0924+0219 in the full (0.4--8.0 keV, red circles), soft (0.4--1.2 keV, green triangles), and hard (1.2--8 keV, blue diamonds) X-ray bands. The first observation (made in 2005) in each panel was shifted by +1,500 days in order to better resolve the next 5 observations  (all made in 2010). 
	        \label{fig:0924_lightcurve}}
\end{figure}

\begin{figure}
	\epsscale{1.0}
	\plotone{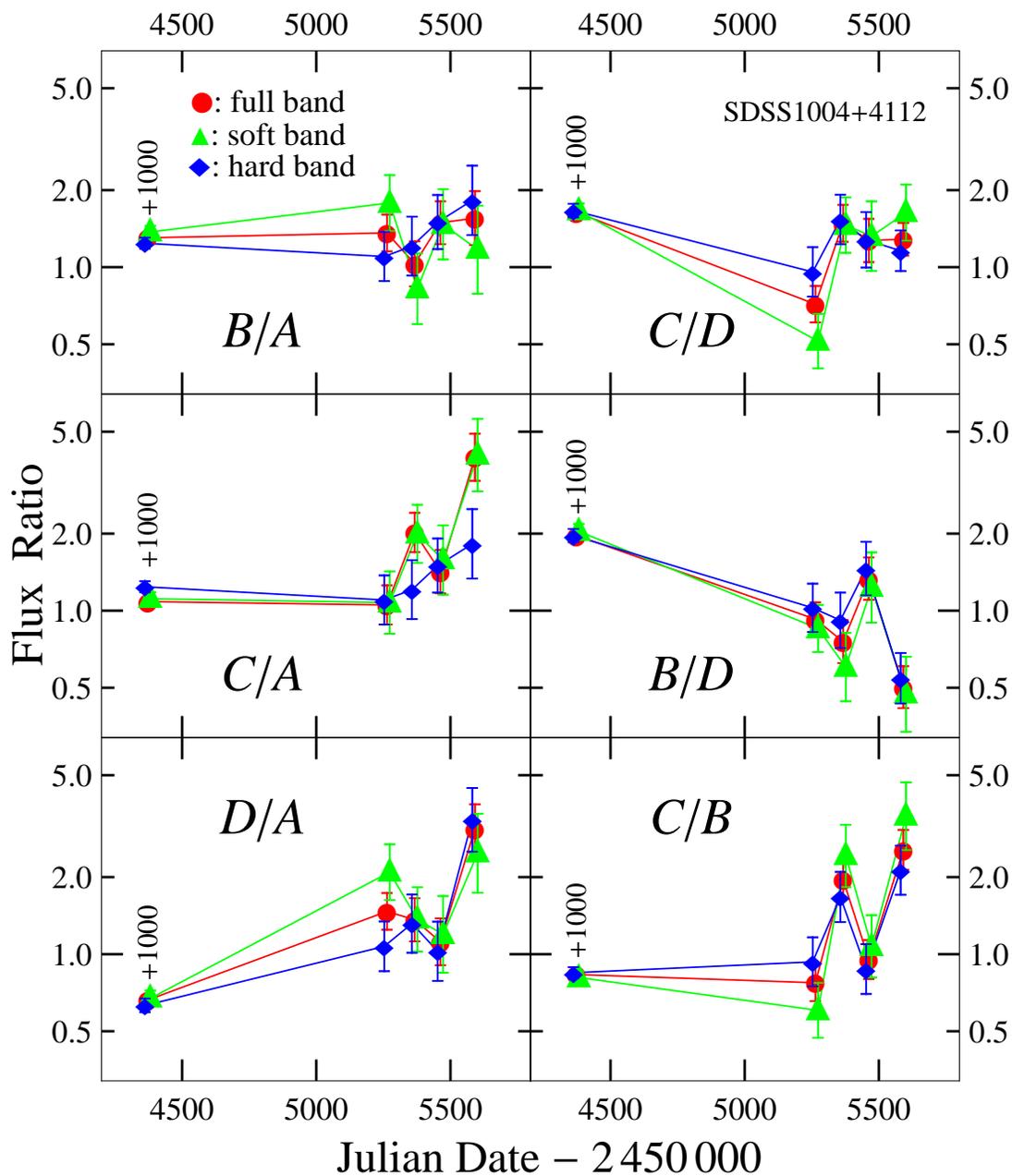}
	\caption{Chromatic X-ray microlensing light curves for SDSS~1004+4112 in the full (0.4--8.0 keV, red circles), soft (0.4--1.1 keV, green triangles), and hard (1.1--8 keV, blue diamonds)  X-ray bands. The first observation (made in 2005) in each panel was shifted by +1,000 days in order to better resolve the next 4 observations  (all made between 2010 and 2011).   
	        \label{fig:1004_lightcurve}}
\end{figure}

Our microlensing analysis is based on the image count rates presented in Tables \ref{tab:0158count}--\ref{tab:0435count}.  In principle, both microlensing by stars in the foreground galaxy and intrinsic variability of the background quasar can cause variability, and intrinsic variability modulated by time delays can mimic microlensing in any given eopch. 
In Q~2237+0305, the time delays are observationally constrained to be less than one day (Dai et al.\ 2003), and the intraday variability of the background quasar is measured to be only  $\sim$1\%  of the total flux (Chartas et al. 2001; Dai et al. 2003). 
The time delays are longer in the other galaxy lenses, (less than 15 days in HE~0435--1223, Kochanek et al. 2006; Courbin et al. 2011; about 15 days in SDSS~0924+0219, Inada et al. 2003;  $\le15$ days in QJ~0158--4325, Morgan et al. 2008;  and $\sim$162 days in HE~1104$-$1805, Morgan et al. 2008). 
The effects of source variability will be most significant in the cluster lens SDSS~1004+4112, where the time delays between some of the images are several years (e.g., image C leads A by $\sim$822 days, image D lags A  by lat least 1250 days, Fohlmeister et al. 2007, 2008). 
We focus on the evolution of the flux ratios between pairs of images to minimize the effects of intrinsic variability.
If we were focused on flux ratios at a single epoch, the time delay is the relevant time scale over which we need to consider the effects of intrinsic variability. This would be a major problem for SDSS~1004+4112 with its long  A/B to C or D delays. 
However, when we focus on microlensing through the time variability of flux ratios, we can prove mathematically that the relevant time scale for being affected by intrinsic variability is the shorter of the time delays and the sampling time scale.
Hence for images C and D of SDSS~1004+4112, we need only worry about the intrinsic variability on times scales of order 90 days rather than the multi-year A/B to C/D time delays. 
Figures \ref{fig:0158_lightcurve}--\ref{fig:0435_lightcurve} show the evolution of the full, soft, and hard band flux ratios for  QJ~0158$-$4325, HE~1104$-$1805, SDSS~0924+0219, SDSS~1004+4112, and HE~0435$-$1223, respectively.
In these figures, we slightly offset the observation dates for the soft and hard bands for clarity.  
We have already discussed  Q~2237+0305 in Chen et al. (2011).  

\subsection{X-ray Microlensing}

In each lens system, we first fit the microlensing light curves (flux ratios) in each of the three bands as constants to test for the existence of microlensing variability. 
For example, in the two-image lens QJ~0158--4325 (Figure~\ref{fig:0158_lightcurve}), we obtain  $\chi^2=9.7$ (dof=5) with a null hypothesis probability of $\rm NP= 0.08$ in the full band,  $\chi^2=6.4$  with an $\rm NP=0.27$ in the soft band,  and  $\chi^2=8.92$ with an $\rm NP=0.11$ in the hard band. 
Thus we weakly detect microlensing in the full X-ray band.
For HE~1104$-$1805 (Figure~\ref{fig:1104_lightcurve}), we obtain $\chi^2=81.4$ (dof=8) with $\rm NP=2.5\times 10^{-14}$ in the full band, $\chi^2=33.4$ with $\rm NP=0.00005$ in the soft band, and $\chi^2=44.9$ with $\rm NP=3.8\times 10^{-7}$ in the hard band. 
Therefore we detect microlensing with high confidence in all three bands for HE~1104$-$1805.
In the four image lenses  SDSS~0924+0219, SDSS~1004+4112, and HE~0435--1223, where we have to test the existence of microlensing signal from each flux ratio combination, we summarize the microlensing statistics in Tables  \ref{tab:0924chi2}--\ref{tab:0435chi2}. 
In SDSS~0924+0219 (Figure \ref{fig:0924_lightcurve} and Table \ref{tab:0924chi2}), among the 6 pairs of flux ratios,  the $B/A$ pair (images A and B  are  relatively brighter) gives the best statistical evidence for microlensing with a  full band probability of  $99.91\%$.  
The second most significant pair, C/B, shows microlensing at $94\%$ confidence for the full band.  
We also detect microlensing in SDSS~1004+4112 with high confidence (Figure \ref{fig:1004_lightcurve} and Table \ref{tab:1004chi2}). 
For example,  considering the B/D pair in the full band, we have a $\chi^2=144.8$ for ${\rm dof}=4$ with a null hypothesis probability $\rm NP=2.6\times 10^{-30},$ in the soft band, we have a $\chi^2=72.2$ with an $\rm NP=7.7\times 10^{-15}$, and in the hard band, a $\chi^2=73.1$ with an $\rm NP=5.0\times 10^{-15}.$  

\begin{figure}
	\epsscale{0.8}
	\plotone{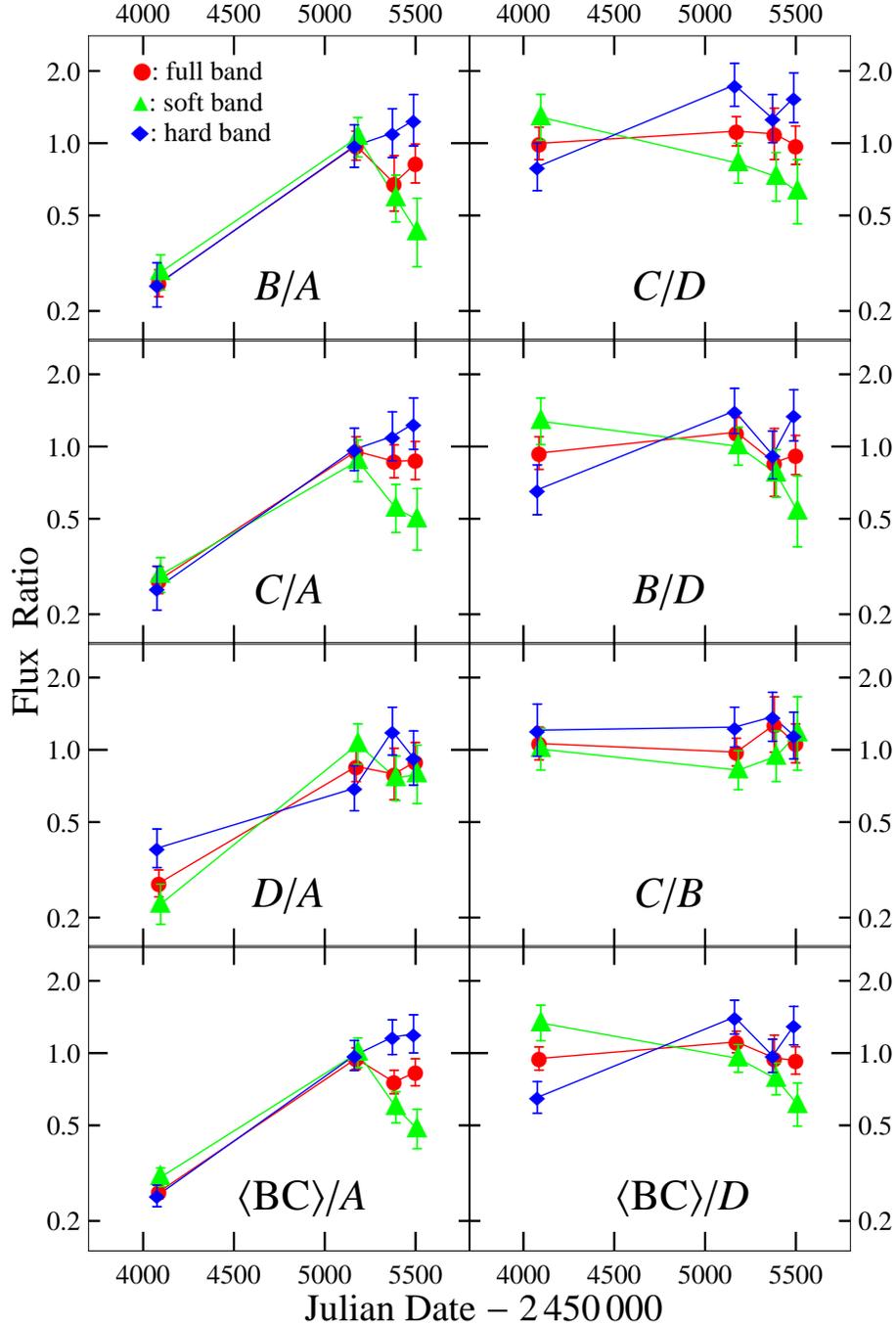}
	\caption{Chromatic X-ray microlensing light curves for HE~0435$-$1223 in the full (0.4--8.0 keV, red circles), soft (0.4--1.3 keV, green triangles), and hard (1.3--8 keV, blue diamonds) X-ray bands. In the last row, panels $\langle BC\rangle/A$ and $\langle BC\rangle/D$ show the light curves of the flux ratios between the intrinsic variability template (constructed from  images B and C assuming their variability is intrinsic) and images A and D.   
	        \label{fig:0435_lightcurve}}
\end{figure}

The microlensing statistics of HE~0435$-$1223 are given in Table~\ref{tab:0435chi2}. 
We observe from Figure~\ref{fig:0435_lightcurve} that the flux ratios between images C and B  are consistent with a constant, with $\chi^2=0.7,$ 0.96 and 0.31 (dof =3) for a model with no variability in the full, soft, and hard bands. 
This suggests that the variability of images B and C is mostly intrinsic during the observations reported in this paper. 
This can also be seen from the similarity between the light curve of B/A and that of C/A, or from the similarity between the light curves C/D and B/D (Figure~\ref{fig:0435_lightcurve}).  
From Figure~\ref{fig:0435_lightcurve} and Table~\ref{tab:0435chi2}, we see that there is strong evidence for microlensing in image A of HE~0435$-$1223. 
For example,  when fitting  B/A ratio as a constant (dof=3), we obtain a $\chi^2=40.8$ with an $\rm NP=7.3\times 10^{-9}$ in the full band, a $\chi^2=17.3$ with an $\rm NP=0.0006$ in the soft band, and a $\chi^2=29.8$ and an $\rm NP=1.5\times 10^{-6}$ in the hard band. 
As for image D, when fitting the light curves of B/D or C/D pair with constants, we find that the light curve is roughly consistent with constant in the full band; however, this is not true when fitting the soft or hard band light curves. 
For example, from the B/D light curve, the full band fit gives  $\chi^2=1.4$ (dof=3) with a one sided probability of 0.30, whereas the soft band fit gives a $\chi^2=6.0$ with an $\rm NP=0.11$ and the hard band fits gives a $\chi^2=6.9$ and an NP value 0.07.  
This suggests that image D of HE~0435$-$1223 is also gravitationally microlensed and that the lensing is energy-dependent with changes in the soft band opposite from those in the hard band (see next section). 
We assume that the variability of images B and C is intrinsic and use their noise-weighted  mean, $\langle BC\rangle$, as an estimate of the intrinsic variability.
We show the flux ratio $\langle BC\rangle$/A between image A and $\langle BC\rangle$ (the intrinsic variability template), and between image D and $\langle BC\rangle$, $\langle BC\rangle$/D, in Figure~\ref{fig:0435_lightcurve}.  
When we fit  $\langle BC\rangle$/A as a constant, we obtain $\chi^2=101.4$ with an NP value of $7.9\times 10^{-22}$ in the full band,  $\chi^2=33.4$ and an $\rm NP=2.7\times 10^{-7}$ in the soft band, and $\chi^2=63.0$ and an $\rm NP=1.4\times 10^{-13}$ in the hard band.  
We therefore detect microlensing in image A  with high confidence. 
This interpretation is confirmed by the better sampled optical light curves presented in Blackburne et al. (2012). 
Fitting the light curve $\langle BC\rangle$/D with a constant, we obtain a $\chi^2=1.5$ with a one sided probability of 0.32 in the full band,  a $\chi^2=9.1$ with an $NP=0.03$ in the soft band, and a $\chi^2=14.1$ with an $NP=0.003$ in the hard band. Therefore, we detect microlensing in image D of HE~0435--1223.

\subsection{X-ray Energy-Dependent Microlensing}\label{sec:chromatic}

Next, we test for differences between the soft and the hard energy bands, although only the observations of Q~2237+0305 were designed to have enough counts for this purpose.  
For the two-image lenses  QJ~0158--4325 and HE~1104--1805, we can simplify compute the $\chi^2$ between the soft and hard band flux ratios.  
In QJ~0158--4325, we find $\chi^2=6.8$  (dof=6) with a null hypothesis probability $\rm NP=0.35$, which suggests no significant chromatic microlensing. 
Among the six observations of QJ~0158$-$4325 (Figure~\ref{fig:0158_lightcurve}), epochs No.\ 1, 3, and 6 show differences in flux ratios between the soft and hard bands, whereas the other three epochs do not.  
In HE~1104$-$1805, we find $\chi^2=7.2$ (dof=9) which also suggests no significant chromatic microlensing. 

For the four-image lenses SDSS~0924+0219, SDSS~1004+4112, and HE~0435--1223, we construct a $\chi^2$ test for energy-dependent microlensing using
\begin{equation}
\chi^2=\sum_{i,j}^{}{\left[\frac{(d^{\rm hard}_{ij}-s^{\rm hard}_i-\mu^{\rm hard}_j-\delta_{ij})^2}{(\sigma^{\rm hard}_{ij})^2}+\frac{(d^{\rm soft}_{ij}-s^{\rm soft}_i-\mu^{\rm soft}_j-\delta_{ij})^2}{(\sigma^{\rm soft}_{ij})^2}\right]},
\end{equation} where $d^{\rm soft}_{ij}$ and $d^{\rm hard}_{ij}$ are the observed count rates in the soft and hard bands at epoch $i$ in image $j,$  $s^{\rm soft}_i$ and $s^{\rm hard}_i$ are the separate source light curves in the soft and hard bands at epoch $i,$ $\mu^{\rm soft}_j$ and $\mu^{\rm hard}_j$ are the mean magnification of the soft and hard bands in image $j,$  and $\delta_{ij}$ is the microlensing variability at epoch $i$ in image $j$ assumed to be the same for the soft and the hard bands. 
If there is an energy dependence (i.e., $\delta^{\rm soft}_{ij}\ne \delta^{\rm hard}_{ij}$), a model with $\delta_{ij}=\delta^{\rm soft}_{ij}=\delta^{\rm hard}_{ij}$ should fit poorly.  For SDSS~1004+4112, we obtain $\chi^2=13.4$ (dof=5) with null hypothesis probability $\rm NP=0.02.$ For SDSS~0924+0219, we obtain $\chi^2=18.6$ (dof=7) with $\rm NP=0.009.$  For HE~0435--1223, we find $\chi^2=21.6$ (dof=3) with $\rm NP=0.00008.$ Therefore,  we detect energy-dependent microlensing at $98\%,$ $99.1\%$ and $99.99\%$ confidence for these three four-image lenses.

Although we have detected \feka\ lines, which probably originate from a different region from the hard X-ray continuum and contaminate the microlensing light curves, the typical equivalent widths of the line are only 0.2--0.3 keV (observed frame) and do not contribute significantly to the hard band X-ray flux. 
This means that the differential microlensing between the soft and hard X-ray bands cannot be attributed to the presence of the iron lines. 
For example, if we use  hard band microlensing light curves excluding the spectral segment containing the \feka\ emission (2.0--2.5 keV observed frame), we  find that the Chen et al. (2011) results for Q~2237+0305 are unchanged.

\begin{figure}
	\epsscale{0.6}
	\plotone{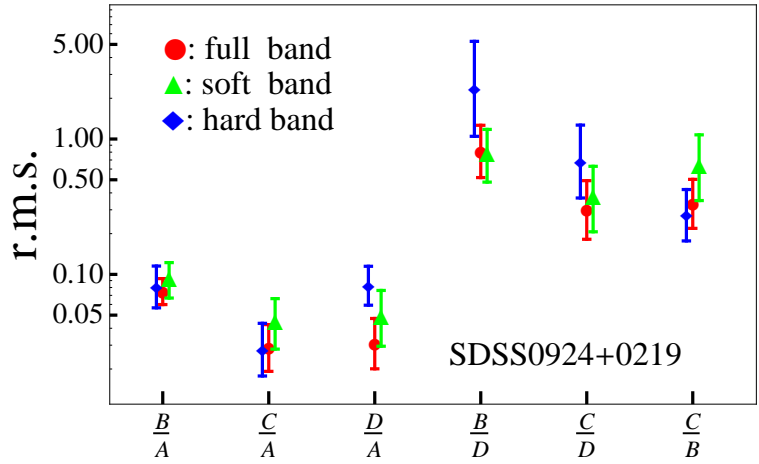}
	\caption{X-ray microlensing variability amplitudes of SDSS~0924+0219 in the full, soft and hard X-ray bands. 
        \label{fig:0924_VarAmp}}
\end{figure}

\begin{figure}
	\epsscale{0.6}
	\plotone{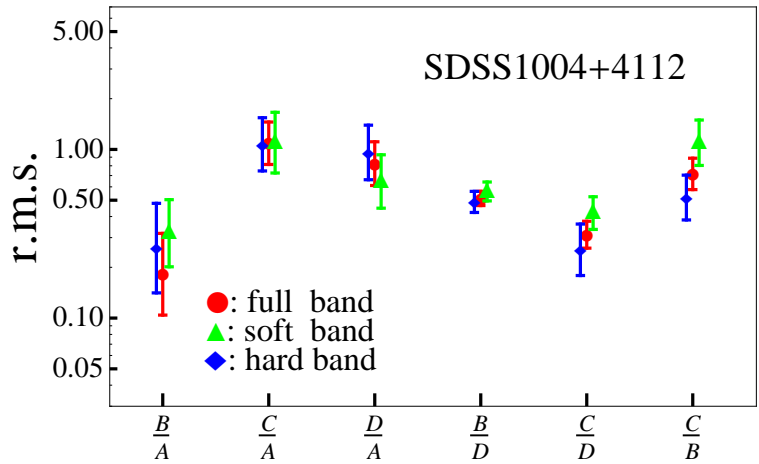}
	\caption{X-ray microlensing variability amplitudes of SDSS~1004+4112 in the full, soft and hard X-ray bands. 
        \label{fig:1004_VarAmp}}
\end{figure}

\begin{figure}
	\epsscale{0.6}
	\plotone{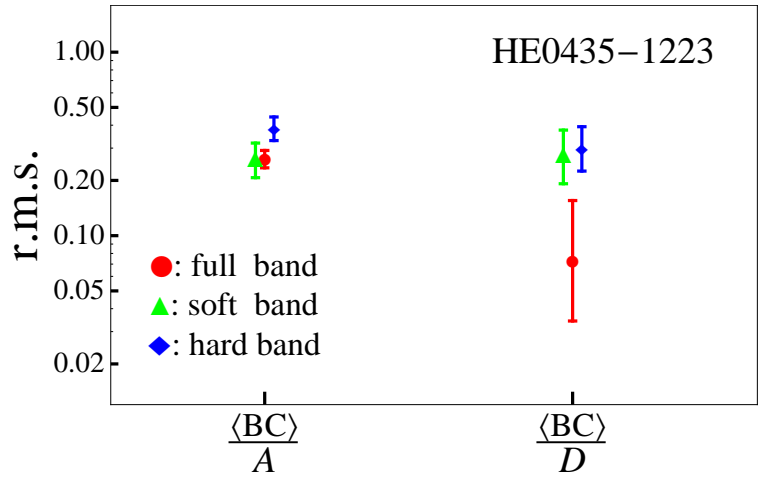}
	\caption{X-ray microlensing variability amplitudes of HE~0435$-$1223 in the full, soft and hard X-ray bands. Here ``$\langle BC\rangle$" indicates the intrinsic variability template constructed from images B and C. 
        \label{fig:0435_VarAmp}}
\end{figure}

Unfortunately, this statistical detection of an energy dependence is difficult to visualize. For example, in Figures \ref{fig:0924_VarAmp}--\ref{fig:0435_VarAmp} we show the RMS flux ratios for various image combinations in SDSS~0924+0219, SDSS~1004+4112, and HE~0435--1223. 
Except for Q~2237+0305 (Chen et al. 2011) and image A of HE~0435$-$1223, any energy dependence is not visually obvious. 
Recall, however, that these observations are limited to integration times shorter than those needed to measure the hard/soft band light curves well.

\subsection{Microlensing of the \feka\ Line}\label{sec:iron}

Microlensing of the \feka\ line has previously been detected in  MG~J0414+0534 (Chartas et al.\ 2002), Q~2237+0305 (Dai et al.\ 2003), H~1413+117 (Chartas et al.\ 2007), and SDSS~1004+4112 (Ota et al.\ 2006). 
For the six lenses with \feka\ detections in the combined spectra (Table~\ref{tab:EW} and Figure~\ref{fig:EW_Eline}), we can test for microlensing by searching for changes in the line (EW or profile) between the images. Unfortunately we can only use the combined spectra rather than examining individual epochs because the S/N of the individual epochs is too low. This time averaging will smooth and reduce any microlensing signal.

In principle, if the continuum is microlensed differently from the lines, the equivalent widths (EWs) of the lines of the different images will differ. 
The effect tends to be one-sided even though microlensing can both magnify or demagnify a source. 
The demagnified regions tend to be smooth and wide, and thus  demagnify large and small regions almost equally, while magnified regions have the caustics that will more strongly magnify  compact emission regions. 
Thus there will be a tendency for lensed quasars to have either larger or smaller EWs than unlensed quasars depending on whether the line emitting region is smaller or larger then the continuum.  
We fit the EWs of the images as constants in each lens system (except SDSS~0924+0219, see Figure \ref{fig:EW}). 
For QJ~0158$-$4325, HE~1104$-$1805,  SDSS~1004+4112, and HE~0435$-$1223, the EWs of the images are well fit by constants of 0.57 keV, 0.38 keV, 0.60 keV, and 0.88 keV, respectively, whereas the EW of image C of Q~2237+0305 differs from those of the other images.  
For Q~2237+0305, we obtain a best fit constant EW of 0.50 keV with $\chi^2=4.87$ (dof = 3) and an NP value of $0.18$, which weakly suggests differential microlensing effects between the X-ray continua and the \feka\ line. 
Overall, we conclude that the flux ratios of the \feka\ lines are comparable to those of the X-ray continua. Of course this null result does not mean that the \feka\ line is not microlensed, rather it implies that the size of the \feka\ emission has to be comparable to the X-ray continuum emitting region.

\begin{figure}\label{fig:EW_Lumi.eps}
	\epsscale{0.8}
	\plotone{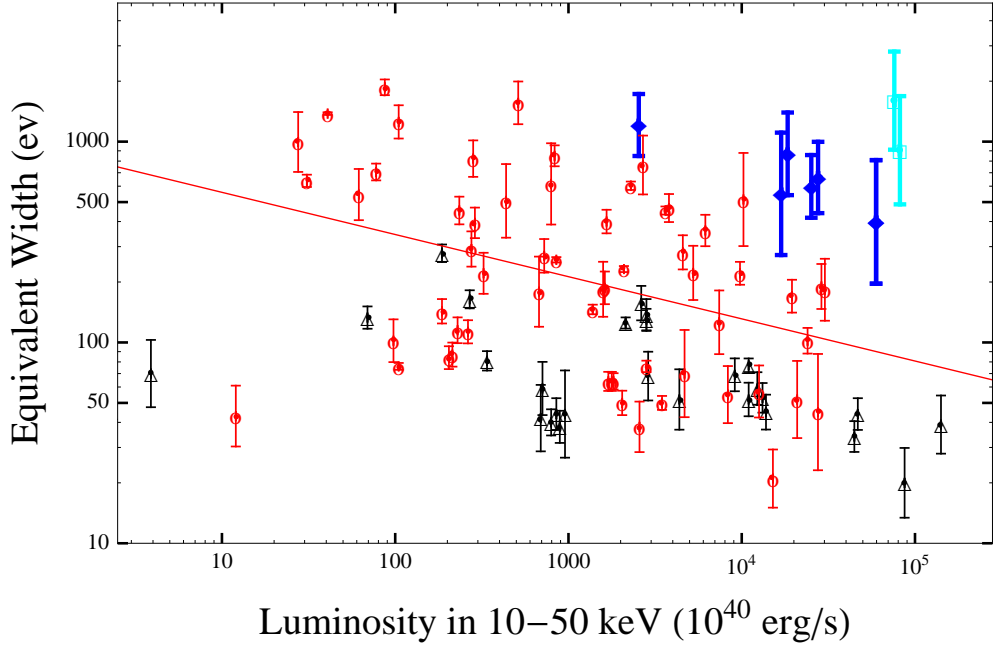}
	\caption{Equivalent  width of  \feka\  lines as a  function of rest frame X-ray luminosity (10--50 keV) for  SDSS~0924+0219, QJ~0158$-$4325, HE~0435$-$1223,  Q~2237+0305,  SDSS~1004+4112, and HE~1104$-$1805 (left to right, blue diamonds). H~1413+117 (Chartas et al. 2007) and MG~J0414+0534 (Chartas et al. 2002) are marked by cyan squares. The black triangles (red circles) are AGN with absorption column densities greater (less) than $10^{22} \>\rm cm^{-2}$ (Fukazawa et al.  2011). The red line is the best fit to the low absorption systems (red points) as a power-law with additional intrinsic scatter in the $\rm EWs.$  
	 \label{fig:EW_Lumi}}
\end{figure}

As another test, we compared the EW--Luminosity relation of these six lenses (Table \ref{tab:EL_EW_Lumi}) with those of 88 nearby Seyfert galaxies observed by {\it Suzaku} (Fukazawa et al.\ 2011). 
The inverse correlation between the EW of the iron line and the X-ray luminosity is referred to as the X-ray Baldwin effect, and was first discovered by Iwasawa \& Taniguchi (1993) for neutral \feka\ lines and X-ray luminosities in the 2--10 keV band.  
Recently, Fukazawa et al.\ (2011) extended the trend to the 10--50 keV band and for ionized \feka\ lines. 
If the iron line emission region is smaller than the continuum and therefore more strongly microlensed, lenses should on average have higher  EWs than unlensed quasars with similar luminosities. 
Figure~\ref{fig:EW_Lumi}  shows the results where we have  divided the local samples into low ($<10^{22}\,\rm cm^{-2}$, red circles) and high ($>10^{22}\,\rm cm^{-2}$, black triangles) absorption systems.
Our six points are marked by blue diamonds, where we have corrected for typical gravitational lensing magnifications of 16 for the four image lenses and 8 for the two image lenses. 
We also include the results for  MG J0414+0534 (Chartas et al.\ 2002) and H~1413+117 (Chartas et al.\ 2007).  
The data from Fukazawa et al.\ (2011) clearly shows the X-ray Baldwin effect of declining EW with luminosity. 
The gravitationally lensed quasars have systematically higher EWs than typical AGN of similar X-ray luminosity (in the 10--50 keV band).  
To test if our data is statistically consistent with typical AGN, we first fit the data in Fukazawa et al.\ (2011) with a power-law model plus intrinsic scatter in equivalent widths, and obtain
\begin{equation}
\log {\rm EW} = (6.81\pm0.51) - (0.21\pm0.07) \log {\rm L} \pm (1.02\pm0.11) 
\end{equation}
 with a reduced $\chi^2=0.94.$ 
 Compared to this relation and its scatter, the 8 lenses have a $\chi^2=25.2$ (dof = 8), which is inconsistent with the distribution of typical AGN even without considering that all eight lenses lie on one side of the relation. 
 The difference cannot be resolved by errors in the magnification correction applied to the luminosity.
 These are uncertain by at most a factor of 2, while a change by a factor of 30 would be needed to explain the differences.  
This suggests that the \feka\ line is microlensed more strongly than the X-ray continua, which in turn implies that the \feka\  emission region is smaller than the X-ray continuum, confirming earlier conclusions (Dai et al. 2003;  Popovi$\rm\acute{c}$ et al. 2003, 2006).

\section{Discussion}

In this paper, we report our \chandra\ monitoring observations of six gravitationally lensed quasars: QJ~0158$-$4325, HE~1104$-$1805, HE~0435$-$1223, SDSS~0924+0219, SDSS~1004+4112, and Q~2237+0305.
We find that simple power-law models plus Gaussian emission lines modified by absorption provide good fits to the spectra. 
We do not detect the predicted reflection component, other than the metal emission lines, due to the low S/N in the hard X-ray band.
We detect intrinsic spectral variability in two epochs for Q~2237+0305, and differential absorption between images in QJ~0158$-$4325,  SDSS~1004+4112, HE~0435$-$1223, and Q~2237+0305, but not in HE~1104$-$1805, or SDSS~0924+0219. 
From the stacked spectra of each lens, we detect \feka\ lines in all lenses and in almost all images. 
In Q~2237+0305, we marginally detect the Ni XXVII K$\alpha$ line in images A and D. 

We clearly detect  X-ray microlensing in all six lenses,  and this can now be analyzed to determine the size of the X-ray emitting regions. 
The first of these analyses, for HE~0435$-$1223, is presented in Blackburne et al. (2012). Of these 6 lenses, the integration times were designed to provide accurate measurements in multiple bands only for Q~2237+0305. For this lens, we clearly see energy-dependent microlensing. Despite the lower S/N of the observations of the remaining 5 lenses, we still succeed in detecting  energy-dependent  microlensing in HE~0435$-$1223, SDSS~1004+4112, and SDSS~0924+0219, but not QJ~0158$-$4325 or HE~1104$-$1805.

In general, we do not see shifts in the \feka\ energy or EW between images for these lenses, except for Q~2237+0305 and HE~0435$-$1223, which implies that the sizes of the iron line emission regions are comparable to those of the X-ray continua.
However, since the \feka\ line is extracted from the time averaged spectrum of each image, any microlensing effect has been smoothed. 
We note, however, that the lenses seem to have significantly higher EWs than a sample of 88 nearby Seyfert galaxies observed by {\it Suzaku} (Fukazawa et al. 2011).
 This suggests that the \feka\ line is more strongly microlensed  than the X-ray continuum, and this implies a smaller emission region for the iron lines.
This comparison is made at fixed luminosity, so it is not simply due to comparing local Seyferts with higher redshift more luminous quasars.
Since the size of the X-ray continuum is already constrained to be small, $\sim$$10\,r_g$ (Morgan et al.\ 2008; Chartas et al.\ 2009; Dai et al.\ 2010; Blackburne \etal 2012), the \feka\ emission must come from the very inner region of the disk. 
The existence of chromatic microlensing  between the soft and hard X-ray bands  and the constraints on the size of the line emission region have important implications for testing general relativity in the strong field regime and the theoretical modeling of quasar accretion disks.  


Although X-ray emission is one of the defining characteristic of AGN, the origin of this emission is still unknown.
Unlike X-ray binaries, the AGN accretion disk itself is not hot enough to generate X-rays.
The X-ray emission is generally attributed to inverse Compton scattering of soft UV photons from the accretion disk by hot electrons from a ``corona'' (see review of Reynolds \& Nowak  2003). 
Although the origin of the hot electrons in the corona is  debated, recent microlensing analyses have constrained the extent of this corona to be of order $10~r_g$ (Morgan et al.\ 2008; Chartas et al.\ 2009; Dai et al.\ 2010; Blackburne \etal 2012).
Combining the results of this paper and Chen et al.\ (2011), we further find that the corona might have a temperature gradient with higher temperature electrons occupying a smaller volume.

Several models have been proposed to heat the electrons.
In the classic ``sandwich'' corona model (Haardt \& Maraschi 1991, 1993) and its extension, the ``patchy'' corona model (Haardt et al.\ 1994), the gravitational energy of the accretion disk is assumed to be transferred into the disk and corona with constant fractions.   
This group of models is in conflict with the microlensing results, since they predict similar sizes for the X-ray and optical emission regions, whereas the microlensing studies uniformly find smaller X-ray emission sizes than optical sizes (Pooley et al.\ 2006; Morgan et al.\ 2008; Chartas et al.\ 2009; Dai et al.\ 2010; Blackburne \etal 2011; Blackburne \etal 2012).
However, simply requiring that the energy transfer process operates  only on scales  $\lesssim 10~r_g$ will solve this problem.
A second group of models assumes that the X-ray emission is composed of a large number of flares triggered by magnetic field re-connections (e.g., Collin et al.\ 2003; Czerny et al.\ 2004; Torricelli-Ciamponi et al.\ 2005; Trzesniewski et al.\ 2011). 
The flare models can be considered as realistic realizations of the ``patchy'' corona model, and have advantages for interpreting the X-ray variability of AGN.  
Another difference from the steady corona models is that the electron distribution is mostly non-thermal in the flare models (Torricelli-Ciamponi et al.\ 2005).
In general, the flare models predict similar X-ray and optical emission sizes for the microlensing analysis, because the corona is composed of a larger number of flares and the flares are randomly distributed over the disk. 
Some recent developments of the flare model specify that an avalanche of flares will be triggered by a single flare with a preferred direction of the triggered flares to the black hole (Trzesniewski et al.\ 2011), which will result in a smaller X-ray emission size at the disk center.
A final set of models heat electrons with shocks.  
For example, Ghisellini et al.\ (2004) proposed the aborted jet model, where a jet launched near the black hole cannot escape the potential well, returns, and collides with other material to form shocks.  These shocks then convert the kinetic energy of the aborted jet into thermal energy. 
This model predicts  compact X-ray emission regions, possibly  consistent with the microlensing results.

Given a disk-corona geometry, there should also be a reflection component  created by the X-ray irradiation of the disk. 
Complicated radiative transfer processes occur during the reflection and the resulting continuum emission peaks at $\sim 50$~keV and produces metal emission lines, such as \feka\ (e.g., George \& Fabian 1991).
Based on the geometry of the model, one would expect the reflection region to be much larger than the corona, because the reflection can occur anywhere in the disk, a disk wind, or a surrounding torus. 
With the discovery of broad, skewed \feka\ lines in many Seyferts (e.g., MCG6--30--15, Tanaka et al.\ 1995), the reflection component near the black hole must contribute significantly to the total emissivity.
In addition, the reflection components in some Seyferts are measured to be constant while the X-ray continuum is variable.  
Fabian \& Vaughan (2003) proposed a light bending model, where the majority of the X-ray emission will illuminate the accretion disk due to the light bending effect by the black hole, producing the constant reflection component, and the escaping X-ray photons produce the X-ray continuum. 
If the X-ray source moves vertically above the accretion disk, it will produce a variable X-ray continuum and almost constant reflection component.
In a competing model, Miller \etal (2008) proposed that the reflection occurs in the disk wind and that the redshifted \feka\ lines are artifacts created by a forest of warm absorbers in the X-ray bands.
Our microlensing results suggest that the \feka\ emission region is compact and comparable in size or  smaller than the X-ray continuum region, which prefers the light bending model over the disk-wind model.

We have composed, for the first time, a modest sample of quasar microlensing light curves in the full, soft, and hard X-ray bands, and a sample of \feka\ measurements in lensed quasars.
Microlensing of the X-ray emission of lensed quasar is now clearly established  in the full band (Morgan \etal 2001; Dai \etal 2003; Blackburne \etal 2006; Morgan \etal 2008; Chartas \etal 2009; Dai \etal 2010; Chen \etal 2011; Blackburne \etal 2011, 2012),  and in the \feka\ lines (Chartas \etal 2002, 2004; Dai \etal 2003; Popovi$\rm\acute{c}$ et al. 2003, 2006). 
There are well-defined methods of analyzing the results to estimate the sizes of the X-ray emission regions (Kochanek 2004; Kochanek \etal 2007; Dai \etal 2010; Blackburne et al. 2012). 
To date, these analyses have yielded only upper bounds on the sizes of the X-ray emission regions because of the sparse X-ray light curves and most observations are too short to study multiple energy bands or the \feka\ emission in detail.
There are, however, no observational or theoretical limits to obtaining the necessary data as long as the \chandra\ observatory continues to operate. Unfortunately, this science will become impossible with \chandra's demise  because no future mission is designed to have the necessary angular resolution.   
  
We acknowledge the support from the NASA/SAO grants GO0-11121A/B/C/D, GO1-12139A/B/C, and GO2-13132A/B/C.  CSK and JAB are supported by NSF grant AST-1009756. CWM is grateful for support by the National Science Foundation under Grant No. AST-0907848.

 \clearpage

\begin{deluxetable}{lllllll}
\tabletypesize{\scriptsize}
\tablecolumns{7}
\tablewidth{0pt}
\tablecaption{Absorption-Corrected Count Rates for QJ~0158$-$4325.    \label{tab:0158count}}
\tablehead{
\colhead{ObsId \tablenotemark{a}}&
\colhead{ 11556 }&
\colhead{ 11557}&
\colhead{ 11558 }&
\colhead{ 11559 } &
\colhead{11560}&
\colhead{11561} \\
\colhead{ Date}&
\colhead{ 2009-11-06 }&
\colhead{ 2010-01-12}&
\colhead{ 2010-03-10 }&
\colhead{  2010-05-23 } &
\colhead{2010-07-28}&
\colhead{2010-10-06} \\
\colhead{Exposure (s)}&
\colhead{ 5026 }&
\colhead{ 5016}&
\colhead{ 5035  }&
\colhead{ 4942 }&
\colhead{4946} &
\colhead{4944}
}
\startdata
 ${\rm A_{full}} $  & $ 28.4{\pm  2.9}$  ${\rm } $  & $ 33.6{\pm  3.3}$  ${\rm } $  & $ 34.4{\pm  3.0}$  ${\rm } $  & $ 35.9{\pm  3.3}$  ${\rm } $  & $ 37.3{\pm  3.5}$  ${\rm } $  & $ 30.7{\pm  2.9}$ \\ 
 ${\rm A_{soft}} $  & $ 13.3{\pm  2.1}$  ${\rm } $  & $ 18.3{\pm  2.4}$  ${\rm } $  & $ 20.5{\pm  3.5}$  ${\rm } $  & $ 20.1{\pm  2.5}$  ${\rm } $  & $ 18.6{\pm  2.5}$  ${\rm } $  & $ 20.3{\pm  2.4}$ \\ 
 ${\rm  A_{hard}} $  & $ 13.5{\pm  1.8}$  ${\rm } $  & $ 14.7{\pm  2.2}$  ${\rm } $  & $ 13.5{\pm  1.8}$  ${\rm } $  & $ 15.8{\pm  2.0}$  ${\rm } $  & $ 17.4{\pm  2.0}$  ${\rm } $  & $ 10.4{\pm  1.5}$ \\ 
 \tableline
 ${\rm B_{full}} $  & $ 10.7{\pm  1.6}$  ${\rm } $  & $ 15.2{\pm  2.1}$  ${\rm } $  & $ 13.3{\pm  1.8}$  ${\rm } $  & $  9.7{\pm  1.6}$  ${\rm } $  & $  9.0{\pm  1.5}$  ${\rm } $  & $ 12.2{\pm  1.8}$ \\ 
 ${\rm B_{soft}} $  & $  7.7{\pm  1.5}$  ${\rm } $  & $  8.5{\pm  1.6}$  ${\rm } $  & $  6.2{\pm  1.8}$  ${\rm } $  & $  5.0{\pm  1.2}$  ${\rm } $  & $  5.4{\pm  1.2}$  ${\rm } $  & $  6.5{\pm  1.3}$ \\ 
 ${\rm B_{hard}}$  & $  3.9{\pm  1.0}$  ${\rm } $  & $  6.5{\pm  1.3}$  ${\rm } $  & $  6.7{\pm  1.3}$  ${\rm } $  & $  3.9{\pm  1.0}$  ${\rm } $  & $  4.1{\pm  1.0}$  ${\rm } $  & $  5.4{\pm  1.1}$ \\ 
 \tableline
 ${\rm bkg_{full}}  $  & $ 2.97$             ${\rm } $  & $ 1.92$             ${\rm } $  & $ 3.13$             ${\rm } $  & $ 4.97$             ${\rm } $  & $ 3.37$             ${\rm } $  & $ 2.66$            \\ 
 ${\rm bkg_{soft}} $  & $ 0.35$             ${\rm } $  & $ 0.35$             ${\rm } $  & $ 0.35$             ${\rm } $  & $ 0.89$             ${\rm } $  & $ 0.18$             ${\rm } $  & $ 0.35$            \\
\enddata
\tablenotetext{a}{All events have standard ASCA grades of 0,2,3,4,6 and have energies between 0.4--8~keV. We split the full X-ray band (0.4--8.0 keV) into a soft X-ray band (0.4--1.3 keV), and a hard X-ray band (1.3--8 keV). The background counts are extracted from an annulus with inner and outer radii of $5''$ and $20'',$ respectively. The units of the image count rates are $10^{-3}\,{\rm s}^{-1}$, and the units of  the background fluxes are $10^{-6}\,{\rm s}^{-1}\rm{arcsec}^{-2}.$ }
\end{deluxetable}

\begin{deluxetable}{llllllllll}
\tabletypesize{\scriptsize}
\setlength{\tabcolsep}{0.03in} 
\tablecolumns{10}
\tablewidth{0pt}
\tablecaption{Absorption-Corrected Count Rates for HE~1104$-$1805.  \label{tab:1104count}}
\tablehead{
\colhead{ObsId \tablenotemark{a}}&
\colhead{ 375  }&
\colhead{ 6918}&
\colhead{ 6917}&
\colhead{ 6919}&
\colhead{6920}&
\colhead{6921}&
\colhead{11553} &
\colhead{11554}&
\colhead{11555}\\
\colhead{ Date}&
\colhead{ 2000-06-10 }&
\colhead{ 2006-02-16 }&
\colhead{ 2006-03-15 }&
\colhead{ 2006-04-09 }&
\colhead{ 2006-10-31}&
\colhead{ 2006-11-08}&
\colhead{ 2010-02-09 }&
\colhead{ 2010-07-12}&
\colhead{ 2010-12-20} \\
\colhead{Exposure (s)}&
\colhead{ 47425 }&
\colhead{ 4959   }&
\colhead{ 4549   }&
\colhead{ 4865   }&
\colhead{ 5007   } &
\colhead{ 4915   }&
\colhead{ 12756 }&
\colhead{ 12758 } &
\colhead{ 12757 }
}
\startdata
${\rm A_{full}   } $ & $ 25.3{\pm  0.8}$  & $  8.4{\pm  1.5}$  & $ 10.3{\pm  1.7}$  & $  7.5{\pm  1.4}$  & $ 16.2{\pm  2.2}$  & $ 13.4{\pm  2.3}$  & $ 10.8{\pm  1.1}$  & $ 23.5{\pm  2.1}$  & $ 11.7{\pm  1.1}$ \\ 
${\rm A_{soft}   } $ & $ 15.1{\pm  0.8}$  & $  3.9{\pm  1.0}$  & $  5.3{\pm  1.3}$  & $  3.8{\pm  1.1}$  & $  9.0{\pm  1.9}$  & $  7.3{\pm  1.6}$  & $  5.6{\pm  0.9}$  & $ 11.2{\pm  1.2}$  & $  5.3{\pm  1.0}$ \\ 
${\rm A_{hard}   } $ & $ 10.3{\pm  0.7}$  & $  4.2{\pm  0.9}$  & $  5.4{\pm  1.2}$  & $  4.0{\pm  1.1}$  & $  5.9{\pm  1.4}$  & $  6.4{\pm  1.3}$  & $  4.2{\pm  0.6}$  & $ 11.5{\pm  1.4}$  & $  5.9{\pm  0.8}$ \\ 
\tableline
${\rm B_{full}   } $ & $ 13.6{\pm  0.6}$  & $ 21.8{\pm  2.8}$  & $ 17.3{\pm  2.5}$  & $ 16.5{\pm  2.3}$  & $  8.7{\pm  1.5}$  & $ 13.0{\pm  2.3}$  & $ 14.5{\pm  1.3}$  & $  9.3{\pm  1.1}$  & $ 14.3{\pm  1.2}$ \\ 
${\rm B_{soft}   } $ & $  8.6{\pm  0.6}$  & $  9.3{\pm  1.8}$  & $ 10.4{\pm  2.0}$  & $  8.9{\pm  1.8}$  & $  6.1{\pm  1.5}$  & $  7.3{\pm  1.7}$  & $  7.7{\pm  1.1}$  & $  4.6{\pm  0.7}$  & $  6.4{\pm  1.1}$ \\ 
${\rm B_{hard}   } $ & $  4.5{\pm  0.4}$  & $ 11.3{\pm  1.7}$  & $  5.8{\pm  1.4}$  & $  6.5{\pm  1.4}$  & $  3.6{\pm  1.0}$  & $  4.8{\pm  1.1}$  & $  7.0{\pm  0.9}$  & $  4.2{\pm  0.7}$  & $  7.2{\pm  1.0}$ \\
\tableline
${\rm bkg_{full} } $ & $  3.14          $  & $  1.99          $  & $  2.65          $  & $  8.11          $  & $  3.94          $  & $  2.68          $  & $  2.06          $  & $  3.18          $  & $  3.01          $ \\
${\rm bkg_{soft} } $ & $  1.02          $  & $  0.22          $  & $  0.72          $  & $  1.80          $  & $  0.88          $  & $  0.45          $  & $  0.09          $  & $  0.43          $  & $  0.94          $ \\
\enddata
\tablenotetext{a} {All events have standard ASCA grades of 0,2,3,4,6 and have energies between 0.4--8~keV. We split the full X-ray band (0.4--8.0 keV) into a soft X-ray band (0.4--1.2 keV), and a hard X-ray band (1.2--8 keV). The background counts are extracted from an annulus with inner and outer radii of $10''$ and $20'',$ respectively. The units of the image count rates are $10^{-3}\,{\rm s}^{-1}$, and the units of  the background fluxes are $10^{-6}\,{\rm s}^{-1}\rm{arcsec}^{-2}.$  }
\end{deluxetable}

\begin{deluxetable}{lllllll}
\tabletypesize{\scriptsize}
\tablecolumns{7}
\tablewidth{0pt}
\tablecaption{Absorption-Corrected Count Rates for SDSS~0924+0219.   \label{tab:0924count}}
\tablehead{
\colhead{ObsId\tablenotemark{a}}&
\colhead{ 5604  }&
\colhead{ 11562 }&
\colhead{ 11563 }&
\colhead{ 11564 }&
\colhead{ 11565 }&
\colhead{11566 }\\
\colhead{ Date}&
\colhead{ 2005-02-24}&
\colhead{ 2010-01-13}&
\colhead{ 2010-03-10 }&
\colhead{ 2010-05-10 } &
\colhead{2010-06-22}&
\colhead{2010-10-05} \\
\colhead{Exposure (s)}&
\colhead{ 17944 }&
\colhead{ 21542}&
\colhead{ 21264  }&
\colhead{ 21642 }&
\colhead{21546} &
\colhead{21543}
}
\startdata
${\rm A_{full}   } $ & $  6.6{\pm  0.8}$  & $ 12.0{\pm  1.3}$  & $  8.0{\pm  0.8}$  & $ 18.5{\pm  1.2}$  & $ 13.6{\pm  1.2}$  & $  9.9{\pm  0.8}$ \\ 
${\rm A_{soft}   } $ & $  3.6{\pm  0.6}$  & $  7.0{\pm  1.0}$  & $  3.9{\pm  0.6}$  & $  9.8{\pm  0.9}$  & $  7.2{\pm  0.9}$  & $  5.8{\pm  0.6}$ \\ 
${\rm A_{hard}   } $ & $  2.6{\pm  0.4}$  & $  4.8{\pm  0.6}$  & $  3.7{\pm  0.5}$  & $  7.8{\pm  0.7}$  & $  5.9{\pm  0.9}$  & $  3.5{\pm  0.5}$ \\ 
\tableline
${\rm B_{full}   } $ & $  1.7{\pm  0.3}$  & $  3.2{\pm  0.5}$  & $  0.9{\pm  0.2}$  & $  2.8{\pm  0.4}$  & $  4.6{\pm  0.5}$  & $  1.9{\pm  0.3}$ \\ 
${\rm B_{soft}   } $ & $  1.3{\pm  0.3}$  & $  1.7{\pm  0.4}$  & $  0.4{\pm  0.1}$  & $  1.6{\pm  0.3}$  & $  2.2{\pm  0.4}$  & $  1.4{\pm  0.3}$ \\ 
${\rm B_{hard}   } $ & $  0.4{\pm  0.1}$  & $  1.2{\pm  0.3}$  & $  0.4{\pm  0.1}$  & $  1.1{\pm  0.2}$  & $  2.1{\pm  0.5}$  & $  0.6{\pm  0.2}$ \\ 
\tableline
${\rm C_{full}   } $ & $  0.8{\pm  0.3}$  & $  0.8{\pm  0.3}$  & $  1.1{\pm  0.3}$  & $  1.2{\pm  0.3}$  & $  0.9{\pm  0.3}$  & $  1.0{\pm  0.3}$ \\ 
${\rm C_{soft}   } $ & $  0.5{\pm  0.2}$  & $  0.8{\pm  0.3}$  & $  0.7{\pm  0.2}$  & $  0.5{\pm  0.2}$  & $  0.5{\pm  0.2}$  & $  0.7{\pm  0.2}$ \\
\enddata
\tablenotetext{a} {We split the full X-ray band (0.4--8.0 keV) into a soft band (0.4--1.2 keV) and a hard band (1.2--8.0 keV).  The background counts are extracted from a concentric circle of inner and outer radii  $8''$ and $15''$. The units of the image count rates are $10^{-3}\,{\rm s}^{-1}$, and the units of  the background fluxes are $10^{-6}\,{\rm s}^{-1}\rm{arcsec}^{-2}.$   }
\end{deluxetable}

\clearpage

\begin{deluxetable}{llllll}
\tabletypesize{\scriptsize}
\tablecolumns{5}
\tablewidth{0pt}
\tablecaption{Absorption-Corrected Count Rates for SDSS~1004+4112.  \label{tab:1004count}}
\tablehead{
\colhead{ObsId\tablenotemark{a}}&
\colhead{ 5794  }&
\colhead{ 11546 }&
\colhead{ 11547 }&
\colhead{ 11548 }&
\colhead{ 11549 }\\
\colhead{ Date}&
\colhead{ 2005-01-01}&
\colhead{ 2010-03-08}&
\colhead{ 2010-06-19 }&
\colhead{ 2010-09-23 }&
\colhead{ 2011-01-30 } \\
\colhead{Exposure (s)}&
\colhead{ 79987 }&
\colhead{ 5957}&
\colhead{ 5960 }&
\colhead{ 5963 }&
\colhead{5959} }
\startdata
${\rm A_{full}   } $ & $ 17.4{\pm  0.5}$  & $ 12.2{\pm  1.5}$  & $  9.1{\pm  1.3}$  & $  8.5{\pm  1.3}$  & $  5.4{\pm  1.0}$ \\ 
${\rm A_{soft}   } $ & $  8.1{\pm  0.3}$  & $  4.8{\pm  1.0}$  & $  4.3{\pm  0.9}$  & $  3.2{\pm  0.8}$  & $  2.3{\pm  0.7}$ \\ 
${\rm A_{hard}   } $ & $  9.2{\pm  0.4}$  & $  7.2{\pm  1.2}$  & $  4.7{\pm  0.9}$  & $  5.2{\pm  1.0}$  & $  3.0{\pm  0.8}$ \\ 
\tableline
${\rm B_{full}   } $ & $ 22.7{\pm  0.6}$  & $ 16.6{\pm  1.8}$  & $  9.4{\pm  1.3}$  & $ 12.7{\pm  1.5}$  & $  8.4{\pm  1.3}$ \\ 
${\rm B_{soft}   } $ & $ 11.1{\pm  0.4}$  & $  8.5{\pm  1.3}$  & $  3.5{\pm  0.8}$  & $  4.8{\pm  1.0}$  & $  2.7{\pm  0.7}$ \\ 
${\rm B_{hard}   } $ & $ 11.4{\pm  0.4}$  & $  7.9{\pm  1.2}$  & $  5.7{\pm  1.0}$  & $  7.7{\pm  1.2}$  & $  5.5{\pm  1.0}$ \\ 
\tableline
${\rm C_{full}   } $ & $ 18.9{\pm  0.5}$  & $ 12.8{\pm  1.6}$  & $ 18.4{\pm  1.9}$  & $ 12.1{\pm  1.5}$  & $ 21.5{\pm  2.0}$ \\ 
${\rm C_{soft}   } $ & $  9.0{\pm  0.4}$  & $  5.1{\pm  1.0}$  & $  8.6{\pm  1.3}$  & $  5.1{\pm  1.0}$  & $  9.3{\pm  1.4}$ \\
\enddata
\tablenotetext{a} { All events have standard ASCA grades of 0,2,3,4,6 and have energies between 0.4--8.0~keV. We split the full X-ray band (0.4--8.0 keV) into a soft X-ray band (0.4--1.1 keV) and a hard X-ray band (1.1--8.0 keV).  The units of the image count rates are $\rm 10^{-3}\>{\rm s}^{-1}$.   }
\end{deluxetable}

\begin{deluxetable}{lllll}
\tabletypesize{\scriptsize}
\tablecolumns{5}
\tablewidth{0pt}
\tablecaption{Absorption-Corrected Count Rates for HE~0435$-$1223.  \label{tab:0435count}}
\tablehead{
\colhead{ObsId\tablenotemark{a}}&
\colhead{ 7761  }&
\colhead{ 11550 }&
\colhead{ 11551 }&
\colhead{ 11552 }\\
\colhead{ Date}&
\colhead{ 2006-12-17 }&
\colhead{ 2009-12-07}&
\colhead{ 2010-07-04  }&
\colhead{ 2010-10-29 } \\
\colhead{Exposure (s)}&
\colhead{ 9994 }&
\colhead{ 12876}&
\colhead{ 12757  }&
\colhead{ 12758 }}
\startdata
 ${\rm A_{full}   } $ & $ 41.5{\pm  2.7}$  & $ 12.9{\pm  1.3}$  & $ 12.1{\pm  1.4}$  & $  7.6{\pm  1.0}$ \\ 
${\rm A_{soft}   } $ & $ 25.8{\pm  2.3}$  & $  7.3{\pm  1.0}$  & $  7.3{\pm  1.1}$  & $  4.5{\pm  0.8}$ \\ 
${\rm A_{hard}   } $ & $ 13.3{\pm  1.4}$  & $  4.7{\pm  0.7}$  & $  3.6{\pm  0.6}$  & $  2.9{\pm  0.5}$ \\ 
\tableline
${\rm B_{full}   } $ & $ 10.8{\pm  1.2}$  & $ 12.6{\pm  1.2}$  & $  8.3{\pm  2.0}$  & $  6.3{\pm  0.8}$ \\ 
${\rm B_{soft}   } $ & $  7.5{\pm  1.1}$  & $  7.7{\pm  1.0}$  & $  4.3{\pm  0.7}$  & $  1.9{\pm  0.5}$ \\ 
${\rm B_{hard}   } $ & $  3.4{\pm  0.6}$  & $  4.6{\pm  0.7}$  & $  4.0{\pm  0.7}$  & $  3.6{\pm  0.6}$ \\ 
\tableline
${\rm C_{full}   } $ & $ 11.5{\pm  1.3}$  & $ 12.3{\pm  1.2}$  & $ 10.5{\pm  1.2}$  & $  6.7{\pm  0.9}$ \\ 
${\rm C_{soft}   } $ & $  7.5{\pm  1.1}$  & $  6.4{\pm  0.9}$  & $  4.0{\pm  0.7}$  & $  2.2{\pm  0.5}$ \\
\enddata
\tablenotetext{a} { We split the full X-ray band (0.4--8.0 keV) into a soft X-ray band (0.4--1.3 keV), and a hard X-ray band (1.3--8.0 keV). The background counts are extracted from an annulus with inner and outer radii of $15''$ and $25'',$ respectively.  The units of the image count rates are $10^{-3}\,{\rm s}^{-1}$, and the units of  the background fluxes are $10^{-6}\,{\rm s}^{-1}\rm{arcsec}^{-2}.$ }
\end{deluxetable}

\clearpage

\begin{deluxetable}{cccclllllllllllllll}
\rotate
\tabletypesize{\scriptsize}
\setlength{\tabcolsep}{0.02in} 
\tablecolumns{19}
\tablewidth{0pt}
\tablecaption{Absorption-Corrected Count Rates for Q~2237+0305.   \label{tab:2237count}}
\tablehead{
\colhead{No.}&
\colhead{ObsId\tablenotemark{a}}&
\colhead{HJD} & 
\colhead{Exposure} & 
\colhead{ $\rm{A_{full} }$ }& 
\colhead{$\rm{A_{soft}}$}&
\colhead{$\rm{A_{hard}}$}&
\colhead{ $\rm{B_{full} }$}& 
\colhead{$\rm{B_{soft}}$}&
\colhead{$\rm{B_{hard}}$}&
\colhead{ $\rm{C_{full} }$}& 
\colhead{$\rm{C_{soft}}$}&
\colhead{$\rm{C_{hard}}$}&
\colhead{ $\rm{D_{full} }$ }& 
\colhead{$\rm{D_{soft}}$}&
\colhead{$\rm{D_{hard}}$}&
\colhead{ $\rm{b_{full} }$}& 
\colhead{$\rm{b_{soft}}$}&
\colhead{$\rm{b_{hard}}$}
}
\startdata
 I  &    431 &    51794 &     30287 &  $  70.2{\pm   2.3}$  &   $  49.2{\pm   2.0}$  &   $  21.5{\pm   1.0}$  &   $  16.0{\pm   1.1}$  &    $  10.9{\pm   1.0}$  &   $   4.7{\pm   0.5}$  &    $  34.1{\pm   1.6}$  &   $  26.3{\pm   1.5}$  & $   8.8{\pm   0.6}$  &   $  17.1{\pm   1.3}$  &    $  13.5{\pm   1.4}$  &  $   4.2{\pm   0.5}$   &  $  2.8$  &  $  0.7$ &  $  2.1$  \\ 
 II &   1632 &    52252 &      9535 &  $  59.7{\pm   4.5}$  &   $  38.5{\pm   4.4}$  &   $  19.0{\pm   1.8}$  &   $  11.1{\pm   1.4}$  &    $   8.9{\pm   1.5}$  &   $   3.0{\pm   0.6}$  &    $  19.1{\pm   1.9}$  &   $  11.3{\pm   1.9}$  & $   7.0{\pm   1.0}$  &   $  13.7{\pm   1.8}$  &    $  11.8{\pm   1.9}$  &  $   3.4{\pm   0.7}$   &  $  2.0$  &  $  0.7$ &  $  1.3$  \\ 
III &   6831 &    53745 &      7264 &  $  23.9{\pm   2.5}$  &   $  14.6{\pm   2.1}$  &   $   8.6{\pm   1.2}$  &   $  22.9{\pm   2.5}$  &    $  11.5{\pm   2.1}$  &   $   8.9{\pm   1.3}$  &    $   6.7{\pm   1.3}$  &   $   3.2{\pm   1.0}$  & $   2.8{\pm   0.7}$  &   $  14.0{\pm   2.2}$  &    $  11.3{\pm   2.4}$  &  $   3.5{\pm   0.9}$   &  $  2.3$  &  $  0.9$ &  $  1.4$  \\ 
IV &   6832 &    53856 &      7936 &  $  46.2{\pm   3.4}$  &   $  27.6{\pm   3.0}$  &   $  18.2{\pm   1.7}$  &   $  27.6{\pm   2.7}$  &    $  15.7{\pm   2.3}$  &   $   9.8{\pm   1.3}$  &    $  23.1{\pm   2.5}$  &   $  14.0{\pm   2.0}$  & $   6.9{\pm   1.1}$  &   $  21.0{\pm   2.6}$  &    $  13.4{\pm   2.4}$  &  $   6.2{\pm   1.1}$   &  $  3.2$  &  $  1.1$ &  $  2.1$  \\ 
V &   6833 &    53882 &      7953 &  $  30.3{\pm   2.7}$  &   $  16.4{\pm   2.3}$  &   $  11.4{\pm   1.6}$  &   $  12.1{\pm   1.7}$  &    $   7.6{\pm   1.6}$  &   $   4.1{\pm   0.8}$  &    $  11.2{\pm   1.6}$  &   $   6.2{\pm   1.3}$  & $   4.3{\pm   0.8}$  &   $   6.7{\pm   1.4}$  &    $   4.5{\pm   1.4}$  &  $   2.1{\pm   0.6}$   &  $  2.3$  &  $  0.4$ &  $  1.9$  \\
\enddata
\tablenotetext{a} { We split the full X-ray band (0.4--8.0 keV) into  a soft X-ray band (0.4--1.3 keV), and a hard X-ray band (1.3--8.0 keV). The background counts are extracted from an annulus with inner and outer radii of $10''$ and $20'',$ respectively.  The units of the image count rates are $10^{-3}\,{\rm s}^{-1}$, and the units of  the background fluxes are $10^{-6}\,{\rm s}^{-1}\rm{arcsec}^{-2}.$  }
\end{deluxetable}

\clearpage

\begin{deluxetable}{ccccccccc}
\tabletypesize{\scriptsize}
\tablecolumns{8}
\tablewidth{0pt}
\tablecaption{Fits to the \chandra $\>$ spectra of  QJ~0158$-$4325.  \label{tab:0158fit}}
\tablehead{
\colhead{Fit\tablenotemark{a}} & 
\colhead{Epoch} & 
\colhead{Image} & 
\colhead{$\Gamma$} & 
\colhead{$\rm{ N_{H} }$} & 
\colhead{Flux\tablenotemark{b}} & 
\colhead{Flux\tablenotemark{c}} & 
\colhead{$\chi^{2}_{\nu}(\nu)$} & 
\colhead{$P(\chi^{2}/{\nu})$\tablenotemark{d}}
\\
\colhead{} &
\colhead{} &
\colhead{} &
\colhead{} &
\colhead{$\times10^{22}$} &
\colhead{$\times10^{-13}$} &
\colhead{$\times10^{-13}$}&
\colhead{} &
\colhead{}\\
\colhead{} &
\colhead{} &
\colhead{} &
\colhead{} &
\colhead{$\rm{cm^{-2}}$} &
\colhead{$\rm{erg~s^{-1}~cm^{-2}}$} &
\colhead{$\rm{erg~s^{-1}~cm^{-2}}$}&
\colhead{} &
\colhead{}
}

\startdata
 $ 1$ & I  & A & $ 1.86^{+ 0.17}_{-0.17} $ & $ 0.10^{+ 0.19}_{-0.10}$ & $ 0.37^{+ 0.04}_{- 0.09}$ & $ 1.02^{+ 0.20}_{- 0.22}$ & $ 1.24( 16)$ & $ 0.23$ \\
 $ 1$ & I & B & $\cdot$                   & $ 0.00^{+ 0.05}_{-0.00}$ & $ 0.16^{+ 0.01}_{- 0.06}$ & $ 0.33^{+ 0.10}_{- 0.10}$ & $\cdot$      & $\cdot$ \\
 \tableline
 $ 2$ & II & A & $ 1.95^{+ 0.23}_{-0.21} $ & $ 0.00^{+ 0.04}_{-0.00}$ & $ 0.48^{+ 0.08}_{- 0.07}$ & $ 0.87^{+ 0.20}_{- 0.17}$ & $ 1.27( 20)$ & $ 0.19$ \\
 $ 2$ & II & B & $\cdot$                   & $ 0.04^{+ 0.08}_{-0.04}$ & $ 0.27^{+ 0.03}_{- 0.05}$ & $ 0.54^{+ 0.13}_{- 0.12}$ & $\cdot$      & $\cdot$ \\
 \tableline
 $ 3$ & III & A & $ 2.00^{+ 0.15}_{-0.15} $ & $ 0.02^{+ 0.13}_{-0.02}$ & $ 0.58^{+ 0.02}_{- 0.15}$ & $ 1.05^{+ 0.19}_{- 0.21}$ & $ 0.82( 19)$ & $ 0.68$ \\
 $ 3$ & III & B & $\cdot$                   & $ 0.03^{+ 0.09}_{-0.03}$ & $ 0.20^{+ 0.03}_{- 0.05}$ & $ 0.36^{+ 0.10}_{- 0.09}$ & $\cdot$      & $\cdot$ \\
 \tableline
 $ 4$ & IV & A & $ 1.79^{+ 0.17}_{-0.17} $ & $ 0.00^{+ 0.06}_{-0.00}$ & $ 0.51^{+ 0.07}_{- 0.07}$ & $ 1.17^{+ 0.23}_{- 0.19}$ & $ 1.30( 19)$ & $ 0.17$ \\
 $ 4$ & IV & B & $\cdot$                   & $ 0.00^{+ 0.09}_{-0.00}$ & $ 0.17^{+ 0.03}_{- 0.03}$ & $ 0.38^{+ 0.11}_{- 0.09}$ & $\cdot$      & $\cdot$ \\
\enddata

\tablenotetext{a}{The source and lens redshifts are $z_s=1.29$ and $z_l=0.317,$ respectively. All models include Galactic absorption with a column density of $\rm{N_{H} = 0.0195\times10^{22}~cm^{-2}}$.  All derived errors are at 68\% confidence. The spectral fits were performed within the energy range 0.4--8.0 keV. In the first seven fits, we fit image A and B with the same power-law indices but different absorption. In fit No. 8, we extract and stack spectra from a circular region of radius $\sim$$3''$ containing both images A and B, and fit it with a simple power-law model modified by absorption. }
\tablenotetext{b}{Flux is estimated in the 0.4--1.3 keV band (observed 1.3 keV corresponds to 3.0 keV in the rest frame).}
\tablenotetext{c}{Flux is estimated in the 1.3--8.0 keV band.}
\tablenotetext{d} {$P(\chi^{2}/{\nu})$ is the probability of exceeding $\chi^{2}$ with ${\nu}$ degrees of freedom.}

\end{deluxetable}

\clearpage

\begin{deluxetable}{ccccccccc}
\tabletypesize{\scriptsize}
\tablecolumns{8}
\tablewidth{0pt}
\tablecaption{Fits to the \chandra $\>$ spectra of  HE~1104$-$1805.   \label{tab:1104fit}}
\tablehead{
\colhead{Fit\tablenotemark{a}} & 
\colhead{Epoch} & 
\colhead{Image} & 
\colhead{$\Gamma$} & 
\colhead{$\rm{ N_{H} }$} & 
\colhead{Flux\tablenotemark{b}} & 
\colhead{Flux\tablenotemark{c}} & 
\colhead{$\chi^{2}_{\nu}(\nu)$} & 
\colhead{$P(\chi^{2}/{\nu})$\tablenotemark{d}}
\\
\colhead{} &
\colhead{} &
\colhead{} &
\colhead{} &
\colhead{$\times10^{22}$} &
\colhead{$\times10^{-13}$} &
\colhead{$\times10^{-13}$}&
\colhead{} &
\colhead{}\\
\colhead{} &
\colhead{} &
\colhead{} &
\colhead{} &
\colhead{$\rm{cm^{-2}}$} &
\colhead{$\rm{erg~s^{-1}~cm^{-2}}$} &
\colhead{$\rm{erg~s^{-1}~cm^{-2}}$}&
\colhead{} &
\colhead{}
}

\startdata
 $ 1$ & I   & A & $ 1.82^{+ 0.05}_{-0.05} $ & $ 0.00^{+ 0.01}_{-0.00}$ & $ 0.34^{+ 0.00}_{-0.03}$ & $ 0.77^{+ 0.05}_{-0.05}$ & $ 0.97(172)$ & $ 0.59$ \\ 
 $ 1$ & I   & B & $     \cdot  $            & $ 0.00^{+ 0.01}_{-0.00}$ & $ 0.18^{+ 0.00}_{-0.01}$ & $ 0.41^{+ 0.03}_{-0.03}$ & $\cdot$     & $\cdot$ \\ 
 \tableline
 $ 2$ & II  & A & $ 1.69^{+ 0.33}_{-0.26} $ & $ 0.28^{+ 1.61}_{-0.28}$ & $ 0.16^{+ 0.02}_{-0.05}$ & $ 0.45^{+ 0.15}_{-0.18}$ & $ 0.60( 11)$ & $ 0.83$ \\ 
 $ 2$ & II  & B & $     \cdot  $            & $ 0.08^{+ 0.22}_{-0.08}$ & $ 0.37^{+ 0.03}_{-0.06}$ & $ 1.03^{+ 0.22}_{-0.22}$ & $\cdot$     & $\cdot$ \\ 
 \tableline
 $ 3$ & III  & A & $ 2.60^{+ 0.69}_{-0.53} $ & $ 1.23^{+ 1.67}_{-1.23}$ & $ 0.51^{+ 0.06}_{-0.05}$ & $ 0.37^{+ 0.13}_{-0.16}$ & $ 1.55(  7)$ & $ 0.15$ \\ 
 $ 3$ & III  & B & $     \cdot  $            & $ 0.13^{+ 0.25}_{-0.13}$ & $ 0.45^{+ 0.04}_{-0.08}$ & $ 0.32^{+ 0.15}_{-0.11}$ & $\cdot$     & $\cdot$ \\ 
 \tableline
 $ 4$ & IV  & A & $ 2.11^{+ 0.41}_{-0.27} $ & $ 1.00^{+ 1.28}_{-1.00}$ & $ 0.29^{+ 0.03}_{-0.03}$ & $ 0.42^{+ 0.16}_{-0.18}$ & $ 0.65(  6)$ & $ 0.69$ \\ 
 $ 4$ & IV  & B & $     \cdot  $            & $ 0.00^{+ 0.14}_{-0.00}$ & $ 0.30^{+ 0.00}_{-0.09}$ & $ 0.44^{+ 0.17}_{-0.13}$ & $\cdot$     & $\cdot$ \\
\enddata
\tablenotetext{a}{The source and lens redshifts are $z_s=2.32$ and $z_l=0.73,$ respectively. All models include Galactic absorption with a column density of $\rm{N_{H} = 0.0462\times10^{22}~cm^{-2}}$.  All derived errors are at 68\% confidence. The spectral fits were performed within the energy range 0.4--8.0 keV. In the first nine fits, we fit image A and B with same power-law indices but different absorption. In fits no. 10--14 (correspond to epoch no. 2 to 6, in which the exposure time is short), we extract  the spectrum from a circular region of radius $\sim$$4''$ containing both images A and B, and fit it with a simple power-law model modified by absorption.   In fit no. 17, we extract and stack the spectra from a circular region containing both images A and B, and fit it with same model. }
\tablenotetext{b}{Flux is estimated in the 0.4--1.2 keV band (observed 1.2 keV corresponds to 4.0 keV in the rest frame).}
\tablenotetext{c}{Flux is estimated in the 1.2--8.0 keV band.}
\tablenotetext{d} {$P(\chi^{2}/{\nu})$ is the probability of exceeding $\chi^{2}$ with ${\nu}$ degrees of freedom.}

\end{deluxetable}

\clearpage

\begin{deluxetable}{ccccccccc}
\tabletypesize{\scriptsize}
\tablecolumns{8}
\tablewidth{0pt}
\tablecaption{Fits to the \chandra $\>$ spectra of  SDSS~0924+0219.  \label{tab:0924fit}}
\tablehead{
\colhead{Fit\tablenotemark{a}} & 
\colhead{Epoch} & 
\colhead{Image} & 
\colhead{$\Gamma$} & 
\colhead{$\rm{ N_{H} }$} & 
\colhead{Flux\tablenotemark{b}} & 
\colhead{Flux\tablenotemark{c}} & 
\colhead{$\chi^{2}_{\nu}(\nu)$} & 
\colhead{$P(\chi^{2}/{\nu})$\tablenotemark{d}}
\\
\colhead{} &
\colhead{} &
\colhead{} &
\colhead{} &
\colhead{$\times10^{22}$} &
\colhead{$\times10^{-13}$} &
\colhead{$\times10^{-13}$}&
\colhead{} &
\colhead{}\\
\colhead{} &
\colhead{} &
\colhead{} &
\colhead{} &
\colhead{$\rm{cm^{-2}}$} &
\colhead{$\rm{erg~s^{-1}~cm^{-2}}$} &
\colhead{$\rm{erg~s^{-1}~cm^{-2}}$}&
\colhead{} &
\colhead{}
}

\startdata
 $ 1$ & I & A+C+D & $ 2.57^{+ 0.52}_{-0.44} $ & $ 0.13^{+ 0.20}_{-0.13}$ & $ 0.12^{+ 0.01}_{- 0.03}$ & $ 0.16^{+ 0.05}_{- 0.05}$ & $ 1.23( 11)$ & $ 0.26$ \\ 
 $ 1$ & I & B     & $\cdot$                   & $ 0.00^{+ 0.09}_{-0.00}$ & $ 0.02^{+ 0.00}_{- 0.01}$ & $ 0.02^{+ 0.01}_{- 0.01}$ & $\cdot$      & $\cdot$ \\  
 \tableline
 $ 2$ & II & A+C+D & $ 2.39^{+ 0.25}_{-0.13} $ & $ 0.00^{+ 0.09}_{-0.00}$ & $ 0.26^{+ 0.03}_{- 0.03}$ & $ 0.31^{+ 0.04}_{- 0.04}$ & $ 0.67( 30)$ & $ 0.92$ \\ 
 $ 2$ & II & B     & $\cdot$                   & $ 0.00^{+ 0.08}_{-0.00}$ & $ 0.05^{+ 0.01}_{- 0.01}$ & $ 0.06^{+ 0.01}_{- 0.01}$ & $\cdot$      & $\cdot$ \\  
 \tableline
 $ 3$ & III & A+C+D & $ 2.23^{+ 0.42}_{-0.37} $ & $ 0.13^{+ 0.20}_{-0.13}$ & $ 0.13^{+ 0.01}_{- 0.02}$ & $ 0.28^{+ 0.06}_{- 0.06}$ & $ 1.09( 17)$ & $ 0.35$ \\ 
 $ 3$ & III & B     & $\cdot$                   & $ 0.39^{+ 0.85}_{-0.39}$ & $ 0.01^{+ 0.00}_{- 0.00}$ & $ 0.04^{+ 0.02}_{- 0.02}$ & $\cdot$      & $\cdot$ \\  
 \tableline
 $ 4$ & IV & A+C+D & $ 2.02^{+ 0.13}_{-0.11} $ & $ 0.00^{+ 0.04}_{-0.00}$ & $ 0.30^{+ 0.02}_{- 0.02}$ & $ 0.62^{+ 0.07}_{- 0.07}$ & $ 0.79( 44)$ & $ 0.84$ \\ 
 $ 4$ & IV & B     & $\cdot$                   & $ 0.00^{+ 0.05}_{-0.00}$ & $ 0.05^{+ 0.01}_{- 0.01}$ & $ 0.10^{+ 0.02}_{- 0.02}$ & $\cdot$      & $\cdot$ \\
\enddata
\tablenotetext{a}{The source and lens redshifts are $z_s=1.524$ and $z_l=0.39,$ respectively. All models include Galactic absorption with a column density of $\rm{N_{H} = 0.0375\times10^{22}~cm^{-2}}$.  All derived errors are at  68\% confidence level. The spectral fits were performed within the energy range 0.4--8.0 keV. In the first seven fits, we fit images A+C+D and B with the same power-law indices but different absorption. In fits 8 and 9, we fit the stacked spectra of A+C+D, and of B separately. In fit No. 10, we extract and stack spectra from a circular region of radius $\sim$$3''$  containing image A, B, C and D, and fit it with a simple power-law model modified by absorption. }
\tablenotetext{b}{Flux is estimated in the 0.4--1.2 keV band (observed 1.2 keV corresponds to 3.0 keV in the rest frame).}
\tablenotetext{c}{Flux is estimated in the 1.2--8.0 keV band.}
\tablenotetext{d} {$P(\chi^{2}/{\nu})$ is the probability of exceeding $\chi^{2}$ with ${\nu}$ degrees of freedom.}

\end{deluxetable}

\clearpage

\begin{deluxetable}{ccccccccc}
\tabletypesize{\scriptsize}
\tablecolumns{9}
\tablewidth{0pt}
\tablecaption{Fits to the the \chandra $\>$ spectra of SDSS~1004+4112.    \label{tab:1004fit}}
\tablehead{
\colhead{Fit\tablenotemark{a}} & 
\colhead{Epoch} & 
\colhead{Image} & 
\colhead{$\Gamma$} & 
\colhead{$\rm{ N_{H} }$} & 
\colhead{Flux\tablenotemark{b}} & 
\colhead{Flux\tablenotemark{c}} & 
\colhead{$\chi^{2}_{\nu}(\nu)$} & 
\colhead{$P(\chi^{2}/{\nu})$\tablenotemark{d}}
\\
\colhead{} &
\colhead{} &
\colhead{} &
\colhead{} &
\colhead{$\times10^{22}$} &
\colhead{$\times10^{-13}$} &
\colhead{$\times10^{-13}$}&
\colhead{} &
\colhead{}\\
\colhead{} &
\colhead{} &
\colhead{} &
\colhead{} &
\colhead{$\rm{cm^{-2}}$} &
\colhead{$\rm{erg~s^{-1}~cm^{-2}}$} &
\colhead{$\rm{erg~s^{-1}~cm^{-2}}$}&
\colhead{} &
\colhead{}
}
\startdata
 $ 1$ & I  & A & $ 1.98^{+ 0.04}_{-0.04} $ & $ 0.03^{+ 0.04}_{-0.03}$ & $ 0.27^{+ 0.01}_{-0.01}$ & $ 0.61^{+ 0.03}_{-0.03}$ & $ 1.13(559)$ & $ 0.02$ \\ 
 $ 1$ & I & B & $\cdot$                   & $ 0.05^{+ 0.01}_{-0.01}$ & $ 0.34^{+ 0.01}_{-0.01}$ & $ 0.82^{+ 0.03}_{-0.03}$ & $\cdot$      & $\cdot$ \\ 
 $ 1$ & I & C & $\cdot$                   & $ 0.10^{+ 0.02}_{-0.02}$ & $ 0.27^{+ 0.01}_{-0.01}$ & $ 0.70^{+ 0.03}_{-0.03}$ & $\cdot$      & $\cdot$ \\ 
 $ 1$ & I & D & $\cdot$                   & $ 0.14^{+ 0.02}_{-0.03}$ & $ 0.14^{+ 0.01}_{-0.01}$ & $ 0.40^{+ 0.02}_{-0.02}$ & $\cdot$      & $\cdot$ \\ 
 \tableline
 $ 2$ & II & A & $ 2.08^{+ 0.17}_{-0.17} $ & $ 0.33^{+ 0.60}_{-0.33}$ & $ 0.15^{+ 0.06}_{-0.06}$ & $ 0.52^{+ 0.10}_{-0.12}$ & $ 0.95( 34)$ & $ 0.54$ \\ 
 $ 2$ & II & B & $\cdot$                   & $ 0.08^{+ 0.09}_{-0.06}$ & $ 0.27^{+ 0.03}_{-0.04}$ & $ 0.59^{+ 0.12}_{-0.11}$ & $\cdot$      & $\cdot$ \\ 
 $ 2$ & II & C & $\cdot$                   & $ 0.00^{+ 0.03}_{-0.00}$ & $ 0.24^{+ 0.01}_{-0.06}$ & $ 0.44^{+ 0.09}_{-0.09}$ & $\cdot$      & $\cdot$ \\ 
 $ 2$ & II & D & $\cdot$                   & $ 0.07^{+ 0.07}_{-0.06}$ & $ 0.27^{+ 0.04}_{-0.04}$ & $ 0.59^{+ 0.11}_{-0.10}$ & $\cdot$      & $\cdot$ \\
\enddata

\tablenotetext{a}{The source and lens redshifts are $z_s=1.734$ and $z_l=0.68,$ respectively.  All models include Galactic absorption with a column density of $\rm{N_{H} = 0.0111\times10^{22}~cm^{-2}}$.  All derived errors are at 68\% confidence level. The spectral fits were performed within the energy range 0.4--8.0 keV. In the first 6 fits, we fit image A,  B, C and D simultaneously with the same power-law indices but different absorption.  In fits 7-10, we fit the stacked images of A, B, C and D separately. In fit 11, we extract and stack spectra from a circular region of radius $\sim$$10''$  containing image A, B, C and D, and fit it with a simple power-law model modified by absorption. }
\tablenotetext{b}{Flux is estimated in the 0.4--1.1 keV band. (observed 1.1 keV corresponds to 3.0 keV in the rest frame).}
\tablenotetext{c}{Flux is estimated in the 1.1--8.0 keV band.}
\tablenotetext{d} {$P(\chi^{2}/{\nu})$ is the probability of exceeding $\chi^{2}$ with ${\nu}$ degrees of freedom.}

\end{deluxetable}

\clearpage

\begin{deluxetable}{ccccccccc}
\tabletypesize{\scriptsize}
\tablecolumns{9}
\tablewidth{0pt}
\tablecaption{Fits to the \chandra $\>$ spectra of HE~0435$-$1223.   \label{tab:0435fit}}
\tablehead{
\colhead{Fit\tablenotemark{a}} & 
\colhead{Epoch} & 
\colhead{Image} & 
\colhead{$\Gamma$} & 
\colhead{$\rm{ N_{H} }$} & 
\colhead{Flux\tablenotemark{b}} & 
\colhead{Flux\tablenotemark{c}} & 
\colhead{$\chi^{2}_{\nu}(\nu)$} & 
\colhead{$P(\chi^{2}/{\nu})$\tablenotemark{d}}
\\
\colhead{} &
\colhead{} &
\colhead{} &
\colhead{} &
\colhead{$\times10^{22}$} &
\colhead{$\times10^{-13}$} &
\colhead{$\times10^{-13}$}&
\colhead{} &
\colhead{}\\
\colhead{} &
\colhead{} &
\colhead{} &
\colhead{} &
\colhead{$\rm{cm^{-2}}$} &
\colhead{$\rm{erg~s^{-1}~cm^{-2}}$} &
\colhead{$\rm{erg~s^{-1}~cm^{-2}}$}&
\colhead{} &
\colhead{}
}

\startdata
 $ 1$ & I & A & $ 2.00^{+ 0.11}_{-0.10} $ & $ 0.08^{+ 0.08}_{-0.08}$ & $ 0.41^{+ 0.04}_{-0.03}$ & $ 0.95^{+ 0.11}_{-0.10}$ & $ 1.02( 58)$ & $ 0.43$ \\
 $ 1$ & I & B & $\cdot$                   & $ 0.00^{+ 0.03}_{-0.00}$ & $ 0.18^{+ 0.02}_{-0.02}$ & $ 0.35^{+ 0.05}_{-0.05}$ & $\cdot$      & $\cdot$ \\
 $ 1$ & I & C & $\cdot$                   & $ 0.00^{+ 0.03}_{-0.00}$ & $ 0.17^{+ 0.02}_{-0.02}$ & $ 0.34^{+ 0.05}_{-0.05}$ & $\cdot$      & $\cdot$ \\
 $ 1$ & I & D & $\cdot$                   & $ 0.12^{+ 0.16}_{-0.10}$ & $ 0.14^{+ 0.02}_{-0.02}$ & $ 0.37^{+ 0.06}_{-0.05}$ & $\cdot$      & $\cdot$ \\
 \tableline
 $ 2$ & II & A & $ 1.93^{+ 0.11}_{-0.10} $ & $ 0.00^{+ 0.12}_{-0.00}$ & $ 0.17^{+ 0.02}_{-0.02}$ & $ 0.36^{+ 0.05}_{-0.05}$ & $ 0.96( 50)$ & $ 0.56$ \\
 $ 2$ & II & B & $\cdot$                   & $ 0.00^{+ 0.05}_{-0.00}$ & $ 0.18^{+ 0.02}_{-0.02}$ & $ 0.39^{+ 0.06}_{-0.05}$ & $\cdot$      & $\cdot$ \\
 $ 2$ & II & C & $\cdot$                   & $ 0.00^{+ 0.05}_{-0.00}$ & $ 0.16^{+ 0.02}_{-0.02}$ & $ 0.36^{+ 0.06}_{-0.05}$ & $\cdot$      & $\cdot$ \\
 $ 2$ & II & D & $\cdot$                   & $ 0.00^{+ 0.04}_{-0.00}$ & $ 0.13^{+ 0.02}_{-0.02}$ & $ 0.28^{+ 0.04}_{-0.04}$ & $\cdot$      & $\cdot$ \\
\enddata
\tablenotetext{a}{The source and lens redshifts are $z_s=1.689$ and $z_l=0.46,$ respectively. All models include Galactic absorption with a column density of $\rm{N_{H} = 0.0511\times10^{22}~cm^{-2}}$.  All derived errors are at 68\% confidence level. The spectral fits were performed within the energy range 0.4--8.0 keV. In the first 5 fits, we fit image A,  B, C and D jointly with the same power-law indices but different absorption.  In fits 6-9, we fit the stacked images of A, B, C and D separately. In fit 10, we extract and stack spectra from a circular region of radius $\sim$$4''$ containing image A, B, C and D, and fit it with a simple power-law model modified by absorption. }
\tablenotetext{b}{Flux is estimated in the 0.4--1.3 keV band. (observed 1.3 keV corresponds to 3.5 keV in the rest frame).}
\tablenotetext{c}{Flux is estimated in the 1.3--8.0 keV band.}
\tablenotetext{d} {$P(\chi^{2}/{\nu})$ is the probability of exceeding $\chi^{2}$ with ${\nu}$ degrees of freedom.}

\end{deluxetable}

\begin{deluxetable}{ccccccccc}
\tabletypesize{\scriptsize}
\tablecolumns{8}
\tablewidth{0pt}
\tablecaption{Fits to the \chandra $\>$ spectra of  Q~2237+0305.   \label{tab:2237fit}}
\tablehead{
\colhead{Fit\tablenotemark{a}} & 
\colhead{Epoch} & 
\colhead{Image} & 
\colhead{$\Gamma$} & 
\colhead{$\rm{ N_{H} }$} & 
\colhead{Flux\tablenotemark{b}} & 
\colhead{Flux\tablenotemark{c}} & 
\colhead{$\chi^{2}_{\nu}(\nu)$} & 
\colhead{$P(\chi^{2}/{\nu})$\tablenotemark{d}}\\
\colhead{} &
\colhead{} &
\colhead{} &
\colhead{} &
\colhead{$\times10^{22}$} &
\colhead{$\times10^{-13}$} &
\colhead{$\times10^{-13}$}&
\colhead{} &
\colhead{}\\
\colhead{} &
\colhead{} &
\colhead{} &
\colhead{} &
\colhead{$\rm{cm^{-2}}$} &
\colhead{$\rm{erg~s^{-1}~cm^{-2}}$} &
\colhead{$\rm{erg~s^{-1}~cm^{-2}}$}&
\colhead{} &
\colhead{}
}

\startdata
 $ 1$ & I  & A & $ 1.88^{+ 0.07}_{-0.07} $ & $ 0.02^{+ 0.02}_{-0.02}$ & $ 0.52^{+ 0.02}_{-0.02}$ & $ 1.37^{+ 0.08}_{-0.08}$ & $ 1.03(103)$ & $ 0.39$ \\ 
 $ 1$ & I & B & $\cdot$                   & $ 0.00^{+ 0.02}_{-0.00}$ & $ 0.10^{+ 0.00}_{-0.02}$ & $ 0.24^{+ 0.03}_{-0.03}$ & $\cdot$      & $\cdot$ \\ 
 $ 1$ & I & C & $\cdot$                   & $ 0.05^{+ 0.02}_{-0.02}$ & $ 0.20^{+ 0.01}_{-0.01}$ & $ 0.57^{+ 0.05}_{-0.04}$ & $\cdot$      & $\cdot$ \\ 
 $ 1$ & I & D & $\cdot$                   & $ 0.02^{+ 0.02}_{-0.02}$ & $ 0.11^{+ 0.01}_{-0.01}$ & $ 0.28^{+ 0.03}_{-0.03}$ & $\cdot$      & $\cdot$ \\ 
 \tableline
 $ 2$ & II & A & $ 1.77^{+ 0.10}_{-0.10} $ & $ 0.02^{+ 0.05}_{-0.02}$ & $ 0.36^{+ 0.03}_{-0.03}$ & $ 1.08^{+ 0.13}_{-0.12}$ & $ 1.13( 50)$ & $ 0.25$ \\ 
 $ 2$ & II & B & $\cdot$                   & $ 0.00^{+ 0.04}_{-0.00}$ & $ 0.09^{+ 0.01}_{-0.01}$ & $ 0.25^{+ 0.05}_{-0.05}$ & $\cdot$      & $\cdot$ \\ 
 $ 2$ & II & C & $\cdot$                   & $ 0.08^{+ 0.08}_{-0.06}$ & $ 0.12^{+ 0.02}_{-0.02}$ & $ 0.45^{+ 0.07}_{-0.07}$ & $\cdot$      & $\cdot$ \\ 
 $ 2$ & II & D & $\cdot$                   & $ 0.02^{+ 0.04}_{-0.02}$ & $ 0.12^{+ 0.02}_{-0.02}$ & $ 0.35^{+ 0.06}_{-0.06}$ & $\cdot$      & $\cdot$ \\
\enddata
\tablenotetext{a}{The source and lens redshifts are $z_s=1.69$ and $z_l=0.0395,$ respectively. All models include Galactic absorption with a column density of $\rm{N_{H} = 0.0551\times10^{22}~cm^{-2}}$.  All derived errors are at 68\% confidence level.  The spectral fits were performed within the energy range 0.4--8.0 keV. In the first 17 fits (first 17 epochs), we fit images A, B, C and D with the same power-law indices but different absorption.  In fits 18, 19 and 20 (corresponding to epochs 18, 19, and 20), we extract spectra from a circular region of radius $\sim$$3''$ containing images A, B, C and D, and fit them with a simple power-law model modified by absorption. In fit 21, we stack the last three observations, and fit the spectra of images A, B, C, D  with the same power-law indices. In fits 22--25, we fit the stacked (20 epochs) spectra of images A, B, C, and D independently. In fit 26, we fit the stacked (20 epochs) spectra of images A, B, C, and D with the same power-law index.  In fit 27, we extract and stack (20 epochs) the spectra from a circular region of radius $\sim$$3''$ containing images A, B, C and D, and fit it with a simple power-law model modified by absorption. }
\tablenotetext{b}{Flux is estimated in the 0.4--1.3 keV band (observed 1.3 keV corresponds to 3.5 keV in the rest frame).}
\tablenotetext{c}{Flux is estimated in the 1.3--8.0 keV band.}
\tablenotetext{d} {$P(\chi^{2}/{\nu})$ is the probability of exceeding $\chi^{2}$ with ${\nu}$ degrees of freedom.}

\end{deluxetable}


\begin{deluxetable}{lcclllllc}
\tabletypesize{\scriptsize}
\tablecolumns{7}
\tablewidth{0pt}
\tablecaption{X-ray Emission Line Properties. \label{tab:EW}}
\tablehead{
\colhead{Lens}&
\colhead{Line}&
\colhead{ Image}&
\colhead{ $\rm E_{ line}$ \tablenotemark{a}}&
\colhead{ $\sigma_{\rm line}$ \tablenotemark{a} }&
\colhead{EW  \tablenotemark{a}} &
\colhead{Significance}\\
\colhead{}&
\colhead{}&
\colhead{}&
\colhead{(keV)}&
\colhead{ (keV) }&
\colhead{ (keV) }&
\colhead{}
}
\startdata
                                    & \feka\  &${ \rm A   } $         & $  6.45^{+0.18}_{-0.13}$  & $  <0.19$  & $  0.64^{+0.37}_{-0.32}$    &  99.5\%       \\ 
QJ~0158$-$4325      & \feka\  &${ \rm B   } $\tablenotemark{b}       & \nodata                                &    \nodata                            &       \nodata                              &   \nodata          \\ 
                                   &  \feka\ & ${ \rm A+B   } $ & $  6.54^{+0.08}_{-0.09}$     &  $ <0.18$  & $  0.41^{+0.21}_{-0.21}$       &   99.9\%            \\ 
\tableline
                                    & \feka\  &${ \rm A   } $      & $  6.84^{+0.09}_{-0.10}$     & $  <0.12$  & $  0.40^{+0.27}_{-0.30}$       &     99.3\%          \\ 
HE~1104$-$1805   & \feka\  &${ \rm B   } $\tablenotemark{b}      & \nodata 				 & \nodata				   & \nodata   				   &       \nodata       \\ 
                                   &  \feka\ & ${ \rm A+B   } $ & $  6.80^{+0.10}_{-0.13}$      & $ <0.29$  & $  0.23^{+0.17}_{-0.13}$   &      99.8\%  \\ 
\tableline
			       & \feka\  &${ \rm A+C+D   }      $\tablenotemark{c}      & $  5.92^{+0.20}_{-0.17}$  & $  <0.37$      & $  0.48^{+0.20}_{-0.30}$   &     99.5\%  \\
			       & \feka\  &${ \rm A+C+D   }      $\tablenotemark{c}      & $  7.12^{+0.07}_{-0.06}$  & $  <0.14$      & $  0.73^{+0.25}_{-0.45}$   &     99.996\%  \\
\enddata
\tablenotetext{a} { $\rm E_{line}$, $\sigma_{\rm line},$ and $\rm EW$ represent respectively the  line position, $\sigma$ parameter of the Gaussian profile, and the equivalent line width (rest frame) with respect to the X-ray continuum.  }
\tablenotetext{b} {The \feka\ line is not detected in this image.  }
\tablenotetext{c}{We detect a redshifted and a blueshifted wing of the iron line for this image.}
\end{deluxetable}



\begin{deluxetable}{lcllllll}
\tabletypesize{\scriptsize}
\tablecolumns{8}
\tablewidth{0pt}
\tablecaption{Microlensing Statistics for SDSS~0924+0219 \label{tab:0924chi2}}
\tablehead{
\colhead{band}&
\colhead{$\rm dof$}&
\colhead{ B/A  }&
\colhead{ C/A}&
\colhead{ D/A }&
\colhead{ B/D } &
\colhead{ C/D}&
\colhead{ C/B} \\
 \colhead{}&
\colhead{}&
\colhead{$\chi^2$(prob)}&
\colhead{$\chi^2$(prob)}&
\colhead{$\chi^2$(prob)}&
\colhead{$\chi^2$(prob)}&
\colhead{$\chi^2$(prob)}&
\colhead{$\chi^2$(prob)}
}
\startdata
 ${\rm  full}           $ & 5   &  20.8 (0.0009)    & 5.3 (0.38)  &  4.7 (0.45)  &  4.6 (0.47)  &  4.9 (0.43)  & 10.5 (0.06)   \\ 
 ${\rm  soft}          $ & 5   & 14.1 (0.015)        & 5.8 (0.32)  &  4.2 (0.47)  &  6.7 (0.25)  &  2.2  (0.18)  & 4.5   (0.48)     \\ 
 ${\rm  hard}        $ & 5   & 7.5 (0.19)             & 6.5 (0.26)  &  8.7 (0.12)   &  7.3 (0.20)  &  5.7 (0.34)   & 9.0   (0.11)    \\ 
\enddata
\end{deluxetable}

\begin{deluxetable}{lcllllll}
\tabletypesize{\scriptsize}
\tablecolumns{8}
\tablewidth{0pt}
\tablecaption{Microlensing Statistics for SDSS~1004+4112 \label{tab:1004chi2}}
\tablehead{
\colhead{band}&
\colhead{$\rm dof$}&
\colhead{ B/A  }&
\colhead{ C/A}&
\colhead{ D/A }&
\colhead{ B/D } &
\colhead{ C/D}&
\colhead{ C/B} \\
\colhead{}&
\colhead{}&
\colhead{$\chi^2$(prob)}&
\colhead{$\chi^2$(prob)}&
\colhead{$\chi^2$(prob)}&
\colhead{$\chi^2$(prob)}&
\colhead{$\chi^2$(prob)}&
\colhead{$\chi^2$(prob)}
}
\startdata
 ${\rm  full}           $ & 4   & 2.6 (0.38)     & 19.8 (0.0005)  &  33.4 ($1.0\times 10^{-6}$)  &  144.8 ($2.6\times 10^{-30}$)  &  44.4 ($5.2\times 10^{-9}$)  & 25.4 (0.00004)   \\ 
 ${\rm  soft}          $ & 4   & 5.3 (0.26)     & 8.6 (0.07)  &  15.7 (0.003)  &  72.2 ($7.7\times 10^{-15}$)  &  46.8  ($1.7\times 10^{-9}$)  &  14.6  (0.005)     \\ 
 ${\rm  hard}        $ & 4   & 1.9 (0.24)     & 11.1 (0.025)  &  16.8 (0.002)   &  73.1 ($5.0\times 10^{-15}$)  &  11.9 (0.018)   & 12.1  (0.017)    \\ 
\enddata
\end{deluxetable}

\begin{deluxetable}{lcllllll}
\tabletypesize{\scriptsize}
\tablecolumns{8}
\tablewidth{0pt}
\tablecaption{Microlensing Statistics for HE~0435$-$1223 \label{tab:0435chi2}}
\tablehead{
\colhead{band}&
\colhead{$\rm dof$}&
\colhead{ B/A }&
\colhead{ C/A}&
\colhead{ D/A }&
\colhead{ B/D } &
\colhead{ C/D}&
\colhead{ C/B}\\
 \colhead{}&
\colhead{}&
\colhead{$\chi^2$(prob)}&
\colhead{$\chi^2$(prob)}&
\colhead{$\chi^2$(prob)}&
\colhead{$\chi^2$(prob)}&
\colhead{$\chi^2$(prob)}&
\colhead{$\chi^2$(prob)}
}
\startdata
 ${\rm  full}           $ & 3   & 40.8 ($7.3\times 10^{-9}$)     & 49.5 ($1.0\times 10^{-10}$)  &  36.4 ($6.3\times 10^{-8}$)  &  1.4 (0.30)  &  0.5 (0.08)  & 0.7 (0.14)   \\ 
 ${\rm  soft}          $ & 3   & 17.3 (0.0006)     & 13.8 (0.003)  &  29.2 ($2.1\times 10^{-6}$)  &  6.0 (0.11)  &  3.9  (0.27)  &  0.96  (0.19)     \\ 
 ${\rm  hard}        $ & 3   & 29.8 ($1.5\times 10^{-6}$)     & 32.9 ($3.4\times 10^{-7}$)  &  13.5 (0.004)   &  6.9 (0.07)  &  7.8 (0.05)   & 0.31  (0.04)    \\ 
\enddata
\end{deluxetable}

\begin{deluxetable}{llllll}
\tabletypesize{\scriptsize}
\tablecolumns{8}
\tablewidth{0pt}
\tablecaption{\feka\ Lines  and X-ray Luminosities \label{tab:EL_EW_Lumi}}
\tablehead{
\colhead{Lens}&
\colhead{$\rm E_{line}$ (keV)}&
\colhead{$\rm EW$ (keV)  }&
\multicolumn{3}{c}{ L  ($\rm 10^{44}\, ergs\, s^{-1}$) \tablenotemark{a}}\\
\cline{4-6}\\
\colhead{} &
\colhead{} &
\colhead{} &
\colhead{0.2--2} &
\colhead{2--10} &
\colhead{10--50} 
}
\startdata
QJ~0158--4325        & $6.31\pm0.23$   & $0.55\pm0.39$&1.89  & 1.51 &  $1.69$  \\ 
SDSS~0924+0219  &  $6.40\pm0.40$  & $0.76\pm0.47$& 0.80 & 0.39 &  $0.25$  \\ 
SDSS~1004+4112  &  $6.37\pm0.17$  & $0.66\pm0.27$& 3.19 & 2.66 &  $2.76$  \\ 
HE~0435--1223       &  $6.53\pm0.10$  & $0.87\pm0.41$& 1.68  &1.51 &  $1.84$  \\ 
Q~2237+0305          &  $6.38\pm0.14$  & $0.60\pm0.22$& 2.50 & 2.37 &  $2.52$ \\
HE~1104--1805       &  $6.90\pm0.30$  & $0.38\pm0.27$& 4.87  &4.78 &  $5.96$  
\enddata
\tablenotetext{a}{X-ray luminosities in 0.2--2 keV, 2--10 keV, and 10--50 keV band (rest frame) integrated from the best fit spectral model corrected for foreground absorption.}
\end{deluxetable}

\clearpage

\clearpage

\clearpage

\clearpage

\clearpage

\clearpage

\clearpage

\clearpage

\clearpage

\clearpage

\clearpage

\clearpage

\clearpage

\clearpage

\clearpage

\end{document}